\documentclass{pasa}%
\usepackage{graphicx}
\usepackage{amsmath}	
\usepackage{amssymb}	
\usepackage{rotating}   
\usepackage{subcaption} 

\extrafloats{60}

\newcommand{\myaffil}[1]{$^{\rm #1}$}
\newcounter{inst}
\newcommand{\inst}[1]{\noindent%
   \refstepcounter{inst}\myaffil{\arabic{inst}\label{#1}}     
   }

\def\polyflux{{\sc Polygon\_Flux}}
\def\polysummary{Polygons drawn over GLEAM images to measure source and background flux densities for }
\def\polysuffix{ }
\def\spectrasummary{Spectral fitting over the GLEAM band }
\def\spectraE{and Effelsberg 2.695\,GHz data for }
\def\spectraM{and MGPS 843\,MHz data for }
\def\spectraA{where $\nu>150$\,MHz for }
\def\spectraN{for }
\def\spectrasuffix{ }
\newcommand{\perbeam}{\,beam\ensuremath{^{-1}}}
\newcommand{\DMunit}{\,cm\ensuremath{^{-3}}pc}
\newcommand{\SBunit}{\,W\,m\ensuremath{^{-2}}\,Hz\ensuremath{^{-1}}\,sr\ensuremath{^{-1}}}

\newcommand{\Fig}{Fig.}

\newcommand{\Sect}{Section}
\newcommand{\Sects}{Sections}

\newcommand{\Tab}{Table}

\newcommand{\eqn}{equation}

\newcommand{\Eqn}{Equation}
\newcommand{\Eqns}{Equations}

\newcommand{\farcm}{\mbox{\ensuremath{.\mkern-4mu^\prime}}}
\newcommand{\fdg}{\mbox{\ensuremath{.\!\!^\circ}}}
\title[New SNRs detected in GLEAM]{New candidate radio supernova remnants detected in the GLEAM survey over $345^\circ < l < 60^\circ$, $180^\circ < l < 240^\circ$}

\author[Hurley-Walker~et~al.]{N.~Hurley-Walker\myaffil{\ref{ICRAR}},
M.~D.~Filipovi\'c\myaffil{\ref{WSU}},
B.~M.~Gaensler\myaffil{\ref{ASTRO3D},\ref{Toronto}},
D.~A.~Leahy\myaffil{\ref{Calgary}},
P.~J.~Hancock\myaffil{\ref{ICRAR},\ref{CAASTRO}},
T.~M.~O.~Franzen\myaffil{\ref{ASTRON}},
A.~R.~Offringa\myaffil{\ref{ASTRON}},
J.~R.~Callingham\myaffil{\ref{ASTRON}},
L.~Hindson\myaffil{\ref{Hert}},
C.~Wu\myaffil{\ref{UWA}},
M.~E.~Bell\myaffil{\ref{UTS}},
B.-Q.~For\myaffil{\ref{ASTRO3D},\ref{UWA}},
M.~Johnston-Hollitt\myaffil{\ref{ICRAR}},
A.~D.~Kapi\'nska\myaffil{\ref{NRAO}},
J.~Morgan\myaffil{\ref{ICRAR}},
T.~Murphy\myaffil{\ref{USyd},\ref{CAASTRO}},
B.~McKinley\myaffil{\ref{ICRAR}},
P.~Procopio\myaffil{\ref{CAASTRO},\ref{UMelb}},
L.~Staveley-Smith\myaffil{\ref{ASTRO3D},\ref{UWA}}
R.~B.~Wayth\myaffil{\ref{ICRAR},\ref{CAASTRO}},
Q.~Zheng\myaffil{\ref{SHAO}} \\
{\small \myaffil{}\,Email: nhw@icrar.org}\\
{\small \inst{ICRAR}\,International Centre for Radio Astronomy Research, Curtin University, Bentley, WA 6102, Australia}\\
{\small \inst{WSU}\,Western Sydney University, Locked Bag 1797, Penrith NSW 2751, Australia}\\
{\small \inst{ASTRO3D}\,ARC  Centre  of  Excellence  for  All  Sky  Astrophysics  in  3  Dimensions  (ASTRO  3D)}\\
{\small \inst{Toronto}\,Dunlap Institute for Astronomy and Astrophysics, 50 St. George St, University of Toronto, ON M5S 3H4, Canada}\\
{\small \inst{Calgary}\,Department of Physics and Astronomy, Science B 605, University of Calgary, 2500 University Dr NW, Calgary, AB T2N 1N4 Canada}\\
{\small \inst{CAASTRO}\,ARC Centre of Excellence for All-sky Astrophysics (CAASTRO)}\\
{\small \inst{ASTRON}\,ASTRON, Netherlands Institute for Radio Astronomy, Oude Hoogeveensedijk 4, 7991 PD, Dwingeloo, The Netherlands}\\
{\small \inst{Hert}\,Centre for Astrophysics Research, School of Physics, Astronomy and Mathematics, University of Hertfordshire, College Lane, Hatfield AL10 9AB, UK}\\
{\small \inst{UWA}\,International Centre for Radio Astronomy Research, University of Western Australia, Crawley 6009, Australia}\\
{\small \inst{UTS}\,University of Technology Sydney, 15 Broadway, Ultimo NSW 2007, Australia}\\
{\small \inst{NRAO}\,National Radio Astronomy Observatory, P.O. Box O, Socorro, NM 87801, USA}\\
{\small \inst{USyd}\,Sydney Institute for Astronomy, School of Physics, The University of Sydney, NSW 2006, Australia}\\
{\small \inst{UMelb}\,School of Physics, The University of Melbourne, Parkville, VIC 3010, Australia}\\
{\small \inst{SHAO}\,Shanghai Astronomical Observatory, 80 Nandan Rd, Xuhui Qu, Shanghai Shi, China, 200000}\\
\\
}

\jid{PASA}
\doi{10.1017/pas.\the\year.xxx}
\jyear{\the\year}

\usepackage{aas_macros}
\usepackage{hyperref}
\hypersetup{colorlinks,citecolor=blue,linkcolor=blue,urlcolor=blue}

\begin{document}

\begin{frontmatter}
\maketitle

\begin{abstract}
We have detected 27~new supernova remnants (SNRs) using a new data release of the GLEAM survey from the Murchison Widefield Array (MWA) telescope, including the lowest surface-brightness SNR ever detected, G\,$0.1-9.7$. Our method uses spectral fitting to the radio continuum to derive spectral indices for 26/27 candidates, and our low-frequency observations probe a steeper-spectrum population than previously discovered. None of the candidates have coincident \textit{WISE} mid-IR emission, further showing that the emission is non-thermal. Using pulsar associations we derive physical properties for six candidate SNR, finding G\,$0.1-9.7$ may be younger than 10\,kyr. 60\,\% of the candidates subtend areas larger than 0.2\,deg$^{2}$ on the sky, compared to $<25$\,\% of previously-detected SNRs. We also make the first detection of two SNRs in the Galactic longitude range $220^\circ$--$240^\circ$.
\end{abstract}

\begin{keywords}
ISM: individual objects: G0.1-9.7, G2.1+2.7, G7.4+0.3, G18.9-1.2, G19.1-3.1, G19.7-0.7, G20.1-0.2, G21.8+0.2, G23.1+0.1, G24.0-0.3, G25.3-1.8, G28.3+0.2, G28.7-0.4, G35.3-0.0, G230.4+1.2, G232.1+2.0, G349.1-0.8, G350.7+0.6, G350.8+5.0, G351.0-0.6, G351.4+0.4, G351.4+0.2, G351.9+0.1, G353.0+0.8, G355.4+2.7, G356.5-1.9, G358.3-0.7, ISM: supernova remnants, radio continuum: ISM, supernovae: general
\end{keywords}
\end{frontmatter}

\section{INTRODUCTION }\label{sec:intro}

 Statistical studies of Galactic supernova rates have predicted the presence of $\approx$1000--2000 supernova remnants (SNRs) \citep{1991ApJ...378...93L,1994ApJS...92..487T}, but searches have found fewer than 300 \citep{2014BASI...42...47G}. This deficit is likely due to the observational selection effects, which discriminate against the identification both of old, faint, large SNRs, and young, small SNRs. The discovery of these ``missing'' SNRs, and hence a characterisation of the full SNR population, is crucial for understanding the production and energy density of Galactic cosmic rays and the overall energy budget of the interstellar medium (ISM).
 
 Historically, most SNRs have been detected using radio observations of their synchrotron radiation, which arises from shock acceleration of ionised particles in the compressed magnetic fields of the expanding supernova shell. SNRs persist for $<10^6$\,yr, until their expansion velocity matches that of the surrounding ISM, and the remnant becomes indistinguishable from its background \citep{1988ApJ...334..252C, 2017ApJS..230....2B}.
 
 Several recent radio surveys have been successful in detecting tens of SNRs at a time: \cite{2006ApJ...639L..25B} detected 35~candidate SNRs in $4\fdg5<l<22\fdg0$, $|b|<1\fdg25$ using the Very Large Array at 330\,MHz; \cite{2006AJ....131.2525H} proposed 41~candidates over $5^\circ<l<32^\circ$, $|b| < 0\fdg8$ via the Multi-Array Galactic Plane Imaging Survey (MAGPIS) survey at 1.4\,GHz, although many of these have since been found to be thermally-emitting \textsc{Hii} regions (see \citealt{2009AJ....138.1615J}, \citealt{2017A+A...605A..58A}, and Hurley-Walker et al. submitted); most recently, \cite{2017A+A...605A..58A} located 76~candidate SNRs over $17\fdg5<l<67\fdg4$ using continuum 1--2\,GHz data from The HI, OH, Recombination (THOR) line survey of the Milky Way.
 
 The Murchison Widefield Array \citep[MWA; ][]{2013PASA...30....7T} is a low-frequency Square Kilometer Array precursor operating in Western Australia. Its observing band of 80--300\,MHz, wide field-of-view, and Southern location make it ideal for searching for radio emission from supernova remnants. This work uses the GaLactic and Extragalactic All-sky MWA \citep[GLEAM; ][]{2015PASA...32...25W} data to blindly search for and detect 27~new candidate SNRs. It is the second of two papers derived from the GLEAM Galactic data release of Hurley-Walker et al. (submitted). \Sect~2 of Paper~\textsc{I} (Hurley-Walker et al. submitted), details the methodology and data used to confirm or invalidate candidate SNR, and also the methods used in this work. \Sect~\ref{sec:pulsars} discusses the utility of pulsar associations in determining distances to our candidate SNRs, \Sect~\ref{sec:new_snr} examines each candidate in detail, \Sect~\ref{sec:discussion} demonstrates some statistics of the detections, and \Sect~\ref{sec:conclusions} concludes with ideas for future work. 

\section{Pulsar association}\label{sec:pulsars}

In order to determine physical characteristics of our detected candidate SNRs, we require a distance measurement. \cite{2015A&ARv..23....3D} discuss various ways to obtain distance estimates for radio-detected SNR, including kinematic measurements of spectral lines (e.g. \textsc{Hi}, \textsc{CO}) emitted or absorbed by the SNR or the surrounding ISM, or combining X-ray temperature measurements with physical modeling of the expansion of the remnant. A method that does not rely on additional observations is to search for nearby pulsars which may have formed at the same time as the SNR. The Australia Telescope National Facility pulsar catalogue v1.59 \citep{2005AJ....129.1993M}\footnote{\href{http://www.atnf.csiro.au/research/pulsar/psrcat}{atnf.csiro.au/research/pulsar/psrcat/}} provides a comprehensive list of known pulsars and is used throughout this work to find potentially associated pulsars.

Pulsar distance calculations make use of the pulsar dispersion measures (DMs) and the Galactic electron density model of \cite{2017ApJ...835...29Y}. The uncertainties on the DMs are often small but the electron density model has unknown errors, so distances are usually quoted without uncertainties. Pulsars are also useful for estimating the age of SNRs; a ``characteristic'' or ``spin-down'' age $\tau$ can be calculated from the pulsar period $P$ and period derivative $\dot{P}$ ($\frac{dP}{dt}$) via:
\begin{equation}
    \tau = \frac{2P}{\dot{P}}
\end{equation}
These ages often have large uncertainties; for instance \cite{2001ApJ...560..371K} find a discrepant factor of 13 between the age derived from self-consistent properties of a young SNR and the characteristic age of the pulsar at its centre, while \cite{2009ApJ...706.1316H} find a factor of more than five difference between PSR~J\,$1747-2958$'s characteristic age and that derived from its motion from its birth site. These factors arise from the non-zero spin period of pulsars at birth, and potentially also from the evolution of their magnetic fields and internal structures over time, particularly for young pulsars.

It is also possible, even probable, that a pulsar and SNR may simply lie along a common line-of-sight and be completely unrelated; \cite{1995MNRAS.277.1243G} simulated pulsar and SNR surveys and found that up to two-thirds of alignments will be purely by chance geometrical alignment. Coincidence may not even occur for truly related systems: \cite{2002ApJ...568..289A} found that $\approx10$\,\% of pulsars younger than 20\,kyr will appear to lie outside of their host remnants, under standard assumptions for SNR expansion and pulsar spin-down. In an ideal case, the proper motion of a pulsar is also known, allowing one to determine whether its path is consistent with birth at the centre of the SNR, but of the 2,659 known pulsars, only 274 have measured proper motion.

To estimate the chance of accidental association through geometric alignment, we calculate the pulsar spatial density as a function of Galactic latitude for the two Galactic longitude regions considered in this paper, and multiply this by the area subtended by a given candidate SNR. Note that even low-probability chance geometric alignments may still be purely by chance, given that we consider 27~candidates. For each SNR, unless we note a pulsar association, we consider all potentially-associated pulsars too old, or there are none in the area.

\section{New candidate SNRs}\label{sec:new_snr}

By-eye inspection of the longitude range of this data release revealed many new candidate SNRs, the majority lying in the longitude range around the Galactic centre. To filter \textsc{Hii} regions and other chance Galactic emission, we ran the flux density measurement and spectral fitting process described in \Sects~2.3 and 2.4 of Paper~\textsc{I}, and visually inspected the \textit{WISE} images from the same region. This produced a list of 27~new candidate SNRs.

To quantify the likelihood of each candidate being a SNR, we use a similar classification scheme to \cite{2006ApJ...639L..25B}, examining whether each candidate has:
\begin{itemize}
\item{a full shell or shell-like morphology (as determined by at least one of GLEAM, E11, or MGPS);}
\item{a conclusively non-thermal spectrum (negative $\alpha$ where $S\propto\nu^\alpha$);}
\item{no correlated emission in the \textit{WISE} 8-, 12- or 22-$\mu$m bands.}
\end{itemize}
Candidates in which we have strong confidence we denote Class~\textsc{I}; candidates in which we are fairly confident, but are not certain, are denoted Class~\textsc{II}; faint, confused, or complex candidates, particularly those where we cannot measure a spectrum, are designated Class~\textsc{III}. There are 16, 9, and 2~candidates in each class, respectively.

 As most of these new candidate SNRs have low surface brightness and steeper spectra than are typical for known SNRs (\Tab~\ref{tab:results}), there are often limited resources on which to draw in the literature; see \Sect~2 of Paper~\textsc{I} for a description of ancillary data available to assist classification of the various SNRs.

In turn, we examine the members of each class of SNR candidates, drawing on other published data where possible. We proceed in order of Galactic longitude, as shown in \Tab~\ref{tab:results}. Where pulsar associations make possible the calculation of physical parameters, we summarise them in \Tab~\ref{tab:phys_results}, including also the chance of accidental association (see \Sect~\ref{sec:pulsars}).

\begin{table*}[]
    \centering
    \begin{tabular}{ccccccccccccc}
    \hline
    
    Name & RA & Dec & a & b & PA & $S_\mathrm{200MHz}$ & $\alpha$ & Ancillary & Morphology & Class  \\
         & (J2000) & (J2000)    & $'$ & $'$  & $^\circ$ & Jy & & data  &  &  \\
         1 & 2 & 3 & 4 & 5 & 6 & 7 & 8 & 9 & 10 & 11 \\
\hline
\hyperref[G0.1-9.7]{G\,$0.1-9.7$} & 18 25 50 & $-33$ 30 & 66 & 66 & 0 & $2.3\pm0.2$ & $-1.1\pm0.1$ & -- & Filled \& partial shell & \textsc{II} \\
\hyperref[G2.1+2.7]{G\,$2.1+2.7$} & 17 40 10 & $-25$ 39 & 72 & 62 & 0 & $ 9\pm 1$ & $-0.19\pm0.06$ & E11 & Shell? & \textsc{II} \\
\hyperref[G7.4+0.3]{G\,$7.4+0.3$} & 18 01 06 & $-22$ 21 & 18 & 14 & 90 & $2.3\pm0.3$ & $-0.8\pm0.2$ & -- & Shell & \textsc{II} \\
\hyperref[G18.9-1.2]{G\,$18.9-1.2$} & 18 30 04 & $-13$ 00 & 68 & 60 & 355 & $9.0\pm0.8$ & $-1.1\pm0.2$ & -- & Shell & \textsc{I} \\
\hyperref[G19.1-3.1]{G\,$19.1-3.1$} & 18 37 19 & $-13$ 41 & 32 & 32 & 0 & $2.4\pm0.3$ & $-0.6\pm0.2$ & E11 & Shell & \textsc{I} \\
\hyperref[G19.7-0.7]{G\,$19.7-0.7$} & 18 29 35 & $-12$ 03 & 28 & 28 & 0 & $7.0\pm0.3$ & $-0.24\pm0.05$ & E11 & Shell & \textsc{I} \\
\hyperref[G20.1-0.2]{G\,$20.1-0.2$} & 18 28 47 & $-11$ 27 & 38 & 38 & 0 & -- & -- & -- & Partial shell & \textsc{III} \\
\hyperref[G21.8+0.2]{G\,$21.8+0.2$} & 18 30 15 & $-09$ 47 & 64 & 42 & 320 & $37\pm 1$ & $-0.61\pm0.05$ & E11 & Filled & \textsc{I} \\
\hyperref[G23.1+0.1]{G\,$23.1+0.1$} & 18 32 43 & $-08$ 38 & 26 & 26 & 0 & $17.3\pm0.4$ & $-0.64\pm0.05$ & E11 & Shell & \textsc{I} \\
\hyperref[G24.0-0.3]{G\,$24.0-0.3$} & 18 36 26 & $-08$ 01 & 48 & 48 & 0 & $41\pm 1$ & $-0.87\pm0.05$ & E11 & Shell & \textsc{I} \\
\hyperref[G25.3-1.8]{G\,$25.3-1.8$} & 18 44 18 & $-07$ 35 & 76 & 94 & 35 & $17.0\pm0.5$ & $-0.45\pm0.03$ & E11 & Shell & \textsc{I} \\
\hyperref[G28.3+0.2]{G\,$28.3+0.2$} & 18 42 22 & $-03$ 58 & 14 & 14 & 0 & $4.2\pm0.3$ & $-0.7\pm0.1$ & -- & Shell & \textsc{I} \\
\hyperref[G28.7-0.4]{G\,$28.7-0.4$} & 18 45 30 & $-03$ 54 & 10 & 10 & 0 & $3.7\pm0.1$ & $-0.51\pm0.06$ & E11 & Shell & \textsc{I} \\
\hyperref[G35.3-0.0]{G\,$35.3-0.0$} & 18 56 02 & $02$ 09 & 26 & 22 & 5 & $12.9\pm0.4$ & $-0.39\pm0.06$ & -- & Partial shell & \textsc{II} \\
\hyperref[G230.4+1.2]{G\,$230.4+1.2$} & 07 28 57 & $-14$ 56 & 54 & 40 & 60 & $3.5\pm0.1$ & $-0.60\pm0.07$ & E11 & Filled & \textsc{I} \\
\hyperref[G232.1+2.0]{G\,$232.1+2.0$} & 07 35 08 & $-16$ 03 & 50 & 76 & 340 & $7.2\pm0.1$ & $-0.58\pm0.02$ & E11 & Filled & \textsc{I} \\
\hyperref[G349.1-0.8]{G\,$349.1-0.8$} & 17 20 24 & $-38$ 31 & 14 & 14 & 0 & $3.7\pm0.1$ & $-0.83\pm0.07$ & MGPS & Shell & \textsc{II} \\
\hyperref[G350.7+0.6]{G\,$350.7+0.6$} & 17 18 53 & $-36$ 17 & 56 & 80 & 43 & $34\pm 1$* & $-0.69\pm0.07$* & -- & Partial shell & \textsc{II} \\
\hyperref[G350.8+5.0]{G\,$350.8+5.0$} & 17 01 52 & $-33$ 40 & 72 & 52 & 35 & $16.5\pm0.4$ & $-0.27\pm0.06$ & -- & Filled & \textsc{II} \\
\hyperref[G351.0-0.6]{G\,$351.0-0.6$} & 17 25 07 & $-36$ 49 & 12 & 12 & 0 & $0.50\pm0.04$ & $-0.64\pm0.09$ & MGPS & Partial shell & \textsc{II} \\
\hyperref[G351.4+0.4]{G\,$351.4+0.4$} & 17 21 31 & $-35$ 53 & 9 & 9 & 0 & $3.35\pm0.09$ & $-0.42\pm0.07$ & MGPS & Shell & \textsc{I} \\
\hyperref[G351.4+0.2]{G\,$351.4+0.2$} & 17 22 45 & $-35$ 59 & 18 & 14 & 20 & $1.8\pm0.1$ & $-0.9\pm0.1$ & MGPS & Partial shell & \textsc{II} \\
\hyperref[G351.9+0.1]{G\,$351.9+0.1$} & 17 24 14 & $-35$ 40 & 20 & 16 & 0 & $4.4\pm0.2$ & $-0.98\pm0.07$ & MGPS & Shell & \textsc{I} \\
\hyperref[G353.0+0.8]{G\,$353.0+0.8$} & 17 24 46 & $-34$ 21 & 96 & 66 & 20 & $16.5\pm0.4$* & $-1.0\pm0.1$* & -- & Partial shell & \textsc{III} \\
\hyperref[G355.4+2.7]{G\,$355.4+2.7$} & 17 23 28 & $-31$ 16 & 22 & 22 & 0 & $1.5\pm0.2$ & $-0.8\pm0.2$ & -- & Filled & \textsc{I} \\
\hyperref[G356.5-1.9]{G\,$356.5-1.9$} & 17 44 55 & $-32$ 54 & 36 & 48 & 40 & $14.9\pm0.3$ & $-0.71\pm0.05$ & -- & Filled & \textsc{I} \\
\hyperref[G358.3-0.7]{G\,$358.3-0.7$} & 17 44 46 & $-30$ 43 & 34 & 42 & 354 & $21.8\pm0.3$* & $-0.8\pm0.1$* & -- & Partial shell & \textsc{III} \\
\hline
    \end{tabular}
    \caption{Measured properties of the SNRs detected in this work. Columns are as follows: (1) Name derived from Galactic co-ordinates via $lll.l\pm b.b$; (2, 3): Right Ascension and declination in J2000 co-ordinates; (4, 5): major and minor axes in arcminutes; (6) position angle CCW from North; (7, 8) 200-MHz flux density and spectral index derived from a spectral fit: a ``*'' indicates that the fit was made only to data where $\nu>150$\,MHz; (9) Ancillary data used for the fit in addition to the GLEAM measurements: E11 indicates the Bonn 11\,cm survey at 2.695\,GHz \protect\citep{1984A&AS...58..197R} and MGPS indicates the Molonglo Galactic Plane Survey at 843\,MHz \protect\citep{1999ApJS..122..207G,2007MNRAS.382..382M,2014PASA...31...42G}; (10) Morphology via visual inspection (``Shell'' indicates a complete ring of enhanced emission; ``partial shell'' indicates part thereof; ``filled'' indicates an elliptical region of enhanced brightness without clear edge-brightening); (11) our confidence in the reality of the candidate from greatest (\textsc{I}) to least (\textsc{III}); see \Sect~\ref{sec:new_snr} for more details.}
    \label{tab:results}
\end{table*}

\begin{table*}[]
    \centering
    \begin{tabular}{cccccccccc}
\hline
    Name     & Associated          &  Coincidence  &   Likelihood  & Distance & $a$ & $b$ & PSR age & SNR age & Stage \\
             & pulsar        &  chance (\%)  &    of assoc.  &    (kpc)   &  (pc) &  (pc) &  (kyr)    &   (kyr)   &       \\
             1 & 2 & 3 & 4 & 5 & 6 & 7 & 8 & 9 & 10 \\
\hline
  \hyperref[G0.1-9.7]{G\,$0.1-9.7$}    & PSR~J\,$1825-33$ & 5         & good         & 1.24     & 24  & 24  & --      & 1--9     & free / S-T \\
  \hyperref[G21.8+0.2]{G\,$21.8+0.2$} & PSR~J\,$1831-0952$ & 95       & good         & 3.68     & 165  & 45  & 128    & 40--120     & radiative \\
  \hyperref[G230.4+1.2]{G\,$230.4+1.2$} & PSR~J\,$0729-1448$ & 4       & good         & 2.68     & 47  & 31  & 35    & 17--48     & S-T \\
  \hyperref[G232.1+2.0]{G\,$232.1+2.0$} & PSR~J\,$0734-1559$ & 3      & good         & --      & --  & --  & 197    & --     & S-T \\
  \hyperref[G356.5-1.9]{G\,$356.5-1.9$} & PSR~J\,$1746-3239$ & 57      & marginal     & --       & --  & --  & 482    & --     & -- \\
  \hyperref[G358.3-0.7]{G\,$358.3-0.7$} & PSR~B\,$1742-30$   & 79      & marginal     & 2.64     & 32  & 26  & 550    & 10--18 & S-T \\
         \hline
    \end{tabular}
    \caption{Physical properties of those SNR candidates for which pulsar associations can be made. References and discussion can be found in the relevant sections for each candidate. Columns are as follows: (1) Name of the candidate; (2) Name of the most likely associated pulsar; (3) Chance of a single pulsar lying inside the shell of the candidate (see \Sect~\ref{sec:pulsars}); (4) Qualitative assessment of the likelihood that the remnant and pulsar are associated; (5) Pulsar distance, if known; (6,7) Derived remnant major and minor axes; (8) Pulsar characteristic age; (9) SNR age derived from $a$, $b$, and relevant expansion equation; (10) Estimated stage of SNR lifecycle; ``free'' indicates free expansion
    (\eqn~1 of Paper~\textsc{I})
    , ``S-T'' indicates adiabatic Sedov-Taylor expansion
    (\eqn~4 of Paper~\textsc{I})
    , while ``radiative'' indicates the radiative phase.}
    \label{tab:phys_results}
\end{table*}

\subsection{G\,0.1-9.7}\label{G0.1-9.7}

G\,$0.1-9.7$ is primarily visible as a bright arc of non-thermal emission with a total flux density of $S_\mathrm{200MHz}=2.3\pm0.2$\,Jy and a steep spectral index of $\alpha=-1.1\pm0.2$ (\Tab~\ref{tab:results}). An increased brightness level is also visible within an ellipse extrapolated from this arc, lending some weight to the potential of this SNR candidate, but as the ellipse is faint and the spectrum of the arc is very steep, we class this only as level \textsc{II}. The ellipse has too low a brightness to obtain a spectral index across the GLEAM band, but using the lowest frequency band we measure a surface brightness of $5.4\times10^{-22}$\SBunit. Extrapolating this to 1\,GHz using $\alpha=-1.1$, we obtain $3\times10^{-23}$\SBunit, making G\,$0.1-9.7$ the lowest surface brightness SNR ever detected.

Within the ellipse of G\,$0.1-9.7$, \cite{2010MNRAS.402..855B} detected the rotating radio transient (RRAT) PSR~J1825-33. They detected seven sequential pulses with a period of $P=1.27$\,s, implying a nulling rate of $\approx97$\,\%, and were unable to estimate $\dot{P}$; the pulsar has not subsequently been re-detected. If PSR~J1825-33 and G\,0.1-9.7 are associated, then the pulsar's distance of $1.24$\,kpc (from DM$=43\pm2$\DMunit) would imply a remnant diameter of 24\,pc. The pulsar position uncertainty is $15'$, so it would lie $3\pm9$\,pc from the center.

The western shock front may indicate initial interaction with the ISM, while the SNR continues to expand freely in the other directions. Using the measured remnant radius and assuming $M_\mathrm{ejecta}=M_\odot$ and $E=10^{51}$\,erg,  \Eqns~1 and 2 of Paper~\textsc{I} yield a SNR age of $\approx1,200$\,years, potentially lying within human recorded history.
Alternatively, the SNR could be in the Sedov-Taylor phase; reversing \Eqn~4 of Paper~\textsc{I} and assuming $n_\mathrm{H}=1$\,cm$^{-3}$ and $E=10^{51}$\,erg, we can estimate the age of the SNR as $\approx$9,000 years. However, the SNR lies far off the Galactic Plane and is potentially in an underdense region of the ISM, which if properly accounted for would reduce the age calculated by this method, so we could consider 9,000\,yr an upper limit.

\begin{figure*}
   \centering
   \includegraphics[width=\textwidth]{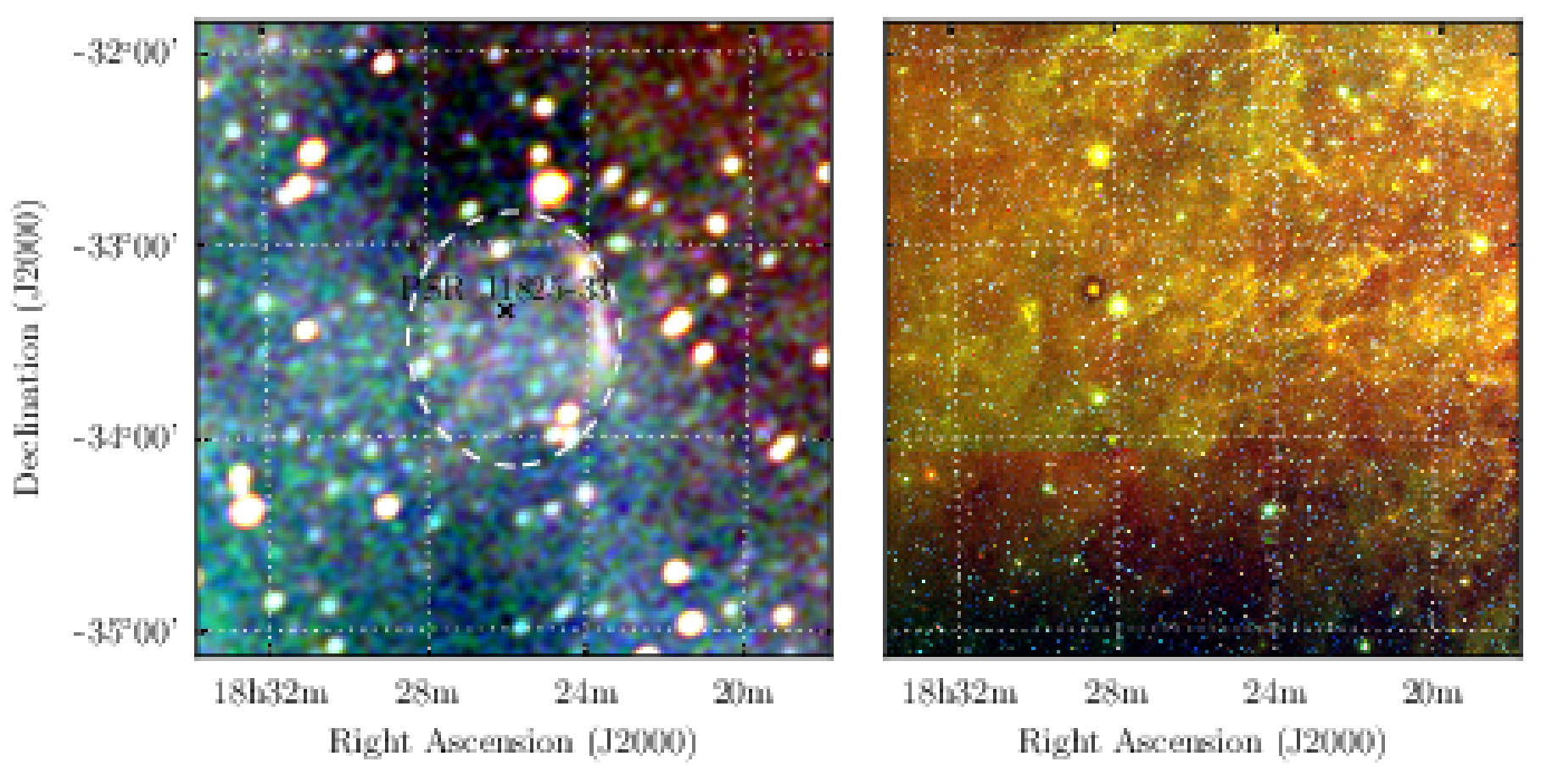}
   \caption{G\,$0.1-9.7$ as observed by GLEAM (left) at 72--103\,MHz (R), 103--134\,MHz (G), and 139--170\,MHz (B), and by \textit{WISE} (right) at 22\,$\mu$m (R), 12\,$\mu$m (G), and 4.6\,$\mu$m (B). The colour scales for the GLEAM RGB cube are -1.0---0.1, -0.5---0.0, and -0.2--0.1\,Jy\perbeam for R, G, and B, respectively. The dashed white ellipse indicates the position of the SNR candidate, and black crosses indicate nearby pulsars.}
    \label{fig:SNR_G0.1-9.7}
\end{figure*}

\subsection{G\,2.1+2.7}\label{G2.1+2.7}

G\,2.1+2.7 is visible as a very faint shell in both the GLEAM and E11 data. It is contaminated by 13 (presumably) extragalactic radio sources which are catalogued by Hurley-Walker et al. (submitted). Of these, 11~sources are bright enough to fit SEDs over the GLEAM band, and subtract them from the data. The total flux density subtracted at 200\,MHz is $S_\mathrm{200MHz}=2.86$\,Jy, with median $\alpha=-0.93$, compared to a (corrected) total remnant flux density of $S_\mathrm{200MHz}= 9\pm 1$\,Jy, and $\alpha=-0.19\pm0.06$. We class this remnant only as Class \textsc{II}.

\begin{figure*}
   \centering
   \includegraphics[width=\textwidth]{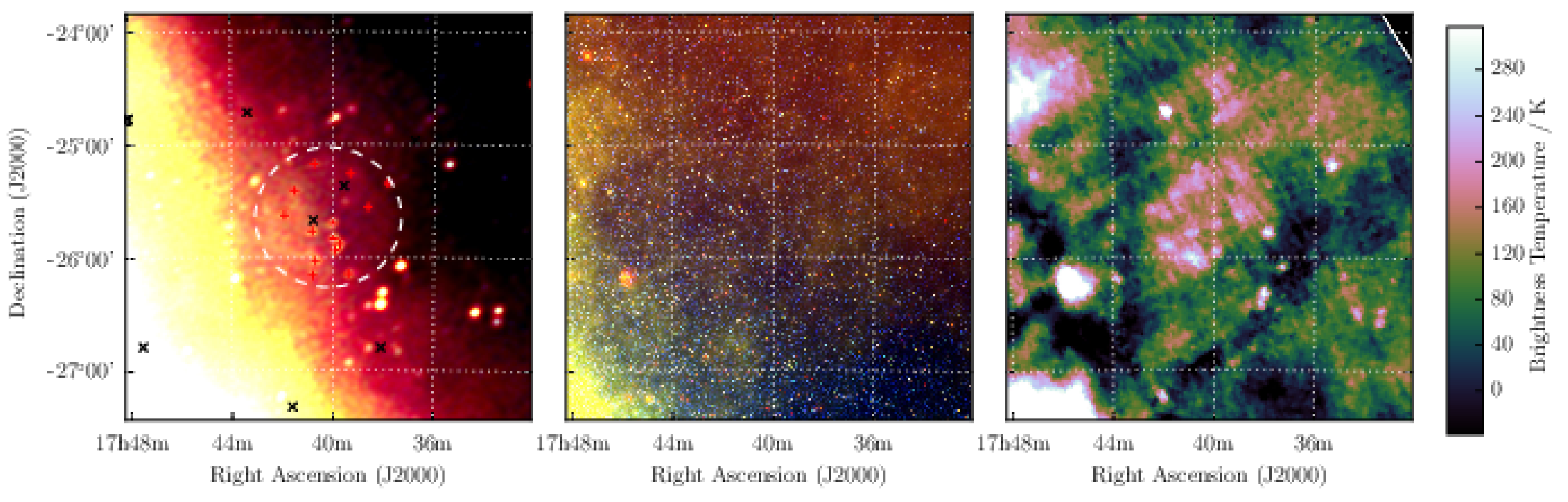}
   \caption{G\,$2.1+2.7$ as observed by GLEAM (left) at 72--103\,MHz (R), 103--134\,MHz (G), and 139--170\,MHz (B), by \textit{WISE} (middle) at 22\,$\mu$m (R), 12\,$\mu$m (G), and 4.6\,$\mu$m (B), and Effelsberg at 2.695\,GHz (right). The colour scales for the GLEAM RGB cube are 0.8--2.8, 0.8--1.5, and 0.0--1.1\,Jy\perbeam for R, G, and B, respectively. Annotations are as in \Fig~\ref{fig:SNR_G0.1-9.7}, and red crosses mark the positions of subtracted extragalactic radio sources (see \Sect~\ref{G2.1+2.7}).}
    \label{fig:SNR_G2.1+2.7}
\end{figure*}

\subsection{G\,7.4+0.3}\label{G7.4+0.3}

\begin{figure*}
   \centering
   \includegraphics[width=\textwidth]{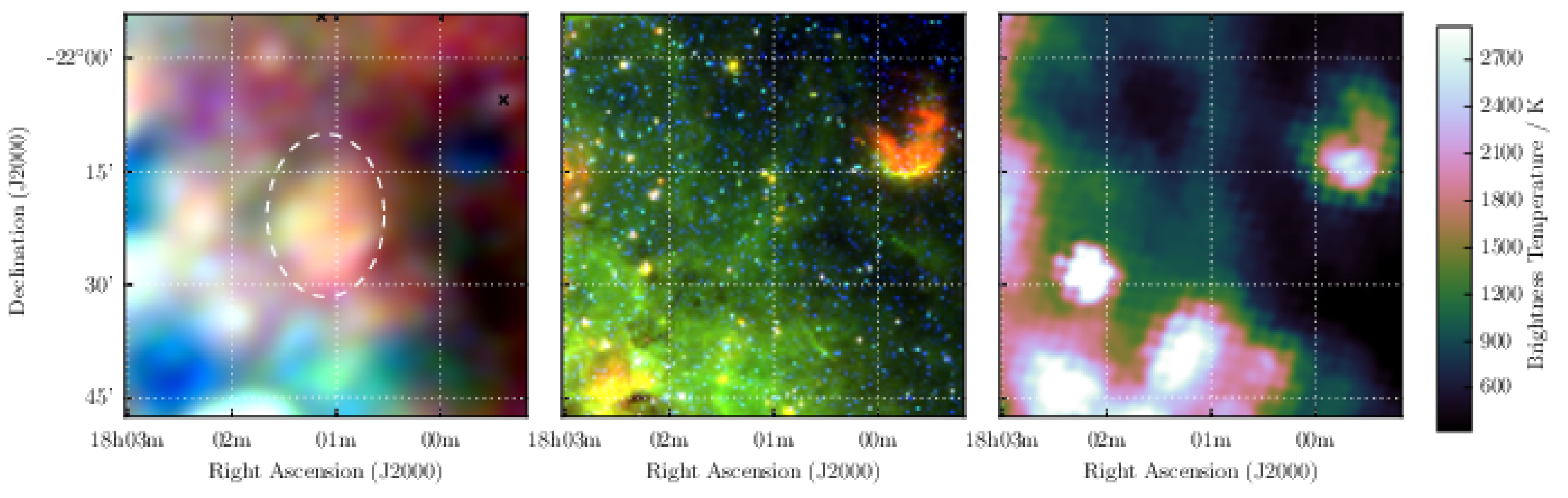}
   \caption{G\,$7.4+0.3$ as observed by GLEAM (left) at 72--103\,MHz (R), 103--134\,MHz (G), and 139--170\,MHz (B), by \textit{WISE} (middle) at 22\,$\mu$m (R), 12\,$\mu$m (G), and 4.6\,$\mu$m (B), and Effelsberg at 2.695\,GHz (right).  The colour scales for the GLEAM RGB cube are 4.7--6.8, 2.5--3.7, and 1.0--1.8\,Jy\perbeam for R, G, and B, respectively. Annotations are as in \Fig~\ref{fig:SNR_G0.1-9.7}. }
    \label{fig:SNR_G7.4+0.3}
\end{figure*}

G\,$7.4+0.3$ appears as a slightly irregular ellipse with a potentially filled shell, but this is difficult to assess given its small size ($\approx16'$) relative to the resolution of GLEAM. There is no corresponding 12- or 22-$\mu$m emission visible in the \textit{WISE} image of this region (\Fig~\ref{fig:SNR_G7.4+0.3}), nor is it bright enough to be seen with E11. It has a total flux density of $2.3\pm0.3$\,Jy and a non-thermal spectral index of $-0.8\pm0.2$, so we conclude that it is a reasonable candidate and denote it Class~\textsc{II}.

\subsection{G\,18.9-1.2}\label{G18.9-1.2}

G\,$18.9-1.2$ appears as a pair of distinct arcs encircling an elliptical region of increased brightness. Superposed is the well-studied SNR G\,18.9-1.1 \citep{1985Natur.314..720F}; the centres of the SNRs are $10'$ apart. We are only able to measure a total flux density for the portion of G\,$18.9-1.2$ which is not confused with G\,$18.9-1.1$, and find $S_\mathrm{200MHz}=9.0\pm0.8$\,Jy, and $\alpha-1.1\pm0.2$. The spectral index of G\,$18.9-1.1$ over the same band is $-0.43\pm0.04$.

Potentially, the two are a single object, with G\,$18.9-1.1$ the central PWNe and G\,$18.9-1.2$ the original progenitor shell, and the apparent difference in central positions the result of the decreased ISM density away from the Galactic plane allowing a highly asymmetric expansion of G\,$18.9-1.2$. However, the filled shell of G\,$18.9-1.1$ is quite clearly delineated, and the jump in spectral indices is quite abrupt, so we postulate that the SNRs are unrelated. Given the clear shell structure and non-thermal spectrum of the SNR, we are confident in the detection and denote it Class~\textsc{I}.

\begin{figure*}
   \centering
   \includegraphics[width=\textwidth]{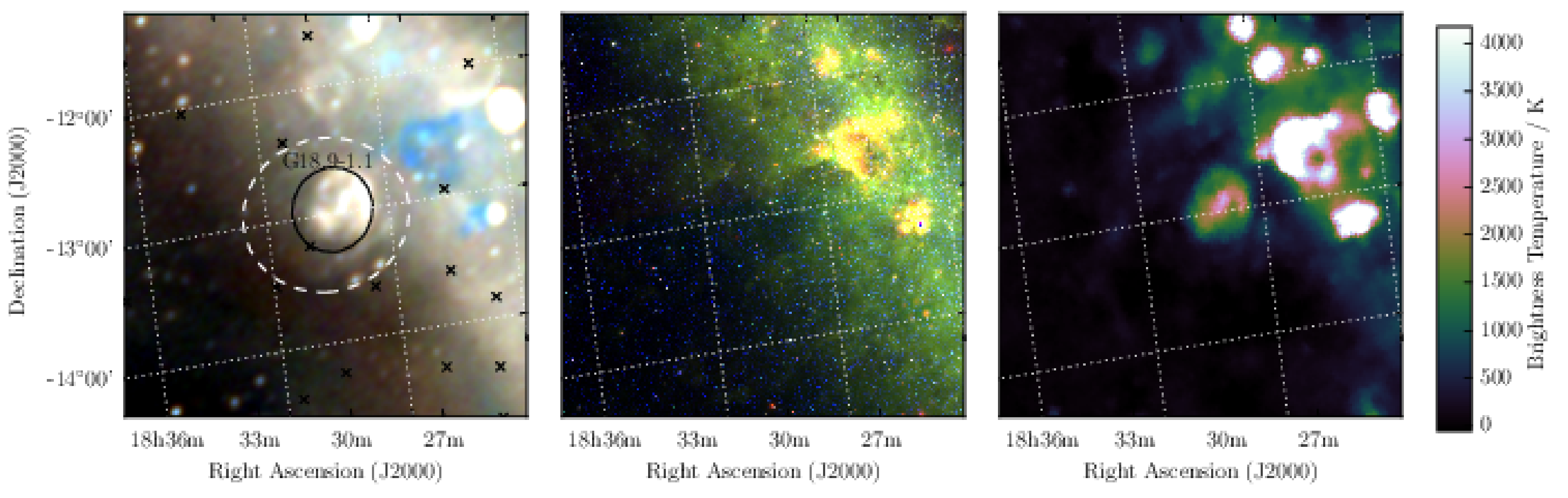}
   \caption{G\,$18.9-1.2$ as observed by GLEAM (left) at 72--103\,MHz (R), 103--134\,MHz (G), and 139--170\,MHz (B), by \textit{WISE} (middle) at 22\,$\mu$m (R), 12\,$\mu$m (G), and 4.6\,$\mu$m (B), and Effelsberg at 2.695\,GHz (right). The colour scales for the GLEAM RGB cube are 0.7--6.7, 0.0--3.3, and -0.1--1.8\,Jy\perbeam for R, G, and B, respectively. Annotations are as in \Fig~\ref{fig:SNR_G0.1-9.7}, and a black ellipse indicates a known SNR intersecting this candidate (see \Sect~\ref{G18.9-1.2}). }
    \label{fig:SNR_G18.9-1.2}
\end{figure*}

\subsection{G\,19.1-3.1}\label{G19.1-3.1}

\begin{figure*}
   \centering
   \includegraphics[width=\textwidth]{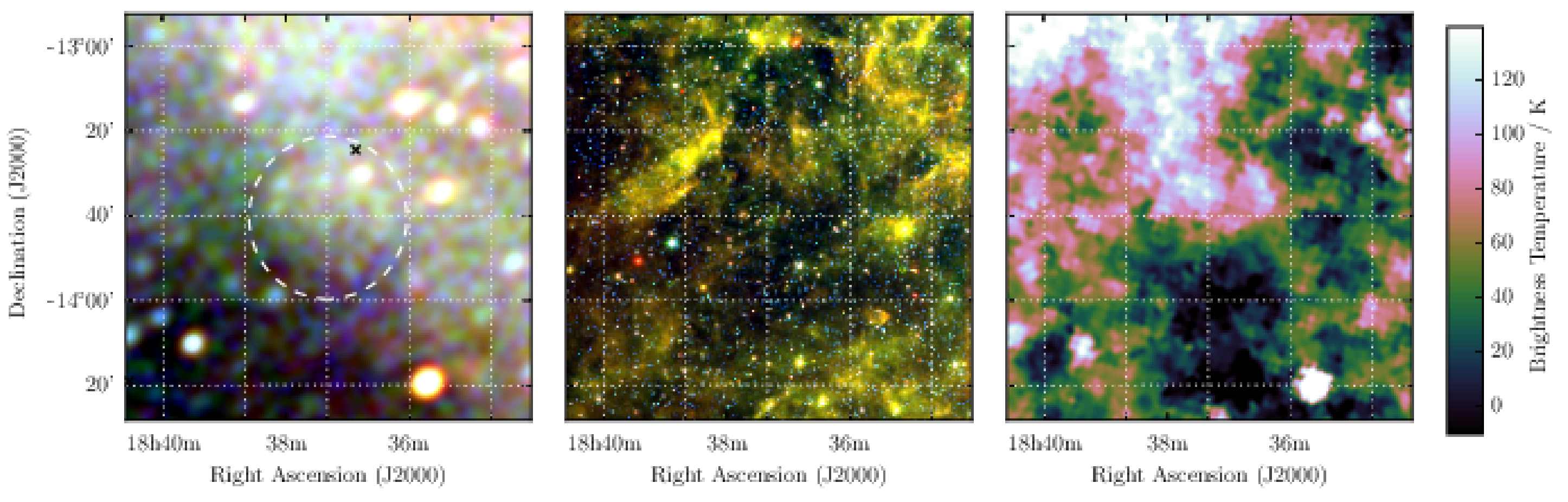}
   \caption{G\,$19.1-3.1$ as observed by GLEAM (left) at 72--103\,MHz (R), 103--134\,MHz (G), and 139--170\,MHz (B), by \textit{WISE} (middle) at 22\,$\mu$m (R), 12\,$\mu$m (G), and 4.6\,$\mu$m (B), and Effelsberg at 2.695\,GHz (right). The colour scales for the GLEAM RGB cube are -0.1--1.3, -0.3--0.4, and -0.2--0.2\,Jy\perbeam for R, G, and B, respectively. Annotations are as in \Fig~\ref{fig:SNR_G0.1-9.7}.}
    \label{fig:SNR_G19.1-3.1}

\end{figure*}

G\,$19.1-3.1$ is relatively faint and small compared to most SNRs in this work (\Tab~\ref{tab:results}), but does appear as a nearly-complete shell with distinct edges on all but the south-east region (\Fig~\ref{fig:SNR_G19.1-3.1}). The (presumably) unrelated extragalactic radio sources GLEAM\,J183811-133837 and GLEAM\,J183646-133024 lie on the edge of the shell; care was taken to exclude them from the measurement region (\Fig~\ref{fig:SNR_G19.1-3.1_poly}). With the GLEAM measurements and an additional measurement from Effelsberg at 2.695\,GHz, we derive $S_\mathrm{200MHz}=2.4\pm0.3$\,Jy and $\alpha=-0.6\pm0.2$.  We denote the candidate Class~\textsc{I} based solely on its intrinsic properties.



\subsection{G\,19.7-0.7}\label{G19.7-0.7}

G\,$19.7-0.7$ appears in the GLEAM data as a complete shell with thicker arcs in the north-west and south-east. The (presumably) unrelated extragalactic radio source GLEAM\,J182921-120914 lies just within the shell; care was taken to avoid including it in the measurements. Including a measurement from Effelsberg, we derive $S_\mathrm{200MHz}=7.0\pm0.3$\,Jy and $\alpha=-0.24\pm0.05$. \cite{2017A+A...605A..58A} also detect this SNR and name it G\,$19.75-0.69$. We are confident that it is a SNR and denote it Class~\textsc{I}. 


\begin{figure*}
   \centering
   \includegraphics[width=\textwidth]{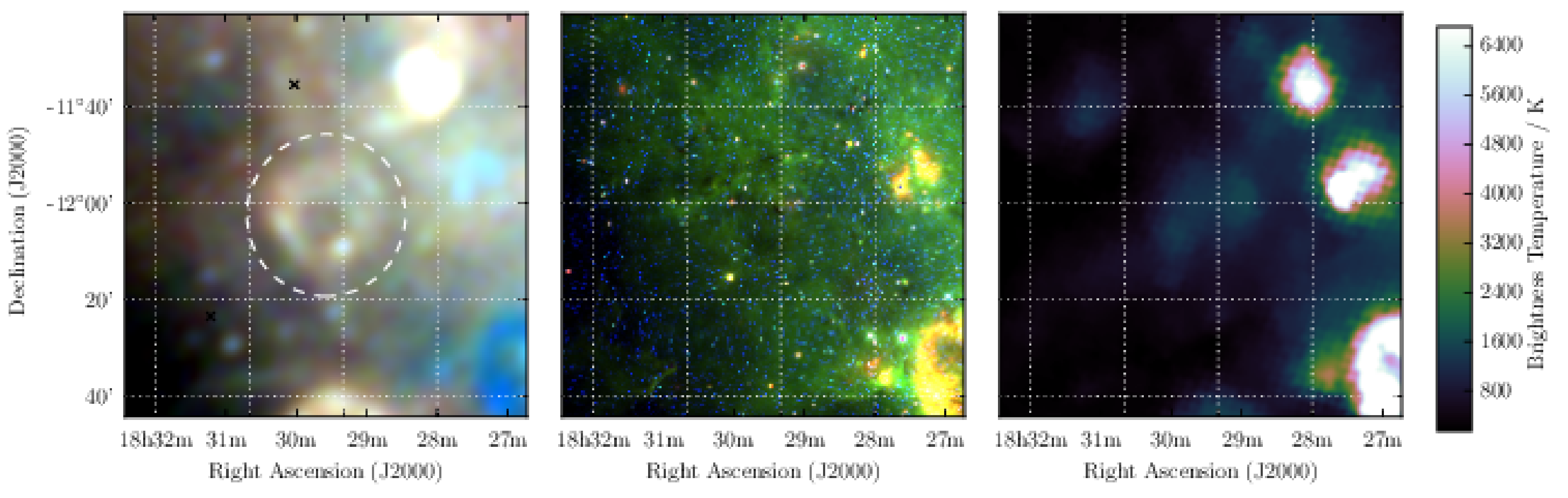}
   \caption{G\,$19.7-0.7$ as observed by GLEAM (left) at 72--103\,MHz (R), 103--134\,MHz (G), and 139--170\,MHz (B), by \textit{WISE} (middle) at 22\,$\mu$m (R), 12\,$\mu$m (G), and 4.6\,$\mu$m (B), and Effelsberg at 2.695\,GHz (right). The colour scales for the GLEAM RGB cube are 2.7--6.8, 1.1--3.3, and 0.4--1.8\,Jy\perbeam for R, G, and B, respectively. Annotations are as in \Fig~\ref{fig:SNR_G0.1-9.7}.}
    \label{fig:SNR_G19.7-0.7}
\end{figure*}

\subsection{G\,20.1-0.2}\label{G20.1-0.2}

\begin{figure*}
   \centering
   \includegraphics[width=\textwidth]{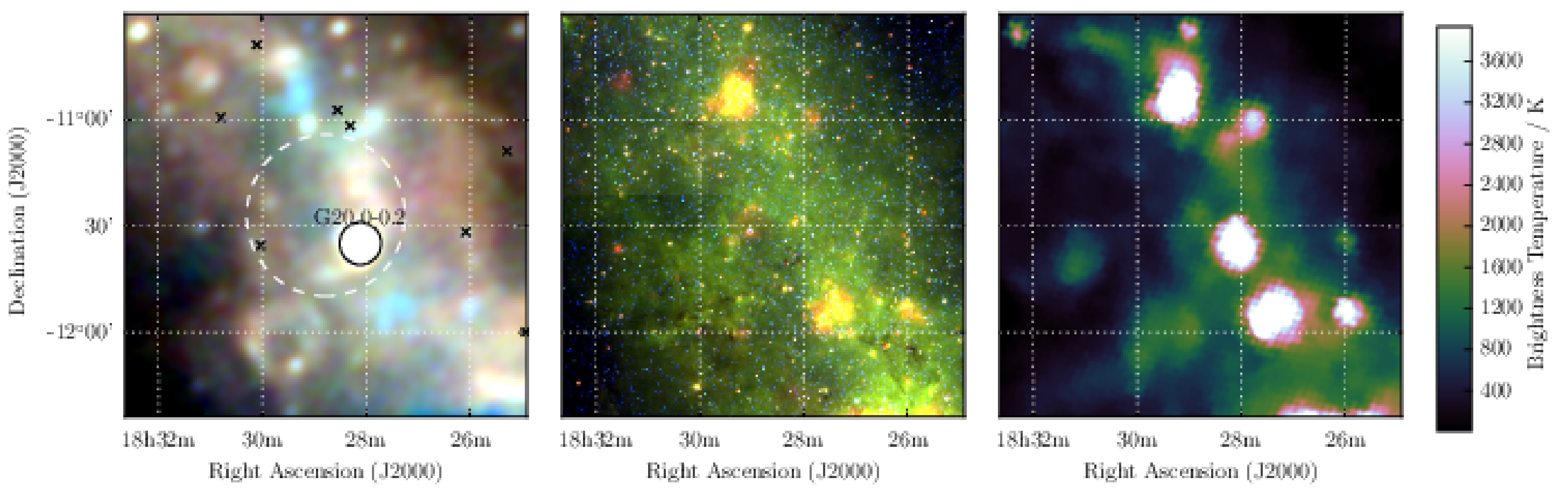}
   \caption{G\,$20.1-0.2$ as observed by GLEAM (left) at 72--103\,MHz (R), 103--134\,MHz (G), and 139--170\,MHz (B), by \textit{WISE} (middle) at 22\,$\mu$m (R), 12\,$\mu$m (G), and 4.6\,$\mu$m (B), and Effelsberg at 2.695\,GHz (right). The colour scales for the GLEAM RGB cube are 3.3--6.7, 1.4--3.4, and 0.5--1.8\,Jy\perbeam for R, G, and B, respectively. Annotations are as in \Fig~\ref{fig:SNR_G0.1-9.7}, and a black ellipse indicates a known SNR intersecting this candidate (see \Sect~\ref{G20.1-0.2}).}
    \label{fig:SNR_G20.1-0.2}
\end{figure*}

G\,$20.1-0.2$ is visible as a partial semi-circular arc bisected by the known SNR G\,$20.0-0.2$. This SNR lies at a distance of 11.2\,kpc \citep{2018AJ....155..204R}; as G\,$20.1-0.2$ has a radius $4\times$ larger, it is unlikely to be at the same distance, as it would be unphysically large. With \textsc{Hii} regions contaminating the centre of the shell, and an extremely confused background (\Fig~\ref{fig:SNR_G20.1-0.2}), we are unable to make reliable flux density measurements of the source. We denote it as merely a potential SNR candidate and Class~\textsc{III}.

\subsection{G\,21.8+0.2}\label{G21.8+0.2}

G\,$21.8+0.2$ is a filled ellipse elongated in the direction of the Galactic Plane. Despite the confusion of the region, we are able to make a reliable flux density measurement by excluding the brighter nearby SNRs from the background measurement. While not visually striking in the Effelsberg data, a measurement can be made, and we thus derive $S_\mathrm{200MHz}=37\pm 1$\,Jy and $\alpha=-0.61\pm0.05$.

Due to its large size, three known pulsars lie within the ellipse; PSR~J1828-1007 and PSR~J1829-1011 lie to the south-west, at the edge of the ellipse, while PSR~J1831-0952 lies closest, 20' southeast from the centroid of the candidate. It has $P\approx67$\,ms and $\dot{P}\approx8\times10^{-15}$\,s\,s$^{-1}$ \citep{2006MNRAS.372..777L}, giving a characteristic age of 130,000\,yr. At a distance of 3.68\,kpc (derived from $DM=247\pm5$\DMunit), an association with the remnant would mean it has major and minor axis sizes of $\approx70\times45$\,pc.

The minor axis size of 45\,pc is consistent with a SNR just entering the radiative phase, where the shocked ISM is able to cool radiatively, and the SNR is driven only by internal pressure instead of the initial kinetic expansion. In this regime, the SNR radius $R$ increases as $t^{2/7}$. The SNR clearly has an elliptical shape, implying an inhomogeneous surrounding ISM. If we make standard assumptions about the SNR energetics and ISM density, and assume adiabatic expansion up to this point, we can use \Eqn~4 of Paper~\textsc{I}
to derive an age estimate of 40,000--120,000 years, for the minor and major axes, respectively. Since the SNR may have been expanding radiatively for some time, these could be underestimates. The upper end of the range is quite close to the pulsar characteristic age. The pulsar natal kick velocity would need to be 170--490\,km\,s$^{-1}$, entirely within the expected range of pulsar kick velocities \citep{2017A&A...608A..57V}.

Given the large angular size of the SNR candidate (subtending 2.3\,deg$^{2}$), we would predict 1--2 pulsars lying inside the ellipse purely by chance, so while the astrophysics of the association are not unreasonable, caution must be advised. Regardless, the SNR candidate is strong based on its intrinsic characteristics and we therefore denote it Class~\textsc{I}.

\begin{figure*}
   \centering
   \includegraphics[width=\textwidth]{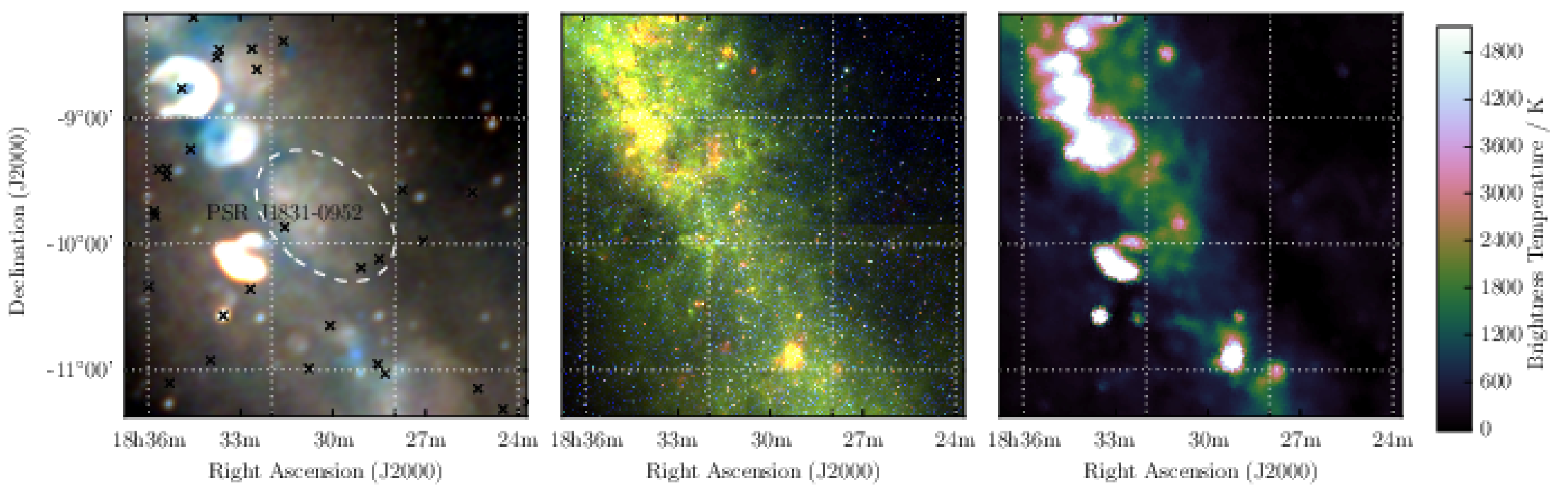}
   \caption{G\,$21.8+0.2$ as observed by GLEAM (left) at 72--103\,MHz (R), 103--134\,MHz (G), and 139--170\,MHz (B), by \textit{WISE} (middle) at 22\,$\mu$m (R), 12\,$\mu$m (G), and 4.6\,$\mu$m (B), and Effelsberg at 2.695\,GHz (right). The colour scales for the GLEAM RGB cube are 2.2--9.4, 0.7--4.8, and 0.2--2.6\,Jy\perbeam for R, G, and B, respectively. Annotations are as in \Fig~\ref{fig:SNR_G0.1-9.7}.}
    \label{fig:SNR_G21.8+0.2}
\end{figure*}

\subsection{G\,23.1+0.1}\label{G23.1+0.1}

\begin{figure*}
   \centering
   \includegraphics[width=\textwidth]{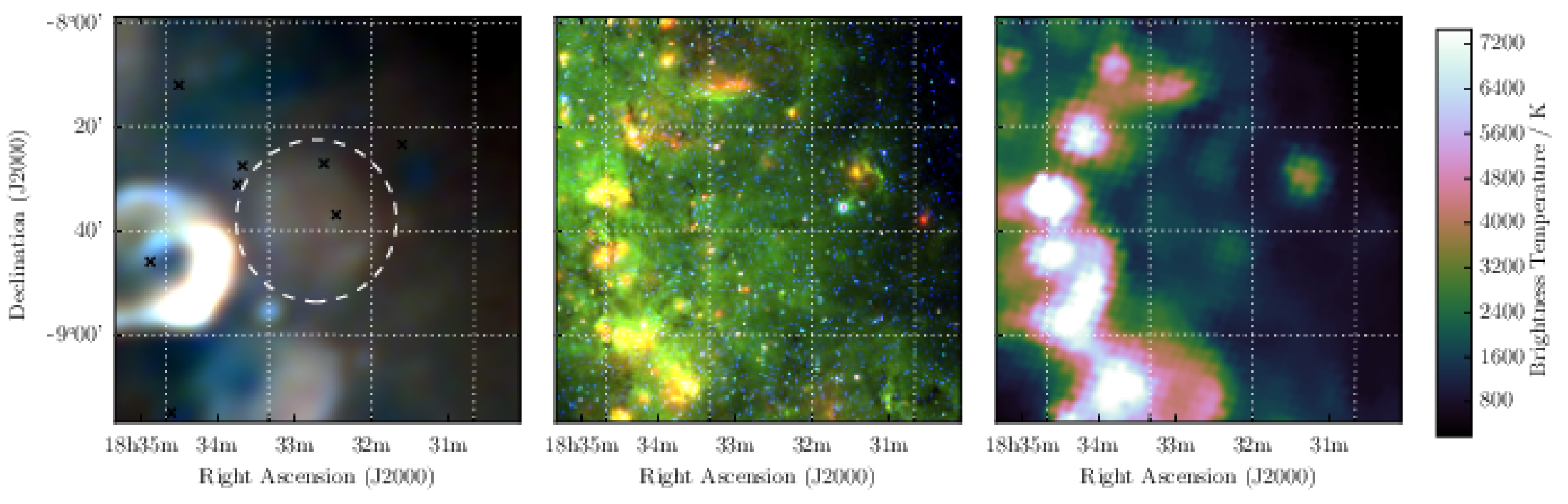}
   \caption{G\,$23.1+0.1$ as observed by GLEAM (left) at 72--103\,MHz (R), 103--134\,MHz (G), and 139--170\,MHz (B), by \textit{WISE} (middle) at 22\,$\mu$m (R), 12\,$\mu$m (G), and 4.6\,$\mu$m (B), and Effelsberg at 2.695\,GHz (right). The colour scales for the GLEAM RGB cube are 3.3--14.5, 1.4--7.7, and 0.6--4.2\,Jy\perbeam for R, G, and B, respectively. Annotations are as in \Fig~\ref{fig:SNR_G0.1-9.7}.}
    \label{fig:SNR_G23.1+0.1}
\end{figure*}

G\,23.1+0.1 is visible as a circular shell with $S_\mathrm{200MHz}=17.3\pm0.4$\,Jy\perbeam and $\alpha=-0.64\pm0.05$ (\Fig~\ref{fig:SNR_G23.1+0.1}). It is also detected in the TeV $\gamma$-rays by the High-Energy Stereoscopic System (HESS) Galactic Plane Survey \citep{2018A&A...612A...1H}, and named HESS\,$1832-085$. This interesting Class~\textsc{I} SNR is also detected by \cite{2017A+A...605A..58A}, and thoroughly discussed in Maxted et al. (submitted).

\subsection{G\,24.0-0.3}\label{G24.0-0.3}

\begin{figure*}
   \centering
   \includegraphics[width=\textwidth]{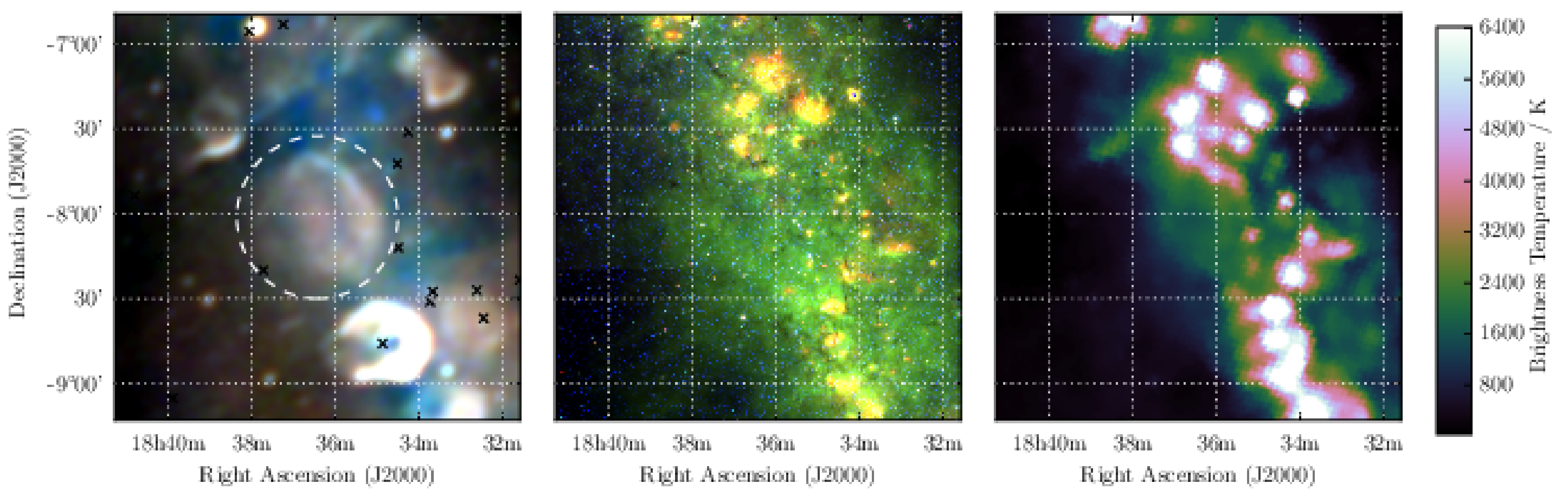}
   \caption{G\,$24.0-0.3$ as observed by GLEAM (left) at 72--103\,MHz (R), 103--134\,MHz (G), and 139--170\,MHz (B), by \textit{WISE} (middle) at 22\,$\mu$m (R), 12\,$\mu$m (G), and 4.6\,$\mu$m (B), and Effelsberg at 2.695\,GHz (right). The colour scales for the GLEAM RGB cube are 2.6--9.2, 0.9--4.9, and 0.3--2.6\,Jy\perbeam for R, G, and B, respectively. Annotations are as in \Fig~\ref{fig:SNR_G0.1-9.7}.}
    \label{fig:SNR_G24.0-0.3}
\end{figure*}

G\,24.0-0.3 is very clearly visible in the GLEAM RGB image (\Fig~\ref{fig:SNR_G24.0-0.3}) and highlights the utility of this view for separating the thermal and non-thermal emission. Despite some contamination from overlapping \textsc{Hii} regions, we are able to obtain $S_\mathrm{200MHz}=41\pm 1$\,Jy and $\alpha=-0.87\pm0.05$; the measurement from Effelsberg has a very large uncertainty so does not constrain the fit. There are hints of edge-brightening in the northwest, and the shell appears to be filled: this could either be from unresolved filamentary structure, or from a central PWNe. The high S/N of this source, clear shell-like morphology, and non-thermal spectrum, suggest strongly that it is a SNR, and we therefore denote it Class~\textsc{I}.


\subsection{G\,25.3-1.8}\label{G25.3-1.8}

G\,25.3-1.8 is spatially the largest candidate discussed in this work, covering 6.2\,deg$^2$. As such it is superposed with the (presumably) unrelated extragalactic radio sources GLEAM\,J184249-075610, GLEAM\,J184309-074224, GLEAM\,J184440-072007, and GLEAM\,J184403-075954, which together comprise $S_\mathrm{200MHz}=6.6$\,Jy, with median $\alpha=-0.95$. Extrapolating each source across the band and subtracting them from the integrated flux density measurements, we fit to the corrected flux densities of G\,25.3-1.8 and find $S_\mathrm{200MHz}=17.0\pm0.5$\,Jy and $\alpha=-0.46\pm0.03$. The SNR is clearly visible and relatively un-confused in the Effelsberg data, which is extremely useful in constraining the fit; we denote the candidate Class~\textsc{I}


\begin{figure*}
   \centering
   \includegraphics[width=\textwidth]{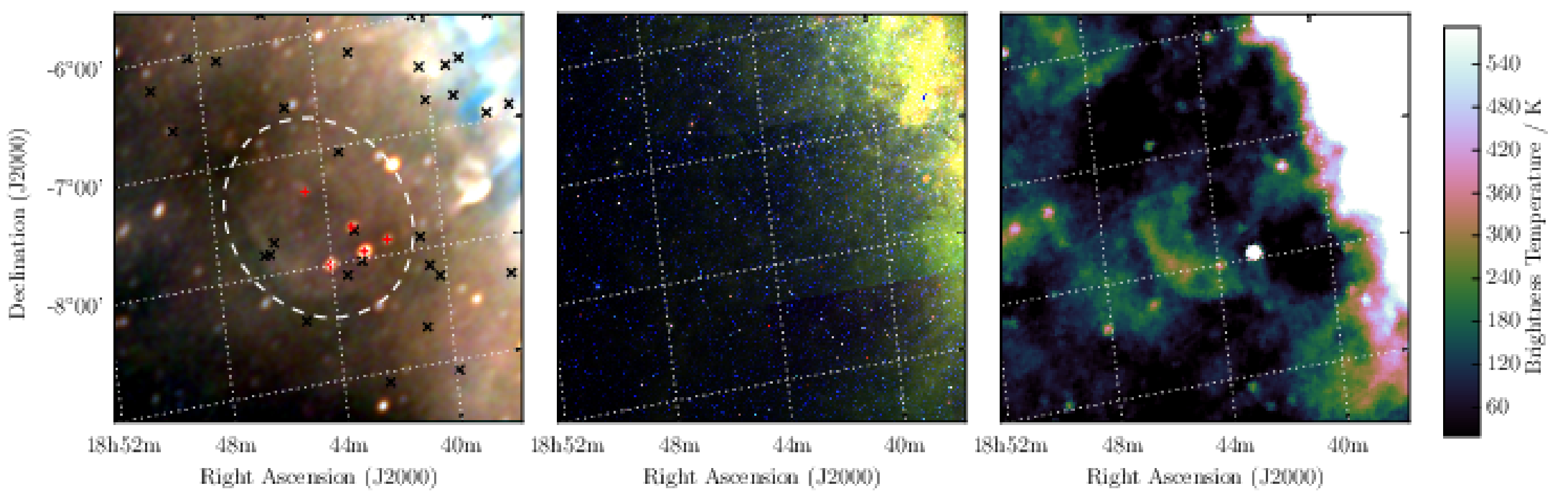}
   \caption{G\,$25.3-1.8$ as observed by GLEAM (left) at 72--103\,MHz (R), 103--134\,MHz (G), and 139--170\,MHz (B), by \textit{WISE} (middle) at 22\,$\mu$m (R), 12\,$\mu$m (G), and 4.6\,$\mu$m (B), and Effelsberg at 2.695\,GHz (right). The colour scales for the GLEAM RGB cube are 0.8--4.9, 0.1--2.6, and 0.0--1.4\,Jy\perbeam for R, G, and B, respectively. Annotations are as in \Fig~\ref{fig:SNR_G0.1-9.7}, and red crosses mark the positions of subtracted extragalactic radio sources (see \Sect~\ref{G25.3-1.8}).}
    \label{fig:SNR_G25.3-1.8}
\end{figure*}

\subsection{G\,28.3+0.2}\label{G28.3+0.2}

\begin{figure*}
   \centering
   \includegraphics[width=\textwidth]{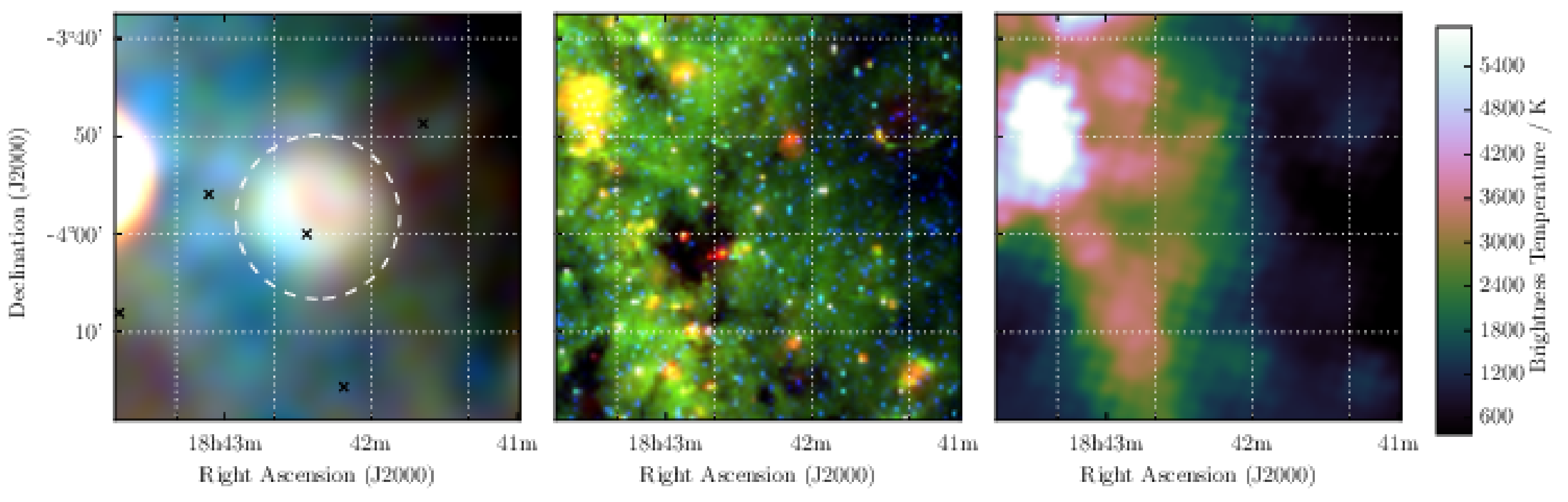}
   \caption{G\,$28.3+0.2$ as observed by GLEAM (left) at 72--103\,MHz (R), 103--134\,MHz (G), and 139--170\,MHz (B), by \textit{WISE} (middle) at 22\,$\mu$m (R), 12\,$\mu$m (G), and 4.6\,$\mu$m (B), and Effelsberg at 2.695\,GHz (right). The colour scales for the GLEAM RGB cube are 3.6--6.5, 1.5--3.4, and 0.7--1.8\,Jy\perbeam for R, G, and B, respectively. Annotations are as in \Fig~\ref{fig:SNR_G0.1-9.7}.}
    \label{fig:SNR_G28.3+0.2}
\end{figure*}

G\,$28.3+0.2$ is compact and circular, and almost unresolved by our observations, and entirely so by Effelsberg (\Fig~\ref{fig:SNR_G28.3+0.2}). Despite the complexity of its surroundings, we are able to derive $S_\mathrm{200MHz}=4.2\pm0.3$\,Jy and $\alpha=-0.7\pm0.1$ purely from the GLEAM measurements (\Tab~\ref{tab:results}). \cite{2017A+A...605A..58A} also detect this candidate and denote it G\,$28.36+0.21$.

PSR~B1839-04 lies very close to the centre of the candidate, but with $P\approx1.8$\,s and $\dot{P}\approx5\times10^{-16}$\,s\,s$^{-1}$ \citep{2004MNRAS.353.1311H}, its characteristic age of 57\,Myr makes a SNR association extremely unlikely. Three other pulsars lie nearby; the youngest appears to be PSR~J1841-0345, with $P\approx204$\,ms and $\dot{P}\approx6\times10^{-14}$\,s\,s$^{-1}$, yielding a characteristic age of 56,000\,yr \citep{2002MNRAS.335..275M}. While this is typical of pulsars associated with SNRs, it is difficult to claim an association given the spatial density of both SNRs and pulsars in this region. The pulsar also lies close to a non-thermal source which is not quite resolved in the GLEAM data. Deeper, higher-resolution observations are required to determine the nature of this source and whether it might be a further candidate SNR. Returning to G\,28.3+0.2, we are confident in its candidacy and denote it Class~\textsc{I}.

\subsection{G\,28.7-0.4}\label{G28.7-0.4}

\begin{figure*}
   \centering
   \includegraphics[width=\textwidth]{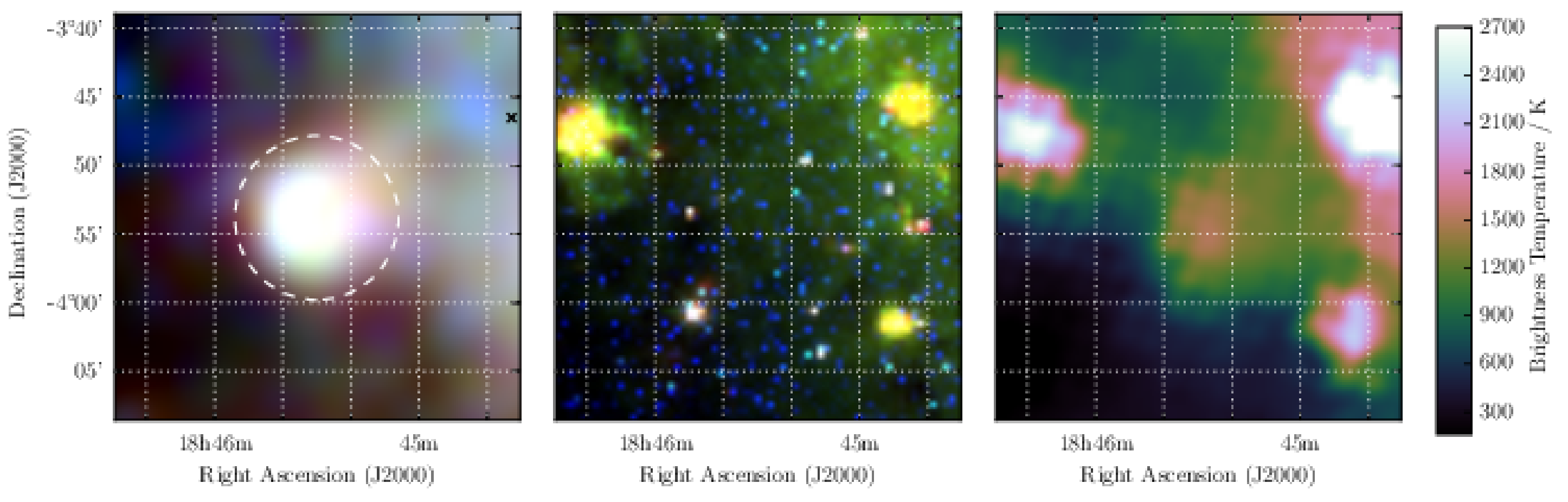}
   \caption{G\,$28.7-0.4$ as observed by GLEAM (left) at 72--103\,MHz (R), 103--134\,MHz (G), and 139--170\,MHz (B), by \textit{WISE} (middle) at 22\,$\mu$m (R), 12\,$\mu$m (G), and 4.6\,$\mu$m (B), and Effelsberg at 2.695\,GHz (right). The colour scales for the GLEAM RGB cube are 3.5--6.3, 1.6--3.0, and 0.7--1.5\,Jy\perbeam for R, G, and B, respectively. Annotations are as in \Fig~\ref{fig:SNR_G0.1-9.7}.}
    \label{fig:SNR_G28.7-0.4}
\end{figure*}

At $10'\times10'$, G\,28.7-0.4 is one of the smallest SNR in the sample, and just visible in the Effelsberg data (\Fig~\ref{fig:SNR_G28.7-0.4}). We thereby derive $S_\mathrm{200MHz}=3.7\pm0.1$\,Jy and $\alpha=-0.51\pm0.06$. No pulsars lie within $2\times$ the shell diameter so we are unable to find an association and derive any physical parameters for this source. \cite{2017A+A...605A..58A} also detect this object, naming it G\,$28.78-0.44$; we are confident that it is a newly-discovered SNR and denote it Class~\textsc{I}.

\subsection{G\,35.3-0.0}\label{G35.3-0.0}

\begin{figure*}
   \centering
   \includegraphics[width=\textwidth]{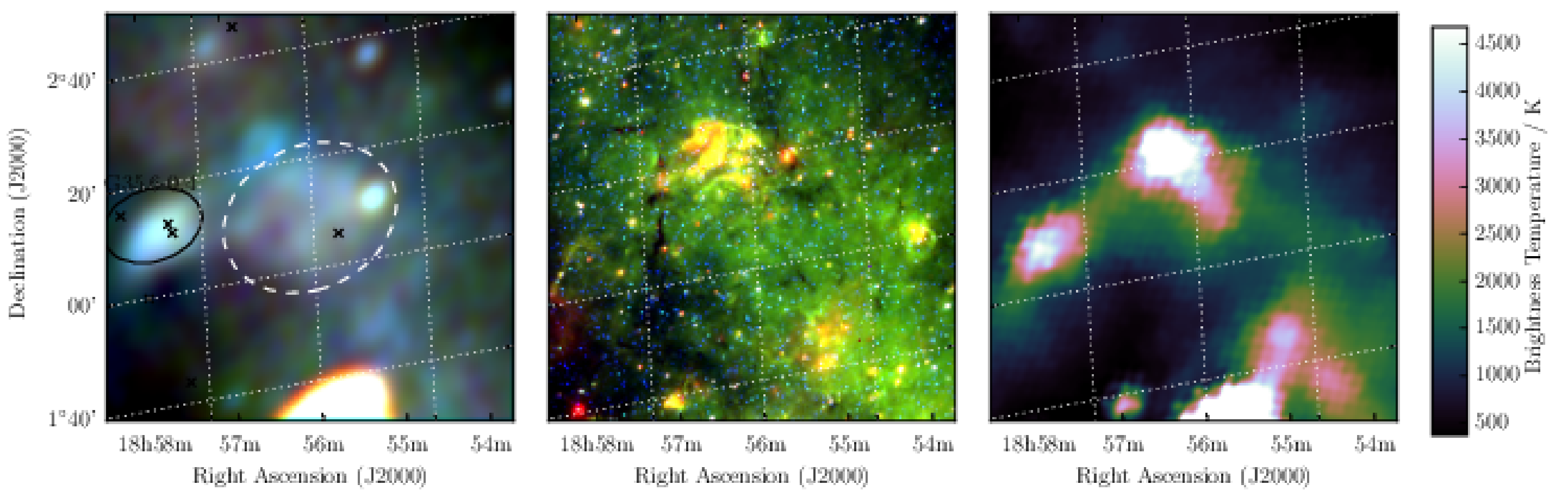}
   \caption{G\,$35.3-0.0$ as observed by GLEAM (left) at 72--103\,MHz (R), 103--134\,MHz (G), and 139--170\,MHz (B), by \textit{WISE} (middle) at 22\,$\mu$m (R), 12\,$\mu$m (G), and 4.6\,$\mu$m (B), and Effelsberg at 2.695\,GHz (right). The colour scales for the GLEAM RGB cube are 2.9--8.7, 1.1--3.8, and 0.4--1.8\,Jy\perbeam for R, G, and B, respectively. Annotations are as in \Fig~\ref{fig:SNR_G0.1-9.7}, and a black ellipse indicates a known nearby SNR (see \Sect~\ref{G35.3-0.0}).}
    \label{fig:SNR_G35.3-0.0}
\end{figure*}

G\,$35.3-0.0$ appears as an irregular elliptical region with a non-thermal spectrum (\Fig~\ref{fig:SNR_G35.3-0.0}). To its immediate east is the SNR\,G\,$35.6-0.4$, and more distantly to the south is the extremely well-studied SNR\,G\,$34.7-0.4$ (W44, 3C\,392). On the northeast edge of the ellipse is a bright \textsc{Hii} region which is clearly visible in the \textit{WISE} data and dominates the Effelsberg image. Care was taken to remove this, and the bright and presumably unrelated radio source to the west, from the measurement of the candidate and background flux densities. From the GLEAM data alone, we derived $S_\mathrm{200MHz}=12.9\pm0.4$\,Jy and $\alpha=-0.39\pm0.06$. Due to its unclear morphology we assign this candidate Class~\textsc{II}, indicating our lowered confidence in a detection. Higher-resolution observations would be useful to disentangle the complexity of this region.


\subsection{G\,230.4+1.2}\label{G230.4+1.2}

\begin{figure*}
   \centering
   \includegraphics[width=\textwidth]{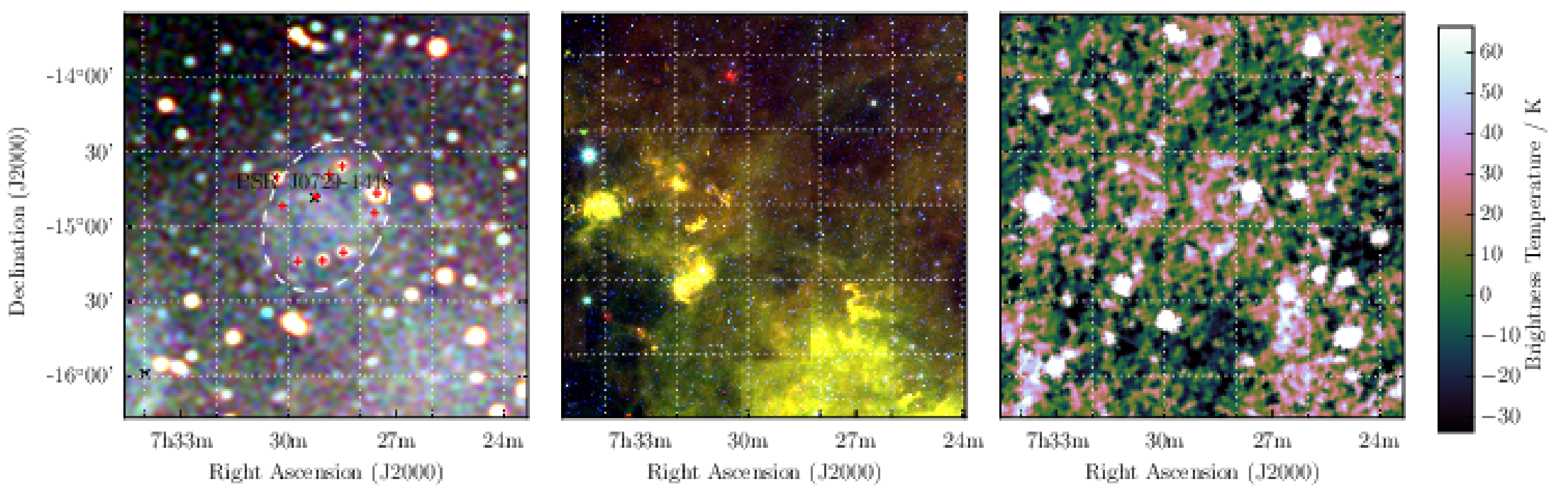}
   \caption{G\,$230.4+1.2$ as observed by GLEAM (left) at 72--103\,MHz (R), 103--134\,MHz (G), and 139--170\,MHz (B), by \textit{WISE} (middle) at 22\,$\mu$m (R), 12\,$\mu$m (G), and 4.6\,$\mu$m (B), and Effelsberg at 2.695\,GHz (right). The colour scales for the GLEAM RGB cube are -0.2--0.4, -0.1--0.1, and -0.1--0.1\,Jy\perbeam for R, G, and B, respectively. Annotations are as in \Fig~\ref{fig:SNR_G0.1-9.7}, and red crosses mark the positions of subtracted extragalactic radio sources (see \Sect~\ref{G230.4+1.2})}
    \label{fig:SNR_G230.4+1.2}
\end{figure*}

G\,$230.4+1.2$ is a very low-S/N irregular ellipse in the outer Galactic plane, just barely visible in the Effelsberg data (\Fig~\ref{fig:SNR_G230.4+1.2}). It is superposed with ten (presumably) unrelated extragalactic radio sources comprising $S_\mathrm{200MHz}=2.7$\,Jy, with median $\alpha=-0.97$. Fitting these sources across the GLEAM band and extrapolating them to the Effelsberg frequency, we subtract them from all measurements and obtain for the candidate $S_\mathrm{200MHz}=3.5\pm0.1$\,Jy and $\alpha=-0.60\pm0.07$.

PSR~J0729-1448 lies within the ellipse, $8\farcm5$ from the centre of the candidate; discovered in the Parkes Multibeam Survey \citep{2002MNRAS.335..275M}, it has $P\approx252$\,ms and $\dot{P}\approx1.13\times 10^{-13}$\,s\,s$^{-1}$, giving it a characteristic age of $35,200$\,yr. Its DM of $91.7$\,cm$^{-3}$\,pc \citep{2013MNRAS.435.1610P} can be used to derive a distance of 2.68\,kpc. If it is associated with the candidate then the ellipse is $47\times31$\,pc in diameter, and the natal kick velocity of the pulsar perpendicular to the line-of-sight would be 180\,km\,s$^{-1}$, which would be fairly typical of pulsar natal velocities \citep{2017A&A...608A..57V}.

Assuming the SNR shares the pulsar's characteristic age, it is likely to be in the Sedov-Taylor phase; for this adiabatic, self-similar expansion, we expect the SNR radius $R$ to be proportional to its age $t$ via
\eqn~4 of Paper~\textsc{I}.
After 35,200 years, a typical SNR in this phase would therefore be $\approx42$\,pc in diameter, which is very close to the value derived from the position association. Given these values, the SNR could also be transitioning to the radiative phase.

Since the chance of finding a known pulsar within this 2\,deg$^2$ candidate is $<5$\,\% for this region, we are fairly confident in this association. The low signal-to-noise of the candidate initially induces caution, but the pulsar association bolsters our confidence, allowing us to denote this candidate Class~\textsc{I}.

\subsection{G\,232.1+2.0}\label{G232.1+2.0}

G\,$232.1+2.0$ appears as an elliptical feature with dimensions $\approx70'\times50'$, and is visible across the entire GLEAM band. It has poorly-defined edges, potentially indicating that it is beginning to merge with the surrounding ISM. Fortunately, it is also visible in the Effelsberg 2.695-GHz data and a joint fit to the data yields $S_\mathrm{200MHz}=7.2\pm0.1$ and $\alpha=-0.58\pm0.02$ (\Tab~\ref{tab:results}).

PSR~J0734-1559 lies 16' from the centre of the candidate. Discovered via its $\gamma$-ray emission by \textit{Fermi-LAT}, it has $P\approx155$\,ms and $\dot{P}=1.25\times 10^{-14}$\,s\,s$^{-1}$, giving it a characteristic age of $197,000$\,yr \citep{2016ApJ...833..271S}. Unfortunately, radio pulsations have not been observed from this pulsar, so its dispersion measure is not known, and no other distance estimates have yet been made.

Similarly to G\,$230.4+1.2$ (\Sect~\ref{G230.4+1.2}), there is $<5$\,\% chance of a pulsar lying within the shell of the candidate purely by chance, so we are reasonably confident in the association, and denote the candidate Class~\textsc{I}.

\begin{figure*}
   \centering
   \includegraphics[width=\textwidth]{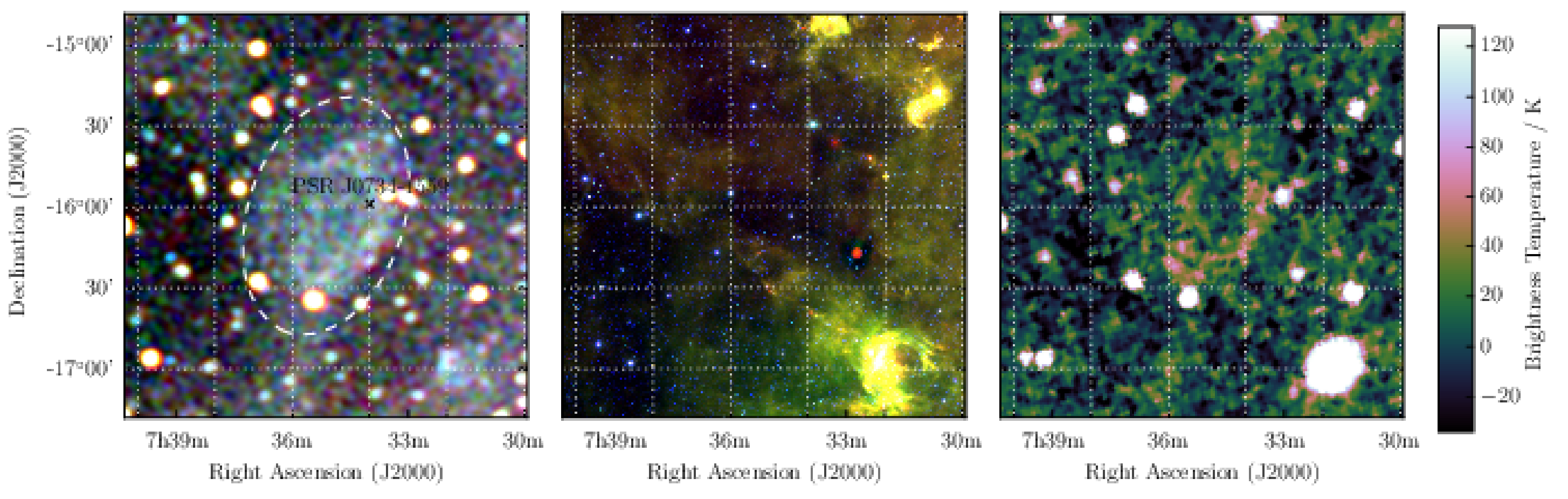}
   \caption{G\,$232.1+2.0$ as observed by GLEAM (left) at 72--103\,MHz (R), 103--134\,MHz (G), and 139--170\,MHz (B), by \textit{WISE} (middle) at 22\,$\mu$m (R), 12\,$\mu$m (G), and 4.6\,$\mu$m (B), and Effelsberg at 2.695\,GHz (right). The colour scales for the GLEAM RGB cube are -0.1--0.3, -0.1--0.1, and 0.0--0.1\,Jy\perbeam for R, G, and B, respectively. Annotations are as in \Fig~\ref{fig:SNR_G0.1-9.7}.}
    \label{fig:SNR_G232.1+2.0}
\end{figure*}

\subsection{G\,349.1-0.8}\label{G349.1-0.8}

\begin{figure*}
   \centering
   \includegraphics[width=\textwidth]{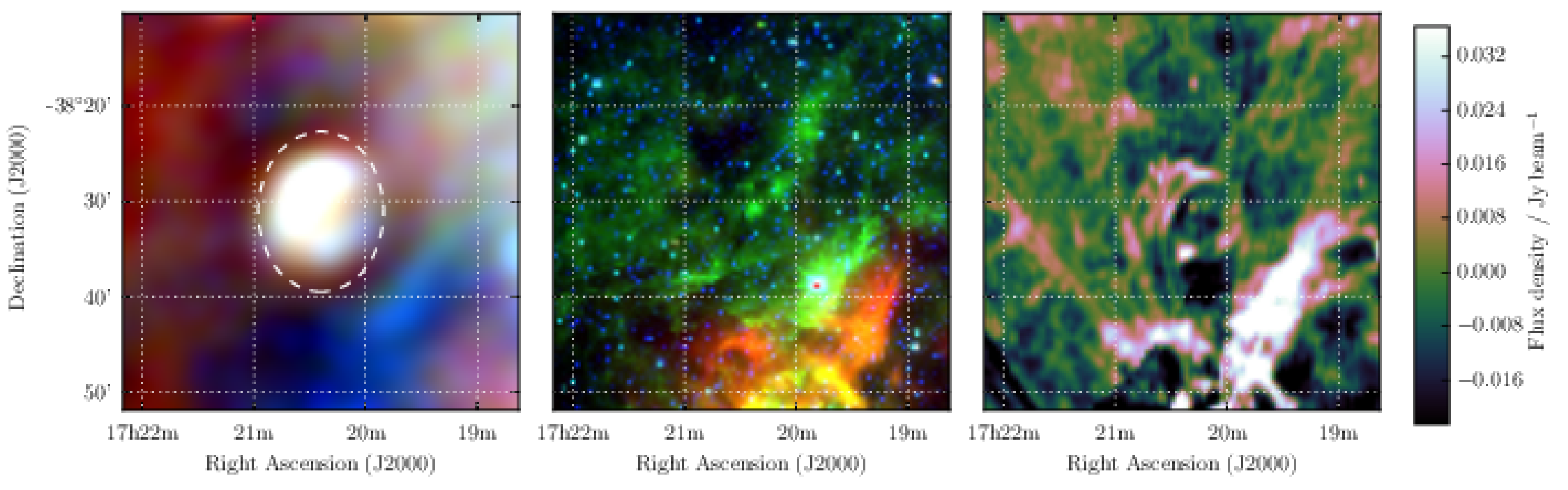}
   \caption{G\,$349.1-0.8$ as observed by GLEAM (left) at 72--103\,MHz (R), 103--134\,MHz (G), and 139--170\,MHz (B), by \textit{WISE} (middle) at 22\,$\mu$m (R), 12\,$\mu$m (G), and 4.6\,$\mu$m (B), and MGPS at 843\,MHz (right). The colour scales for the GLEAM RGB cube are 1.6--4.5, 1.2--2.5, and 0.5--1.2\,Jy\perbeam for R, G, and B, respectively.}
    \label{fig:SNR_G349.1-0.8}
\end{figure*}

G\,$349.1-0.8$ is a compact remnant which is poorly-resolved by the GLEAM data, and has an unclear structure in MGPS due to artefacts from nearby bright sources (\Fig~\ref{fig:SNR_G349.1-0.8}). Despite this we are able to measure $S_\mathrm{200MHz}=3.7\pm0.1$\,Jy and $\alpha=-0.83\pm0.07$ by careful selection of the regions from which to obtain integrated flux densities and background measurements. There are no nearby known pulsars so no physical parameters can be derived. Given the extremely unclear morphology, we denote this source Class~\textsc{II}.

\subsection{G\,350.7+0.6}\label{G350.7+0.6}

\begin{figure*}
   \centering
   \includegraphics[width=\textwidth]{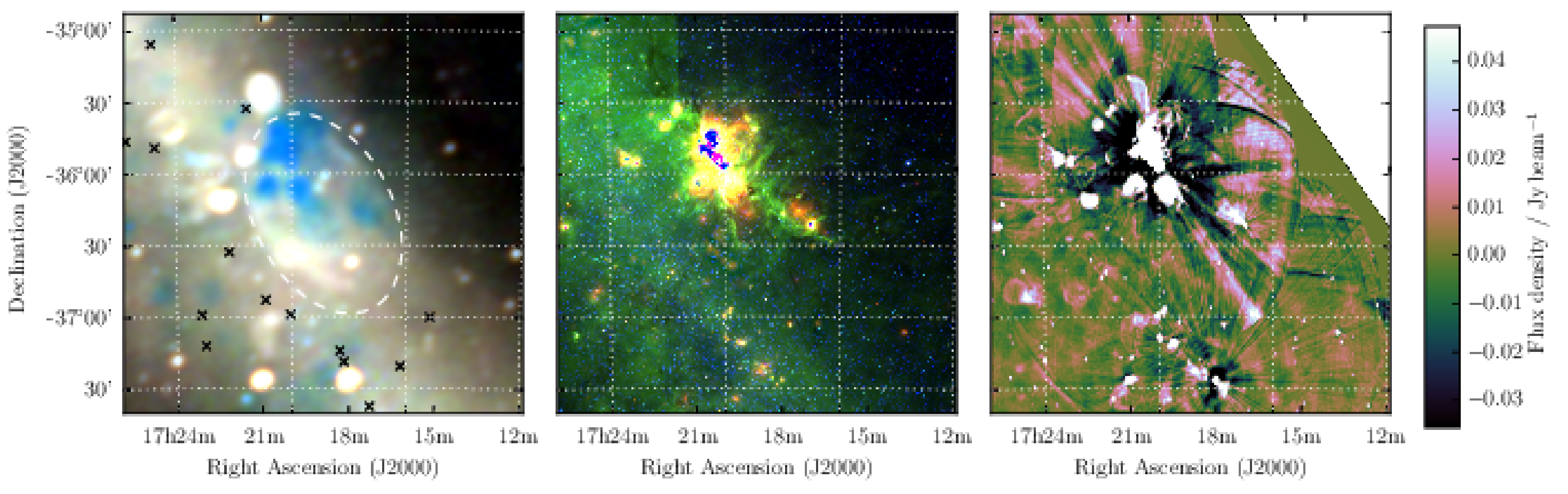}
   \caption{G\,$350.7+0.6$ as observed by GLEAM (left) at 72--103\,MHz (R), 103--134\,MHz (G), and 139--170\,MHz (B), by \textit{WISE} (middle) at 22\,$\mu$m (R), 12\,$\mu$m (G), and 4.6\,$\mu$m (B), and MGPS at 843\,MHz (right). The colour scales for the GLEAM RGB cube are 1.4--8.0, 0.5--4.0, and 0.2--2.0\,Jy\perbeam for R, G, and B, respectively. Annotations are as in \Fig~\ref{fig:SNR_G0.1-9.7}.}
    \label{fig:SNR_G350.7+0.6}
\end{figure*}

G\,350.7+0.6 subtends $80'\times43'$, co-aligned with the Galactic Plane (\Fig~\ref{fig:SNR_G350.7+0.6}). Its northwestern half is superposed with a large number of \textsc{Hii} regions. We measure the less contaminated regions, excluding a single bright radio source to the south-west, and enclosing about half of the shell extent, while avoiding the \textsc{Hii} region complex and compact sources for the background measurement. From these data we derive $S_\mathrm{200MHz}=64\pm 1$\,Jy and $\alpha=-0.9\pm0.1$ across the GLEAM band, and denote the candidate Class~\textsc{II}.

\subsection{G\,$350.8+5.0$}\label{G350.8+5.0}

\begin{figure*}
   \centering
   \includegraphics[width=\textwidth]{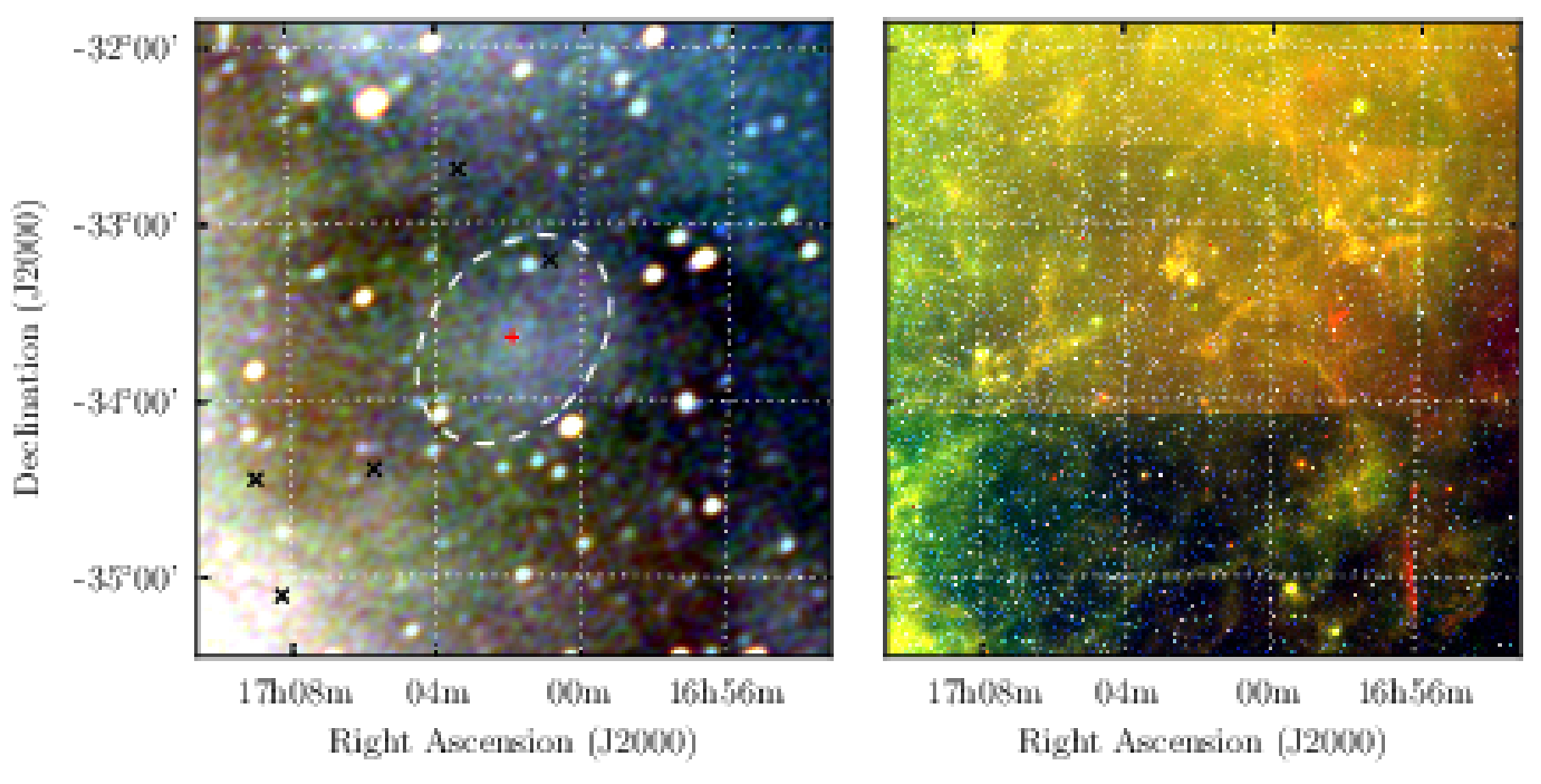}
   \caption{G\,$350.8+5.0$ as observed by GLEAM (left) at 72--103\,MHz (R), 103--134\,MHz (G), and 139--170\,MHz (B), and by \textit{WISE} (right) at 22\,$\mu$m (R), 12\,$\mu$m (G), and 4.6\,$\mu$m (B). The colour scales for the GLEAM RGB cube are 1.4--8.0, 0.5--4.0, and 0.2--2.0\,Jy\perbeam for R, G, and B, respectively. Annotations are as in \Fig~\ref{fig:SNR_G0.1-9.7}.}
    \label{fig:SNR_G350.8+5.0}
\end{figure*}

G\,$350.8+5.0$ lies outside of most Galactic Plane surveys, and appears as a faint filled ellipsoid with no sign of edge-brightening (\Fig~\ref{fig:SNR_G350.8+5.0}). A compact source lies near the centre of the ellipsoid, appearing in the catalogue of Hurley-Walker et al (submitted) as GLEAM\,J170154-333857, with $S_\mathrm{200MHz}=0.20\pm0.02$\,Jy and $\alpha=-1.4\pm0.2$. Subtracting this source from our measurements of G\,$350.8+5.0$, we find $S_\mathrm{200MHz}=16.5\pm0.4$ and $\alpha=-0.27\pm0.06$. This is unusually flat-spectrum for a SNR, although there is a large uncertainty on the measurement. Potentially the steep-spectrum compact source is a pulsar, and the entire SNR is a PWNe. Follow-up observations toward GLEAM\,J170154-333857 to search for radio or $\gamma$-ray pulsations may be useful. We denote this unusual candidate Class~\textsc{II}.

\subsection{G\,351.0-0.6}\label{G351.0-0.6}

\begin{figure*}
   \centering
   \includegraphics[width=\textwidth]{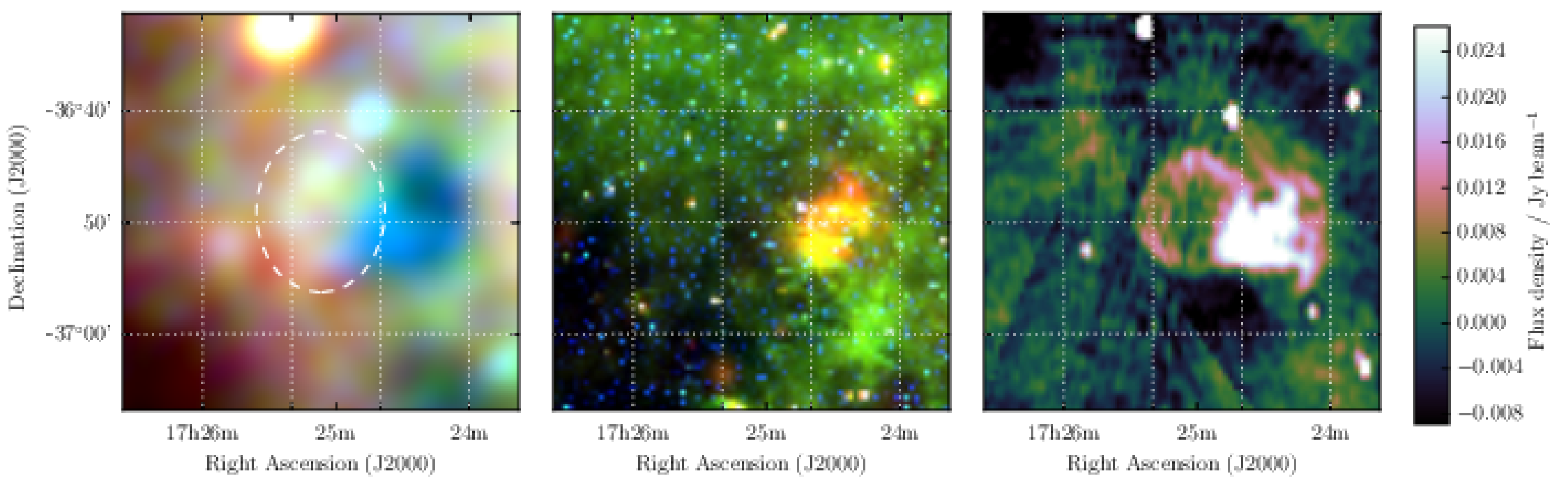}
   \caption{G\,$351.0-0.6$ as observed by GLEAM (left) at 72--103\,MHz (R), 103--134\,MHz (G), and 139--170\,MHz (B), by \textit{WISE} (middle) at 22\,$\mu$m (R), 12\,$\mu$m (G), and 4.6\,$\mu$m (B), and MGPS at 843\,MHz (right). The colour scales for the GLEAM RGB cube are 3.7--5.3, 1.8--2.5, and 0.7--1.1\,Jy\perbeam for R, G, and B, respectively. Annotations are as in \Fig~\ref{fig:SNR_G0.1-9.7}.}
    \label{fig:SNR_G351.0-0.6}
\end{figure*}

G\,$351.0-0.6$ appears as a partial shell half-occluded by the \textsc{Hii} region 
IRAS\,17210-3646. It is more clearly resolved in the MGPS images although only the GLEAM data allows clear separation of the thermal and non-thermal components (\Fig~\ref{fig:SNR_G351.0-0.6}). For the non-occluded part of the shell, we measure $S_\mathrm{200MHz}=0.50\pm0.04$\,Jy and $\alpha=-0.64\pm0.09$ across the GLEAM and MGPS data, and class the candidate as Class~\textsc{II}.

\subsection{G\,351.4+0.4}\label{G351.4+0.4}

\begin{figure*}
   \centering
   \includegraphics[width=\textwidth]{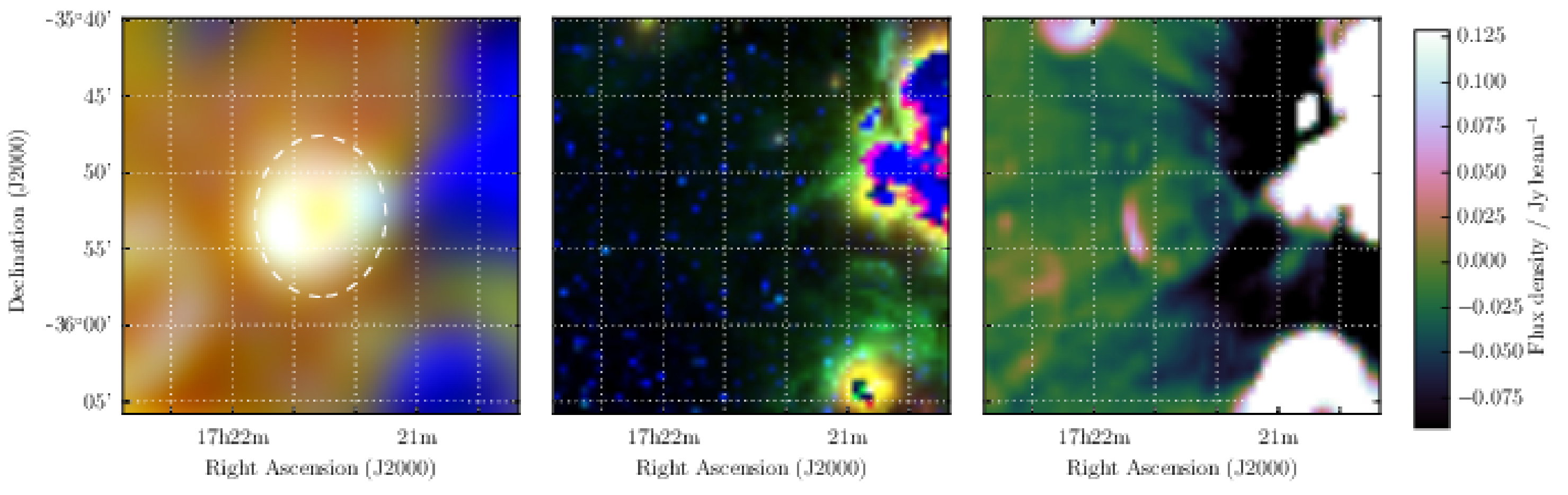}
   \caption{G\,$351.4+0.4$ as observed by GLEAM (left) at 72--103\,MHz (R), 103--134\,MHz (G), and 139--170\,MHz (B), by \textit{WISE} (middle) at 22\,$\mu$m (R), 12\,$\mu$m (G), and 4.6\,$\mu$m (B), and MGPS at 843\,MHz (right). The colour scales for the GLEAM RGB cube are 1.7--9.1, 2.5--4.5, and 1.4--2.4\,Jy\perbeam for R, G, and B, respectively. Annotations are as in \Fig~\ref{fig:SNR_G0.1-9.7}.}
    \label{fig:SNR_G351.4+0.4}
\end{figure*}

G\,$351.4+0.4$ is spatially the smallest candidate detected in this work, and is clearly visible as a full shell in MGPS (\Fig~\ref{fig:SNR_G351.4+0.4}). However, perhaps due to artefacts from the \textsc{Hii} complex to the west, the MGPS flux density is measured as 0.34\,Jy when we would expect $\approx1$\,Jy from the GLEAM spectrum. The fit is therefore driven strongly by the GLEAM data, and we calculate $S_\mathrm{200MHz}=3.35\pm0.09$\,Jy and $\alpha=-0.42\pm0.07$ (\Tab~\ref{tab:results}). 

\subsection{G\,351.4+0.2}\label{G351.4+0.2}

\begin{figure*}
   \centering
   \includegraphics[width=\textwidth]{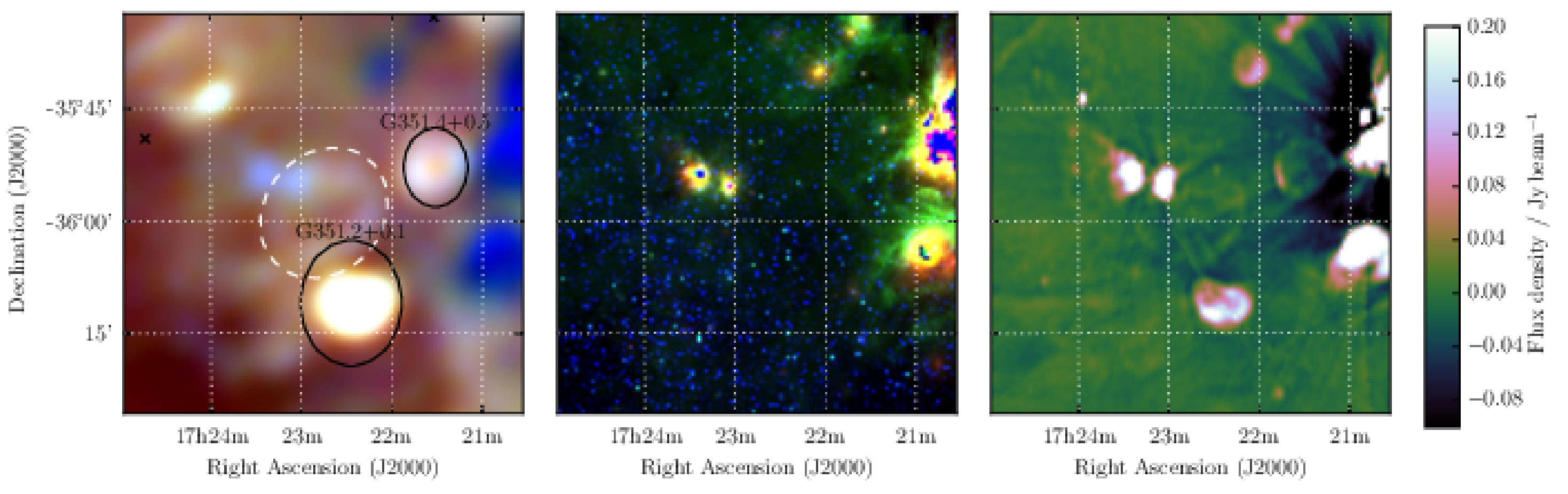}
   \caption{G\,$351.4+0.2$ as observed by GLEAM (left) at 72--103\,MHz (R), 103--134\,MHz (G), and 139--170\,MHz (B), by \textit{WISE} (middle) at 22\,$\mu$m (R), 12\,$\mu$m (G), and 4.6\,$\mu$m (B), and MGPS at 843\,MHz (right). The colour scales for the GLEAM RGB cube are 2.9--9.6, 2.5--4.9, and 1.0--2.4\,Jy\perbeam for R, G, and B, respectively. Annotations are as in \Fig~\ref{fig:SNR_G0.1-9.7}, and black ellipses indicate known nearby SNR (see \Sect~\ref{G351.4+0.2}). The blue region bisecting the NE of the candidate's shell is the \textsc{Hii} region L89b\,351.590+00.183.}
    \label{fig:SNR_G351.4+0.2}
\end{figure*}

G\,$351.4+0.2$ is just to the southeast of G\,$351.4+0.4$ (\Sect~\ref{G351.4+0.4}) and superposed with the \textsc{Hii} region L89b\,351.590+00.183 \citep{1989ApJS...71..469L}. Artefacts from this object reduce the reliability of the MGPS image, but the shell is still clearly visible therein (\Fig~\ref{fig:SNR_G351.4+0.2}). We use the GLEAM and MGPS measurements to determine $S_\mathrm{200MHz}=1.8\pm0.1$\,Jy and $\alpha=-0.9\pm0.1$. There are no pulsars within $2\times$ the diameter of the candidate so we cannot derive any physical properties, and denote the candidate Class~\textsc{II}, due to the confusing non-thermal emission visible to the south and east in the GLEAM image.

\subsection{G\,351.9+0.1}\label{G351.9+0.1}

G\,$351.9+0.1$ is fairly distinct as a shell in the GLEAM images, and can also be seen in MGPS despite some artefacts from the nearby \textsc{Hii} region L89b\,351.590+00.183 \citep{1989ApJS...71..469L}. Within this relatively un-confused environment, we are able to use both the GLEAM and MGPS data to derive $S_\mathrm{200MHz}=4.1\pm0.1$\,Jy and $\alpha=-0.99\pm0.08$. We are uncertain as to whether the bright source in the south-west part of the shell is associated with the SNR or not, and do not attempt to subtract it. Based on its morphology and spectrum, we assign the SNR Class~\textsc{II}.

\begin{figure*}
   \centering
   \includegraphics[width=\textwidth]{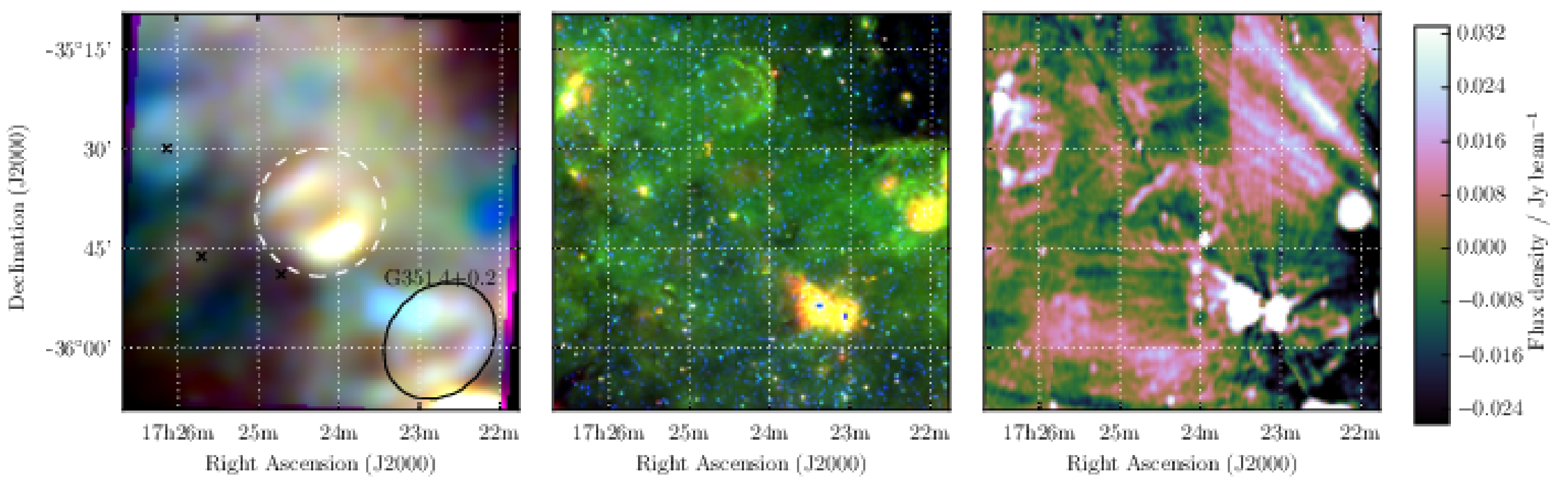}
   \caption{G\,$351.9+0.1$ as observed by GLEAM (left) at 72--103\,MHz (R), 103--134\,MHz (G), and 139--170\,MHz (B), by \textit{WISE} (middle) at 22\,$\mu$m (R), 12\,$\mu$m (G), and 4.6\,$\mu$m (B), and MGPS at 843\,MHz (right). The colour scales for the GLEAM RGB cube are 5.0--8.1, 2.6--4.1, and 1.1--1.9\,Jy\perbeam for R, G, and B, respectively. Annotations are as in \Fig~\ref{fig:SNR_G0.1-9.7}, and a black ellipse indicates a known nearby SNR (see \Sect~\ref{G351.9+0.1}).}
    \label{fig:SNR_G351.9+0.1}
\end{figure*}

\subsection{G\,353.0+0.8}\label{G353.0+0.8}

G\,$353.0+0.8$ is one of the most unusual candidates in the sample as it is almost completely occluded by a large \textsc{Hii} region complex. Restricting ourselves to only the westernmost edge of the shell, and avoiding the \textsc{Hii} region during background measurement, we find $S_\mathrm{200MHz}=16.5\pm0.4$\,Jy and $\alpha=-1.0\pm0.1$ across the GLEAM band. There are some brighter spots toward the edges of the shell where it appears to be in contact with the \textsc{Hii} region, but at this low resolution any physical interpretation is impossible.
Given how unusual this object is, we denote it Class~\textsc{III}.

\begin{figure*}
   \centering
   \includegraphics[width=\textwidth]{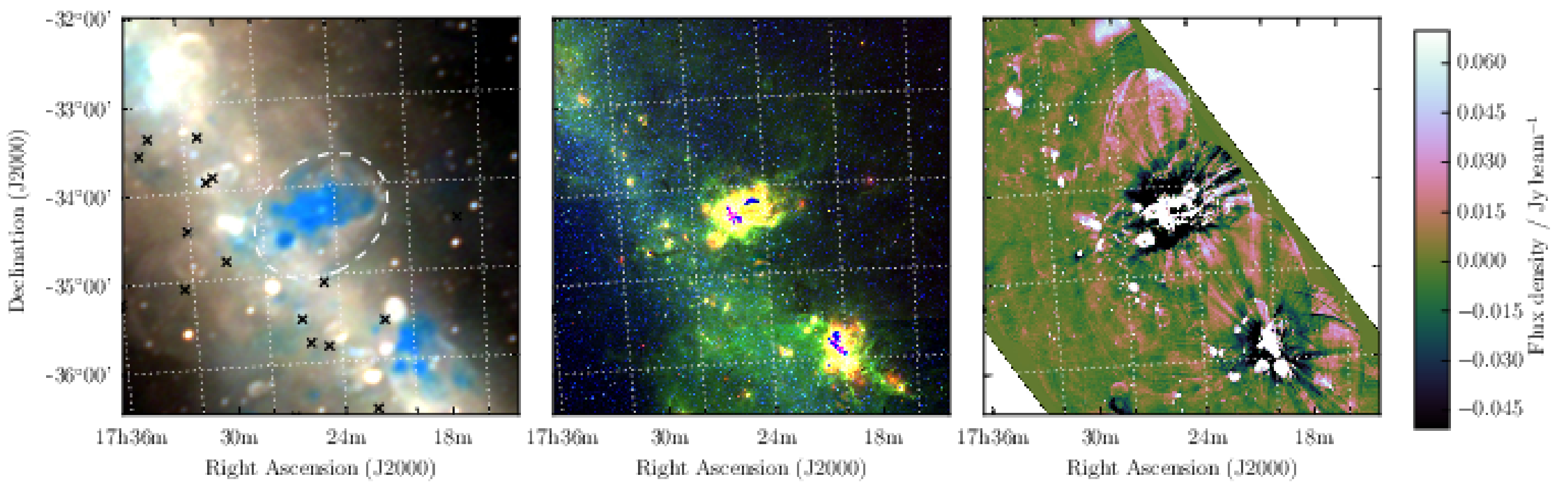}
   \caption{G\,$353.0+0.8$ as observed by GLEAM (left) at 72--103\,MHz (R), 103--134\,MHz (G), and 139--170\,MHz (B), by \textit{WISE} (middle) at 22\,$\mu$m (R), 12\,$\mu$m (G), and 4.6\,$\mu$m (B), and MGPS at 843\,MHz (right). The colour scales for the GLEAM RGB cube are 1.1--8.2, 0.3--4.3, and 0.1--2.1\,Jy\perbeam for R, G, and B, respectively.}
    \label{fig:SNR_G353.0+0.8}
\end{figure*}

\subsection{G\,355.4+2.7}\label{G355.4+2.7}

G\,$355.4+2.7$ appears as an irregular elliptical patch of increased brightness, with a non-thermal spectrum; from the GLEAM measurements alone we derive $S_\mathrm{200MHz}=1.5\pm0.2$\,Jy and $\alpha=-0.8\pm0.2$. With no pulsar association and no coverage from other Galactic Plane surveys, we cannot comment further on this candidate, but suggest that its morphology and spectrum are enough to denote it Class~\textsc{I}.

\begin{figure*}
   \centering
   \includegraphics[width=\textwidth]{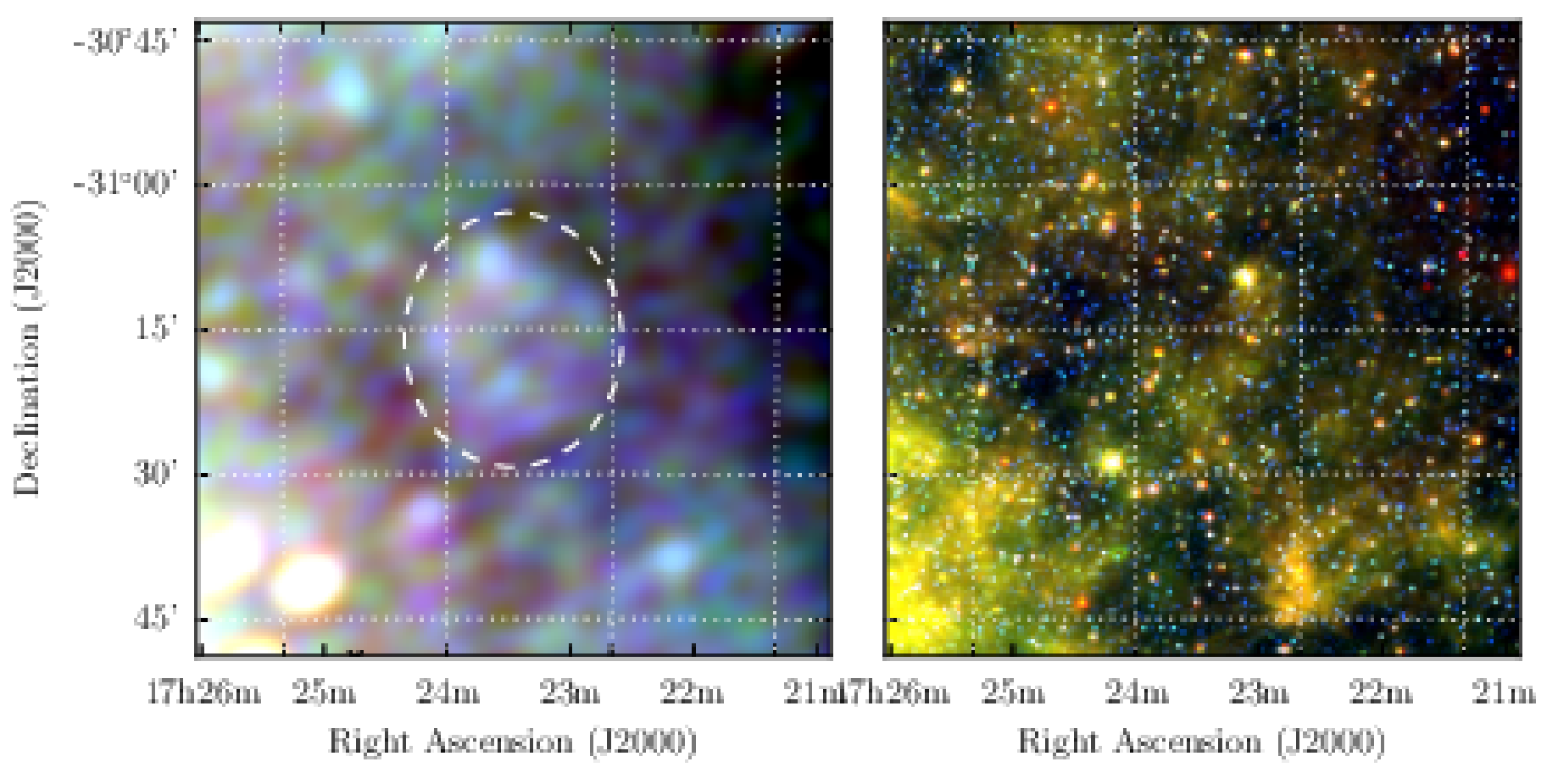}
   \caption{G\,$355.4+2.7$ as observed by GLEAM (left) at 72--103\,MHz (R), 103--134\,MHz (G), and 139--170\,MHz (B), and by \textit{WISE} (right) at 22\,$\mu$m (R), 12\,$\mu$m (G), and 4.6\,$\mu$m (B). The colour scales for the GLEAM RGB cube are 2.0--3.6, 0.8--1.4, and 0.2--0.5\,Jy\perbeam for R, G, and B, respectively.}
    \label{fig:SNR_G355.4+2.7}
\end{figure*}

\subsection{G\,356.5-1.9}\label{G356.5-1.9}

\begin{figure*}
   \centering
   \includegraphics[width=\textwidth]{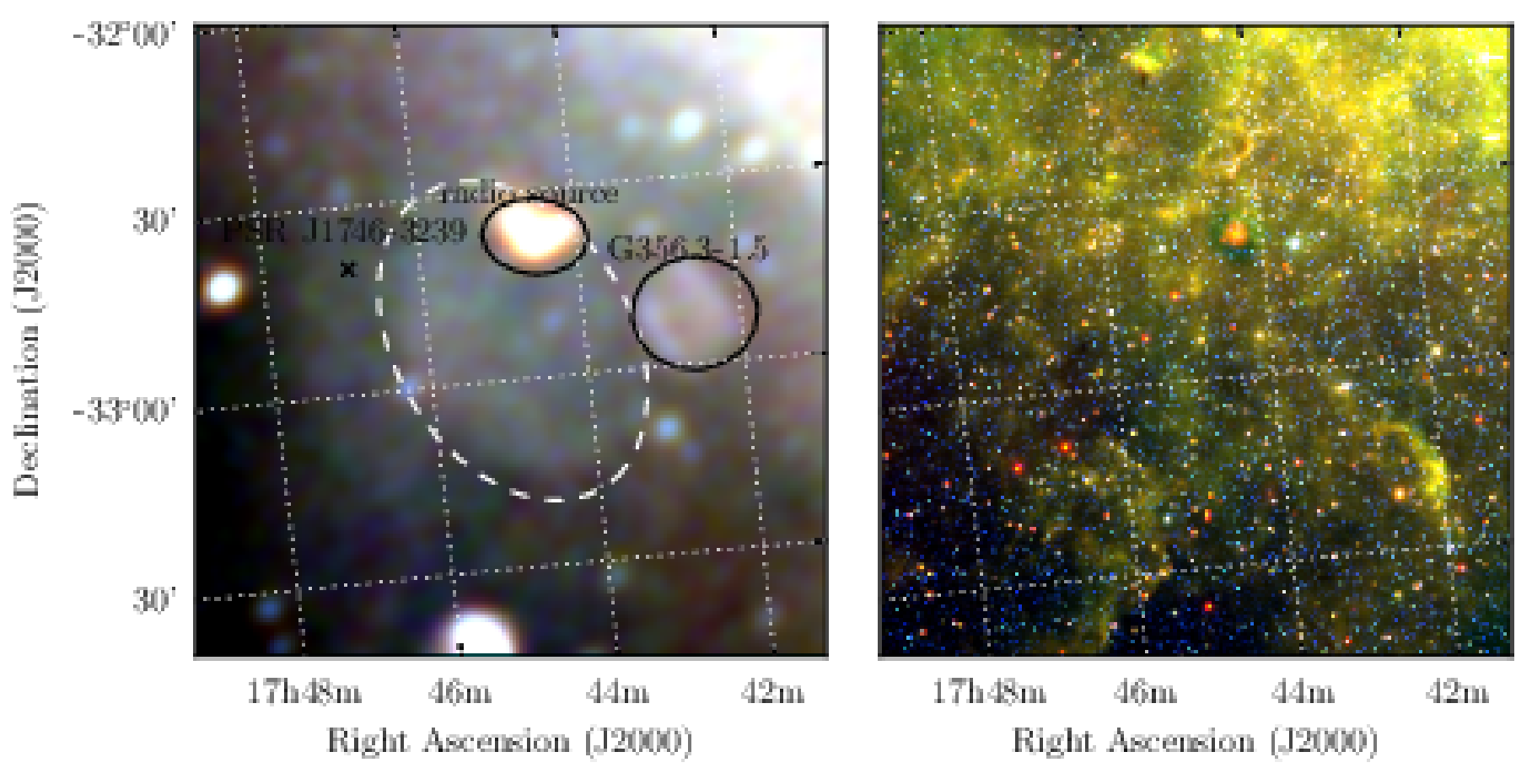}
   \caption{G\,$356.5-1.9$ as observed by GLEAM (left) at 72--103\,MHz (R), 103--134\,MHz (G), and 139--170\,MHz (B), and by \textit{WISE} (right) at 22\,$\mu$m (R), 12\,$\mu$m (G), and 4.6\,$\mu$m (B). The colour scales for the GLEAM RGB cube are 1.9--6.1, 0.6--2.7, and 0.2--1.1\,Jy\perbeam for R, G, and B, respectively. Annotations are as in \Fig~\ref{fig:SNR_G0.1-9.7}, and black ellipses indicate known nearby objects (see \Sect~\ref{G356.5-1.9}).}
    \label{fig:SNR_G356.5-1.9}
\end{figure*}

G\,$356.5-1.9$ is shown as a faint ellipse of uniform brightness, appearing brighter at higher Galacitic latitude (SE on \Fig~\ref{fig:SNR_G356.5-1.9}) due to the gradient of Galactic synchrotron background. It is east of the known SNR G\,$356.3-1.5$ \citep{1994MNRAS.270..847G}, and is superposed to the northwest with a roughly triangular non-thermal radio source. It is unclear whether this source is associated with the infrared source IRAS\,17412-3236, which is potentially an infrared bubble \citep[MGE\,356.7168-01.7246 in][]{2010AJ....139.1542M}, although it has an unusual bipolar structure in the radio \citep[Figure 4 of ][]{2016MNRAS.463..723I}; or, with the SNR candidate IGR\,J$17448-3232$ detected by \cite{2009ApJ...701..811T} using the \textit{Chandra} X-ray Observatory at 0.3--1.0\,keV; or, with some other chance radio emission along the line-of-sight.

In any case, care was taken to avoid this object when calculating integrated flux densities for G\,$356.5-1.9$. A fit over the flux density measurements yielded $S_\mathrm{200MHz}=14.9\pm0.3$\,Jy and $\alpha=-0.71\pm0.05$; while the surface brightness is low, the large size of the candidate yields a high total S/N.

PSR~J1746-3239 lies just outside of G\,$356.5-1.9$; with $P\approx$200\,ms and $\dot{P}\approx6.6\times10^{-15}$ it has a characteristic age of 482,000\,yr, which is on the high side for a potential association. The pulsar was discovered by \cite{2012ApJ...744..105P} via its $\gamma$-ray emission; despite an extensive campaign, the authors did not detect a radio counterpart, so its DM is unknown and a distance estimate cannot be made. In any case, G\,$356.5-1.9$ is denoted Class~\textsc{I} due to its elliptical morphology and non-thermal spectrum.

\subsection{G\,358.3-0.7}\label{G358.3-0.7}

\begin{figure*}
   \centering
   \includegraphics[width=\textwidth]{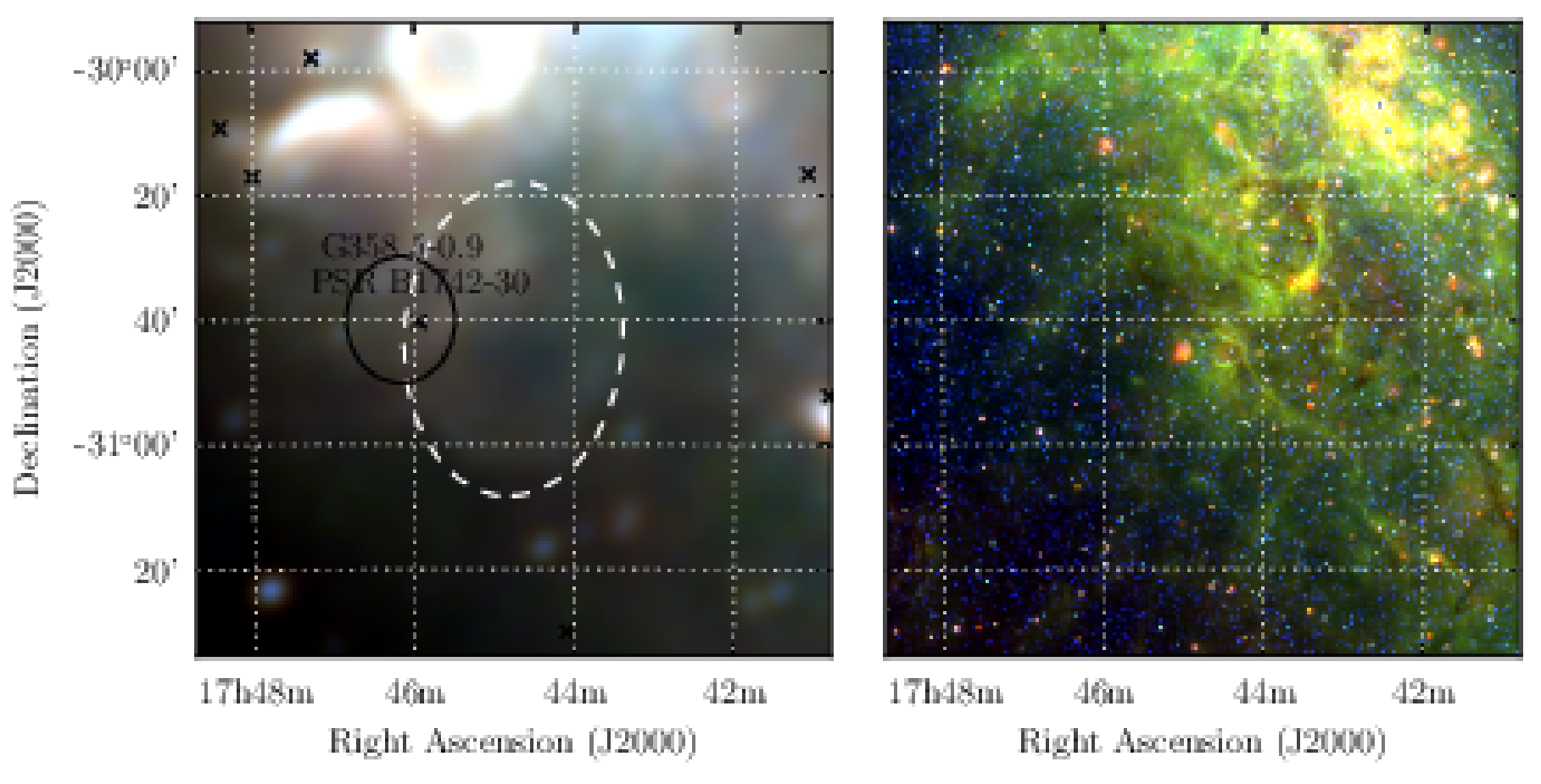}
   \caption{G\,$358.3-0.7$ as observed by GLEAM (left) at 72--103\,MHz (R), 103--134\,MHz (G), and 139--170\,MHz (B), and by \textit{WISE} (right) at 22\,$\mu$m (R), 12\,$\mu$m (G), and 4.6\,$\mu$m (B). The colour scales for the GLEAM RGB cube are 4.6--22.2, 1.8--11.6, and 0.7--5.4\,Jy\perbeam for R, G, and B, respectively. Annotations are as in \Fig~\ref{fig:SNR_G0.1-9.7}, and black ellipses indicate known nearby objects (see \Sect~\ref{G358.3-0.7}).}
    \label{fig:SNR_G358.3-0.7}
\end{figure*}

G\,$358.3-0.7$ is one of the faintest SNR in the sample, and appears to be contaminated with thermal emission in the centre of the shell. It is somewhat confused to the northeast with the known SNRs G\,$359.0-0.9$ and G\,$358.5-0.9$ \citep{1994MNRAS.270..847G}, as well as other non-thermal filaments in the region which may indicate further unknown SNRs (\Fig~\ref{fig:SNR_G358.3-0.7}). Restricting ourselves to the southwest section of the shell which is clear enough to measure, we find $S_\mathrm{200MHz} = 21.8\pm0.3$\,Jy and $\alpha=-0.8\pm0.1$ across the GLEAM band.

The nearby pulsar PSR~B1742-30 \citep{1973ApL....15..169K} has $P\approx370$\,ms and $\dot{P}\approx10^{-14}$\,s\,s$^{-1}$, yielding a characteristic age of 550,000\,yr \citep{2016MNRAS.460.4011L}. The DM measured by \cite{2004MNRAS.352.1439H} is 88.373\DMunit~ and this yields a distance estimate of 2.64\,kpc\footnote{\protect\cite{2016MNRAS.460.4011L} state a distance of 0.20\,kpc for this pulsar but it is unclear where this value arose.}. An association would imply that the candidate has diameter $26\times32$\,pc, placing it in the Sedov-Taylor phase of adiabatic expansion. Reversing
\eqn~4 of Paper~\textsc{I}
and making standard assumptions about the SNR energy and ISM density, we can use the SNR radius to predict the age, finding $t\approx10,000$--18,000\,yr, not very consistent with the pulsar characteristic age.

\cite{2016MNRAS.460.4011L} also measure a range of values for the proper motion of the pulsar, with $\dot{\alpha}=11.9\pm16$--$24\pm6$\,mas\,yr$^{-1}$ and $\dot{\delta}=30\pm11$--$75\pm49$\,mas\,yr$^{-1}$ depending on the derivation method used. Ignoring the listed uncertainties and using purely the range of values, this gives possible pulsar motion directions with angles varying from 10--$40^\circ$ (CCW from North), the latter being marginally consistent with the geometry shown in \Fig~\ref{fig:SNR_G358.3-0.7}. Using \citeauthor{2016MNRAS.460.4011L}'s best estimate of the total proper motion of $33\pm11$\,mas\,yr$^{-1}$ (but not their distance estimate), we calculate that the pulsar would need to have travelled for $\approx27$,000 years to arrive at its current location, again inconsistent with the pulsar characteristic age, although not far from the predicted expansion age, especially given the uncertainties.

We conclude that the association is somewhat unlikely, especially given the complexity of the region; a better estimate of the pulsar's proper motion would be necessary to use this as evidence of association. The candidate itself is denoted Class~\textsc{III}.

\section{Discussion}\label{sec:discussion}

\begin{figure*}
   \centering
   \includegraphics[width=0.5\textwidth]{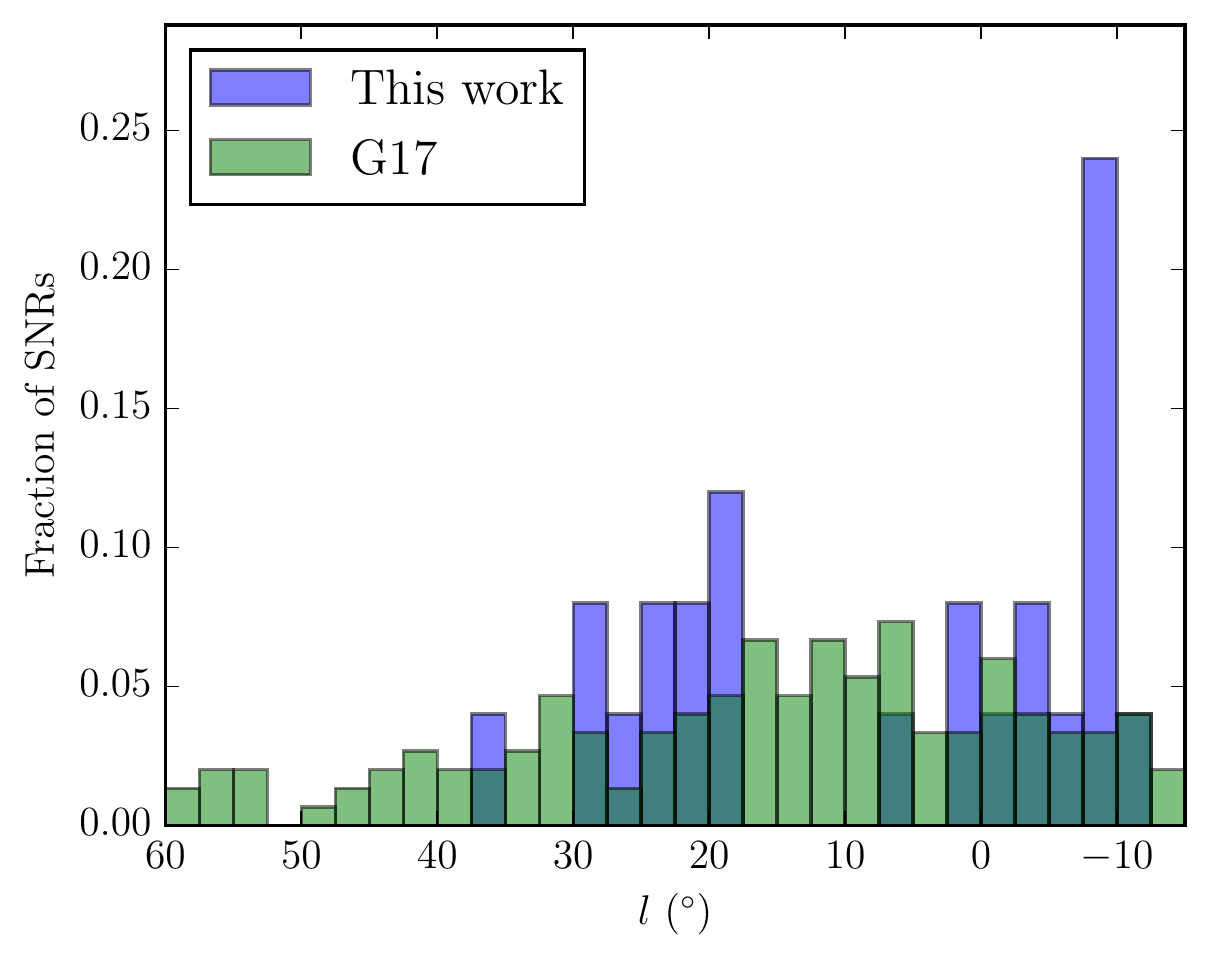}~\includegraphics[width=0.5\textwidth]{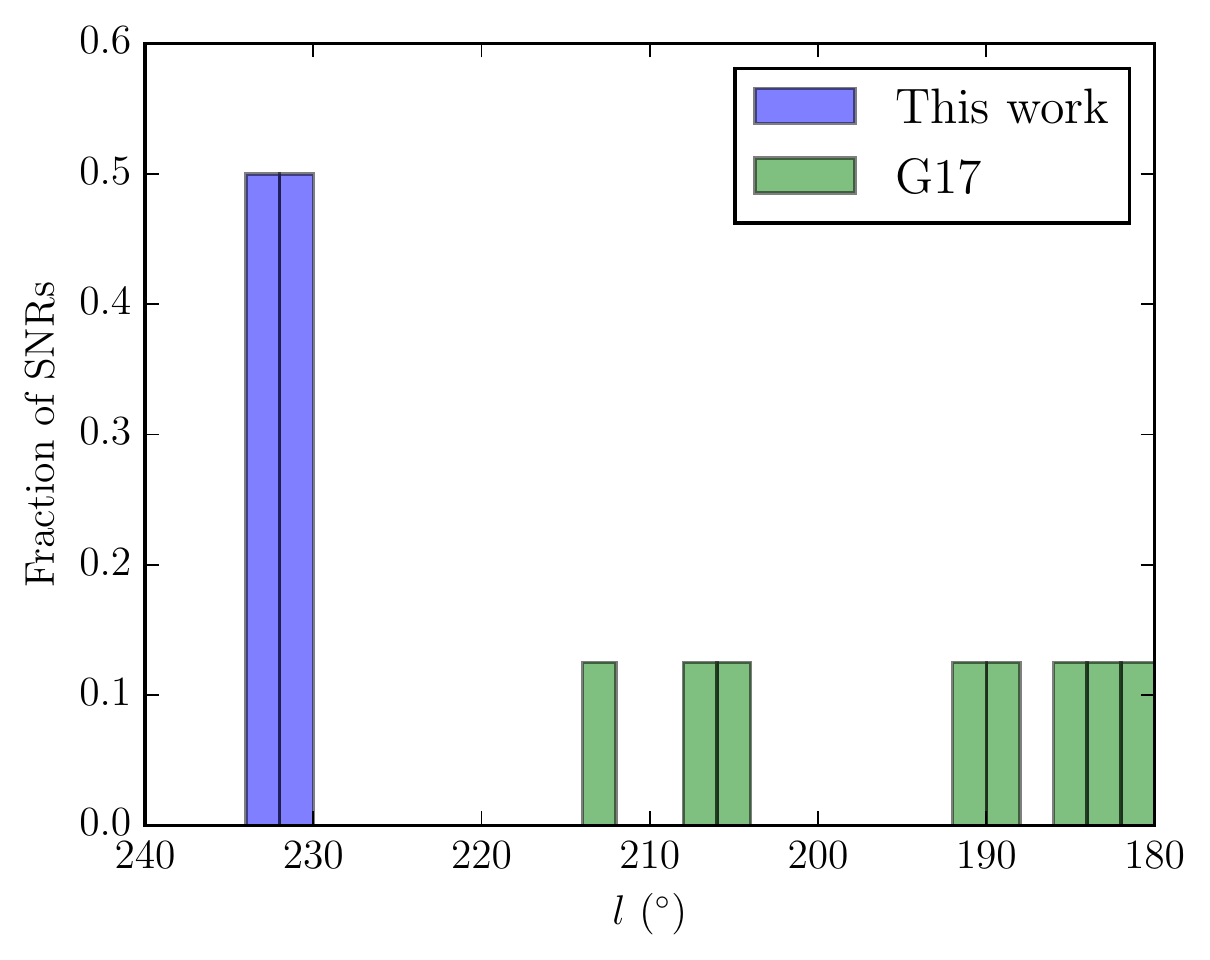}
   \caption{Histograms of Galactic longitude $l$, comparing the SNR candidates discovered in this work with the known SNRs catalogued by G17, normalised by height, for each panel. The left panel the range $345^\circ < l < 60^\circ$ and the right panel shows $180^\circ < l < 240^\circ$.}
    \label{fig:l}
\end{figure*}

\begin{figure}
   \centering
   \includegraphics[width=0.5\textwidth]{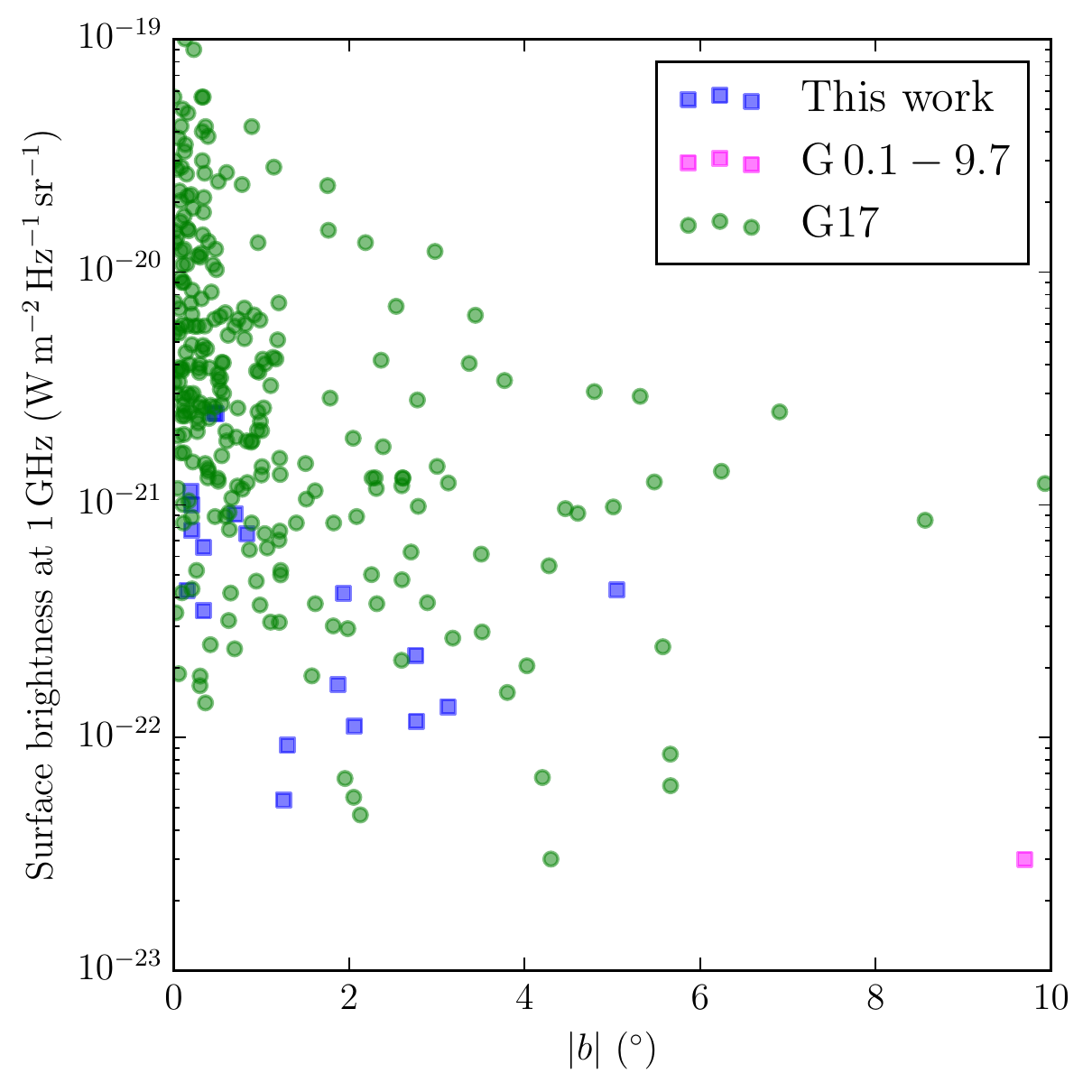}
   \caption{Surface brightness of SNR candidates with respect to absolute Galactic latitude $b$, for the SNR candidates discovered in this work and the known SNRs catalogued by G17. The surface brightness of G\,$0.1-9.7$ was derived by measuring the brightness of a central region of the ellipse at 72\,MHz and extrapolating to 1\,GHz (see \Sect~\ref{G0.1-9.7}).}
    \label{fig:sb_b}
\end{figure}

We compare the newly-detected SNRs with the G17 catalog of known SNRs. \Fig~\ref{fig:l} shows histograms of the Galactic longitude of known and new SNRs. The sensitivity of GLEAM drops toward high $l$, and we find no new SNRs beyond $l>40^\circ$, as image artefacts make interpretation more difficult. We find no new SNRs in the range $3^\circ<l<18^\circ$, in part due to confusion from the amount of Galactic emission, but perhaps also because this is a region that has already been extensively searched for SNRs. Examining \Fig~\ref{fig:sb_b}, we find that lowered confusion at high $|b|$ makes detection of faint shells much easier, and the wide field-of-view of the instrument automatically yields a large survey area. These high-latitude SNRs are likely to be nearby ($<2$\,kpc), and make superb probes of their local ISM, which is otherwise difficult to examine. 60\,\% of the candidates subtend areas larger than 0.2\,deg$^{2}$ on the sky, compared to $<25$\,\% of previously-detected SNRs (\Fig~\ref{fig:aa}). 

We discover two new candidates in a region previously empty of SNRs: G\,$230.4+1.2$ and G\,$232.1+2.0$, each with low 1\,GHz-surface brightnesses of $\approx10^{-22}$\SBunit, amongst the faintest ever detected. A population of low surface brightness objects at $180^\circ < l < 240^\circ$ would be easy to explore with more integration time due to the very low confusion levels, but clearly high sensitivity levels are required. This longitude range has no massive star formation regions, potentially indicating these SNR formed from relatively old B-type stars which have migrated from their stellar nurseries.

\begin{figure*}
   \centering
   \includegraphics[width=0.5\textwidth]{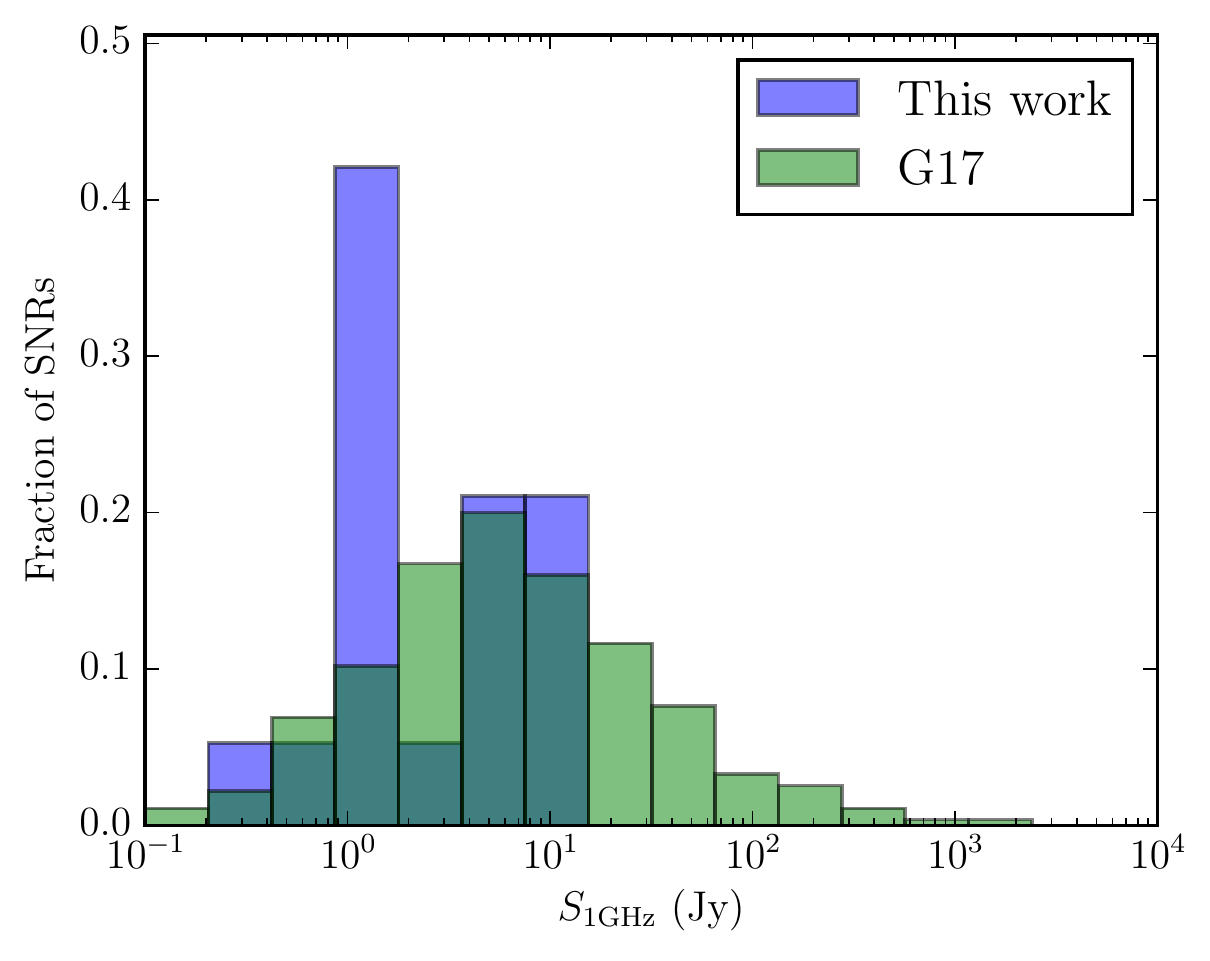}~\includegraphics[width=0.5\textwidth]{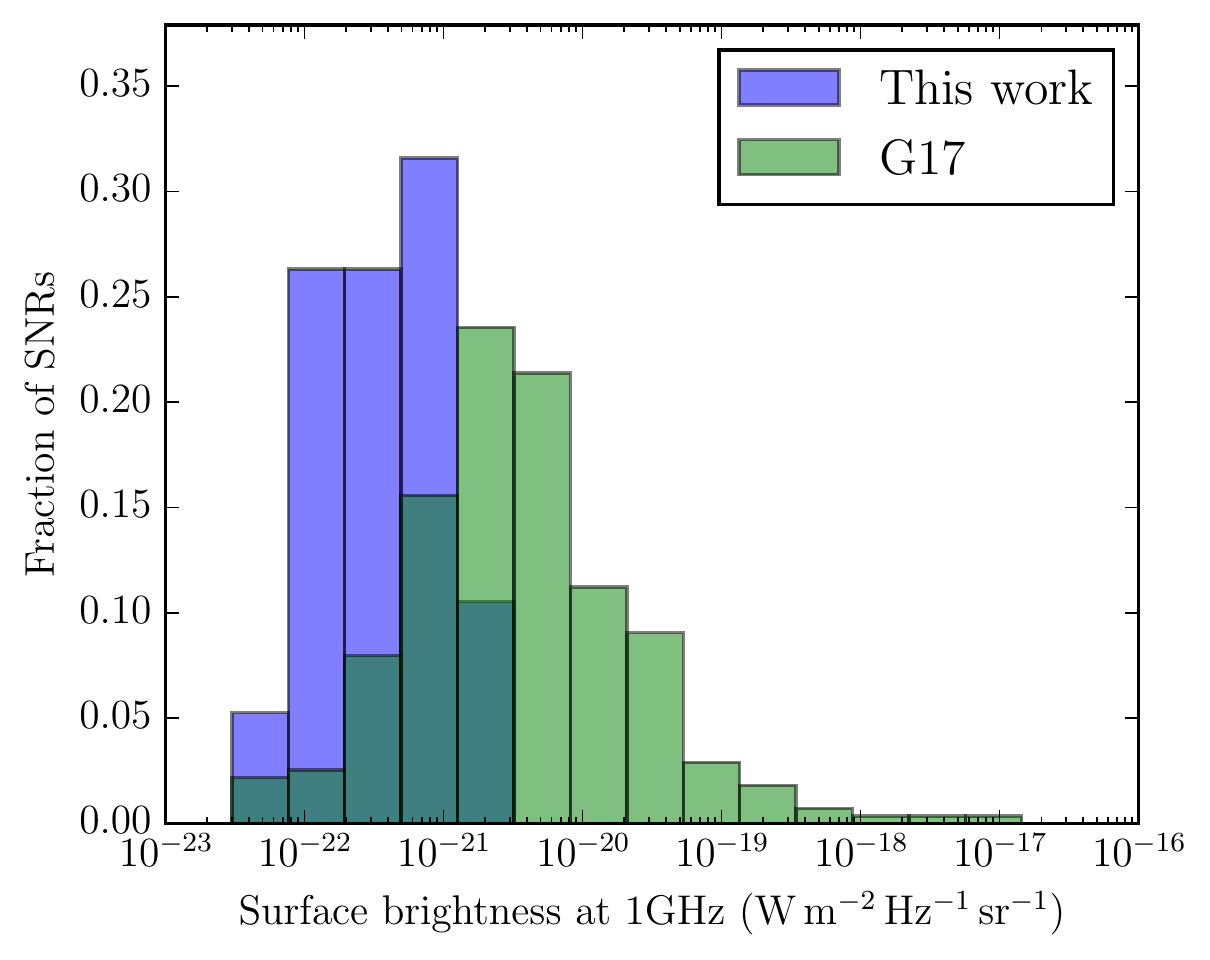}
   \caption{Histograms comparing the SNR candidates discovered in this work with the known SNRs catalogued by G17, normalised by height, for each panel. The left panel shows the 1-GHz flux density and the right panel shows the surface brightness. For the candidates discovered in this work, partial objects are excluded, and the 1-GHz values were derived from the fitted values of $S_\mathrm{200MHz}$ and $\alpha$ shown in \Tab~\ref{tab:results}.}
   \label{fig:fs}
\end{figure*}

\begin{figure*}
   \centering
   \includegraphics[width=0.5\textwidth]{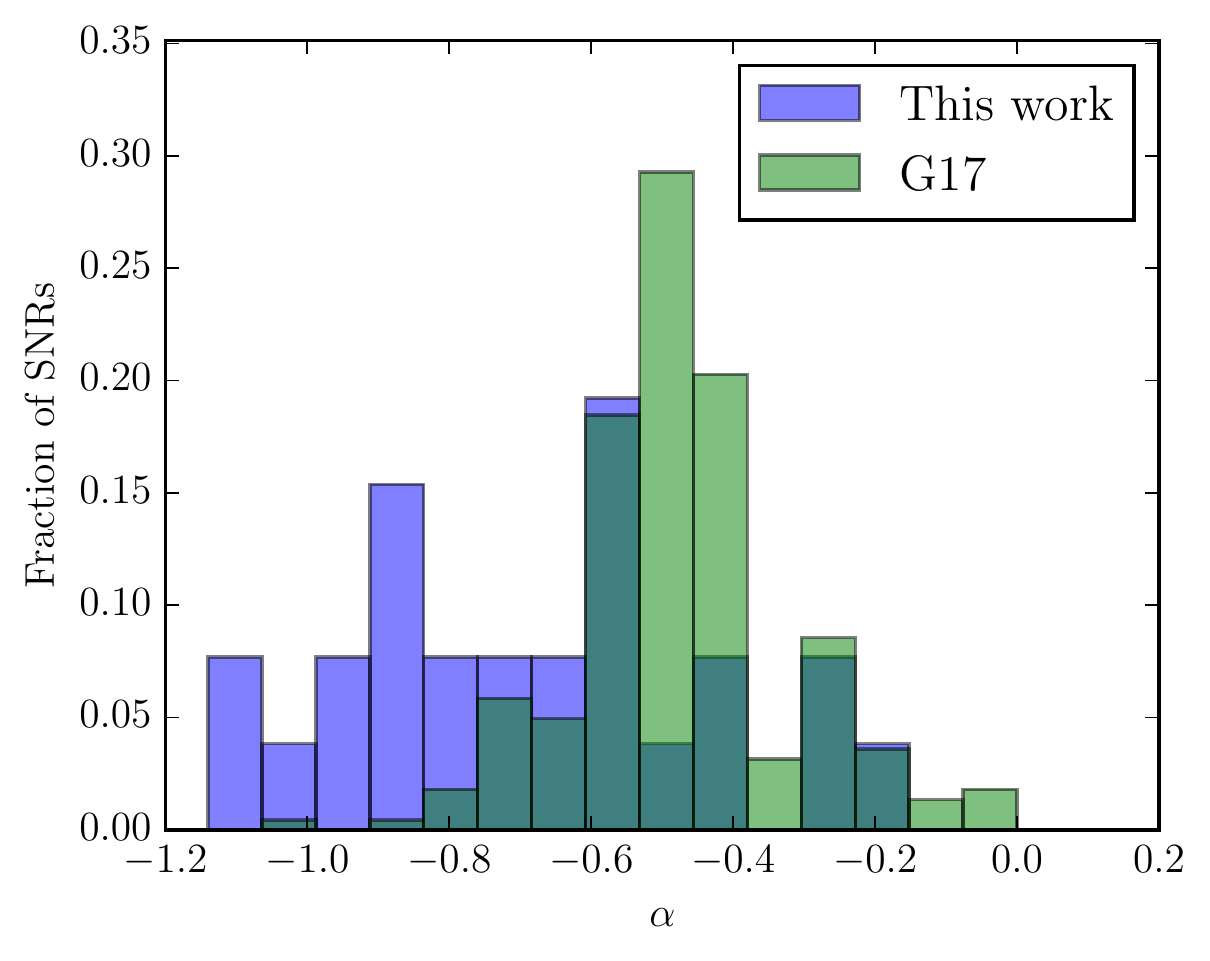}~\includegraphics[width=0.5\textwidth]{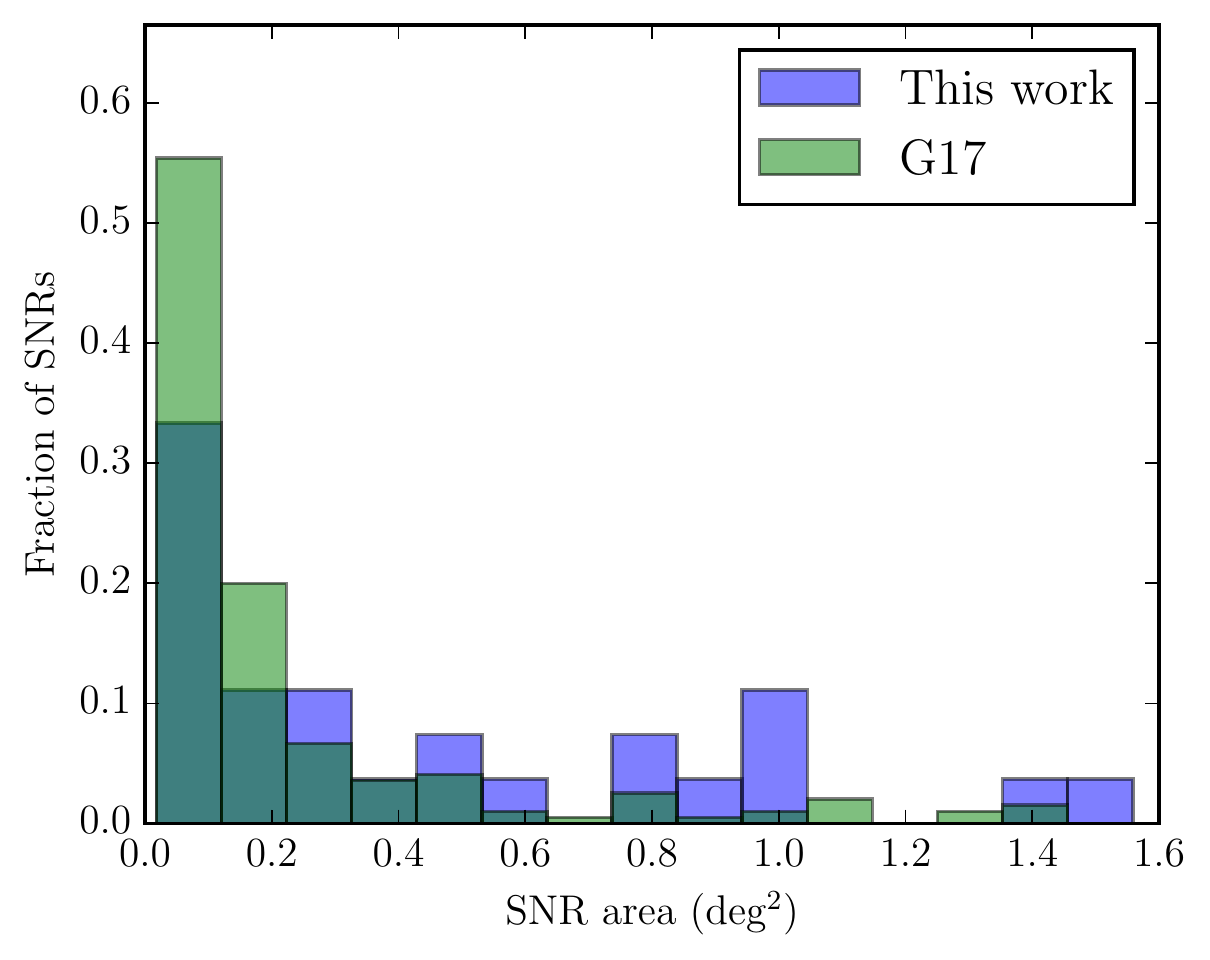}
      \caption{Histograms comparing the SNR candidates discovered in this work with the known SNRs catalogued by G17, normalised by height, for each panel. The left panel shows the spectral index $\alpha$ and the right panel shows the subtended area in square degrees. Partial candidates were \textit{not} excluded and their extrapolated areas were used based on their major and minor axes (\Tab~\ref{tab:results}).}
    \label{fig:aa}
\end{figure*}

\Fig~\ref{fig:fs} shows histograms of the 1-GHz flux density and surface brightness of the candidates, compared to known SNRs. The total flux densities are often not significantly lower than the known population, but it is clear from the second panel that they tend to have much lower brightness. \Fig~\ref{fig:aa} shows that the population also have steeper spectra, and subtend larger areas. We may be seeing SNRs with very little active energy injection, with an older population of electrons, resulting in steeper synchrotron spectra. The fact that the survey is at low frequencies and is searching previously-observed areas does bias our results toward finding steeper-spectrum objects. Looking at \Tab~\ref{tab:results}, there does not seem to be any significant correlation of spectral index with morphology.

We did not detect any of our candidates in the relatively shallow RASS data, and there were no chance alignments with pointed observations using the $30"\times30"$ fields-of-view of \textit{XMM-Newton} and \textit{Chandra}. The upcoming all-sky X-ray extended ROentgen Survey with an Imaging Telescope Array \citep[eROSITA; ][]{2011MSAIS..17..159C} will be $20-30\times$ more sensitive than RASS. Measurement of, or limits on, the X-ray emission of these candidates will allow us to more accurately determine their ages and energetics, which can then confirm or deny the hypothesis that the electron populations in the steep-spectrum sources are older.

Edge-brightening in some of the SNRs may indicate strong interactions with the local ISM. At these locations, we expect to find molecular clouds hosting prominent CO emission, \citep[see e.g.][]{2005ApJ...631..947M,2012MNRAS.422.2230M}. Upcoming surveys such as The Mopra Southern Galactic Plane CO Survey
\citep{2018PASA...35...29B} will be able to detect these clouds and provide SNR distance estimates, as well as probing the interactions between the clouds and shock fronts. Older supernova shocks are more likely to host broad-line spectral line emission, revealing complex velocity gradients across the ISM. Using \textsc{Hi} emission and absorption to probe near the SNRs will map the local dynamics of the ISM, possible on $\approx30"$-scales using the upcoming high-sensitivity Galactic Australian Square Kilometer Array Pathfinder survey \citep[GASKAP; ][]{2013PASA...30....3D}.

The discovery of 27~new SNRs increases the known number by $\approx10$\,\%. The Galactic longitude range $240^\circ < l < 345^\circ$ hosts many massive star-forming regions and is likely to host at least as many candidate SNRs again (Johnston-Hollitt in prep). While our survey method is biased toward detecting steep-spectrum objects, this population was not previously known to exist, indicating that these objects may make up a large fraction of the SNRs yet to be found. The other large fraction yet to be discovered is likely to be SNRs of small apparent size, at large distances, and thus will only be found by high-resolution surveys able to probe through the Galactic disk \citep[e.g. the 76~new candidate SNRs discovered by][]{2017A+A...605A..58A}.

Further discoveries at low Galactic latitudes will not be gained by integrating for longer with low-resolution surveys, due to confusion. The upcoming GLEAM-eXtended (GLEAM-X; Hurley-Walker in prep) survey will cover the entire southern sky at $1'$ resolution and the same wide bandwidth as GLEAM, which will considerably reduce confusion and allow better morphological characterisation of compact candidates. At high Galactic latitudes, extragalactic source confusion introduces a noise floor at $\approx1$\,mJy\perbeam at 200\,MHz ($\approx7\times10^{-24}$\SBunit extrapolated to 1\,GHz with $\alpha=-0.5$). This is $\approx7\times$ the RMS noise of the extragalactic GLEAM survey so further low-resolution integration may yet yield further candidates, but with diminishing returns. A low-frequency telescope configuration including both short and long baselines yielding sub-arcmin resolution and sensitivity to $1^\circ$--$10^\circ$ structures would be best-placed to yield further candidates.

\section{Conclusions}\label{sec:conclusions}

We have detected 27~new SNRs using a new data release of the GLEAM survey from the MWA telescope, including the lowest surface-brightness SNR ever detected, G\,$0.1-9.7$. Our method uses spectral fitting to the radio continuum to derive spectral indices for 26/27 candidates, and our low-frequency observations probe a steeper-spectrum population than previously discovered, as well as SNR of large angular extent. None of the candidates have coincident \textit{WISE} mid-IR emission, further showing that the emission is non-thermal.

We make the first detection of two SNRs in the Galactic longitude range $220^\circ$--$240^\circ$: follow-up observations of the pulsars at the hearts of these SNRs would allow derivation of physical parameters, increasing our knowledge of these SNRs far from known star-forming regions.

For our low-resolution survey, confusion at low Galactic latitudes makes new detections challenging. However the wide bandwidth of the MWA allows us to discriminate thermal and non-thermal emission, uncovering large SNRs in complex regions, previously missed by high-frequency narrowband surveys. A combination of higher-resolution data from the upgraded MWA and the inclusion of all GLEAM data should yield further detections of as-yet unknown SNRs.

\begin{acknowledgements}
We thank Gavin Rowell, Loren Anderson, and Jennifer West for useful feedback on the draft manuscript, and the anonymous referee for their helpful comments which improved the paper. This scientific work makes use of the Murchison Radio-astronomy Observatory, operated by CSIRO. We acknowledge the Wajarri Yamatji people as the traditional owners of the Observatory site. Support for the operation of the MWA is provided by the Australian Government (NCRIS), under a contract to Curtin University administered by Astronomy Australia Limited. We acknowledge the Pawsey Supercomputing Centre which is supported by the Western Australian and Australian Governments. We acknowledge the work and support of the developers of the following following python packages: Astropy \citep{TheAstropyCollaboration2013}, Numpy \citep{vaderwalt_numpy_2011}, and Scipy \citep{Jones_scipy_2001}. We also made extensive use of the visualisation and analysis packages DS9\footnote{\href{http://ds9.si.edu/site/Home.html}{ds9.si.edu/site/Home.html}} and Topcat \citep{Taylor_topcat_2005}. This work was compiled in the very useful free online LaTeX editor Overleaf.
This publication makes use of data products from the \textit{Wide-field Infrared Survey Explorer}, which is a joint project of the University of California, Los Angeles, and the Jet Propulsion Laboratory/California Institute of Technology, funded by the National Aeronautics and Space Administration.
\end{acknowledgements}

\begin{appendix}

\section{Regions used to determine SNR flux densities}

These plots follow the format of
\Fig~1 of Paper~\textsc{I}
, indicating where the polygons were drawn in \polyflux~ to measure SNRs. The top two panels of each figure show the GLEAM 170--231\,MHz images; the lower two panels show the RGB cube formed from the 72--103\,MHz (R), 103--134\,MHz (G), and 139--170\,MHz (B) images. As described in \Sect~2.3 of Paper~\textsc{I}, the annotations on the right two panels consist of: white polygons to indicate the area to be integrated in order to measure the SNR flux density; blue dashed lines to indicate regions excluded from any background measurement; the light shaded area to show the region that is used to measure the background, which is then subtracted from the final flux density measurement. Figures proceed in the same order as in \Sect~\ref{sec:new_snr}.
SNRs for which no GLEAM spectra was extracted are excluded from this list.

\begin{figure}
    \centering
    \includegraphics[width=0.5\textwidth]{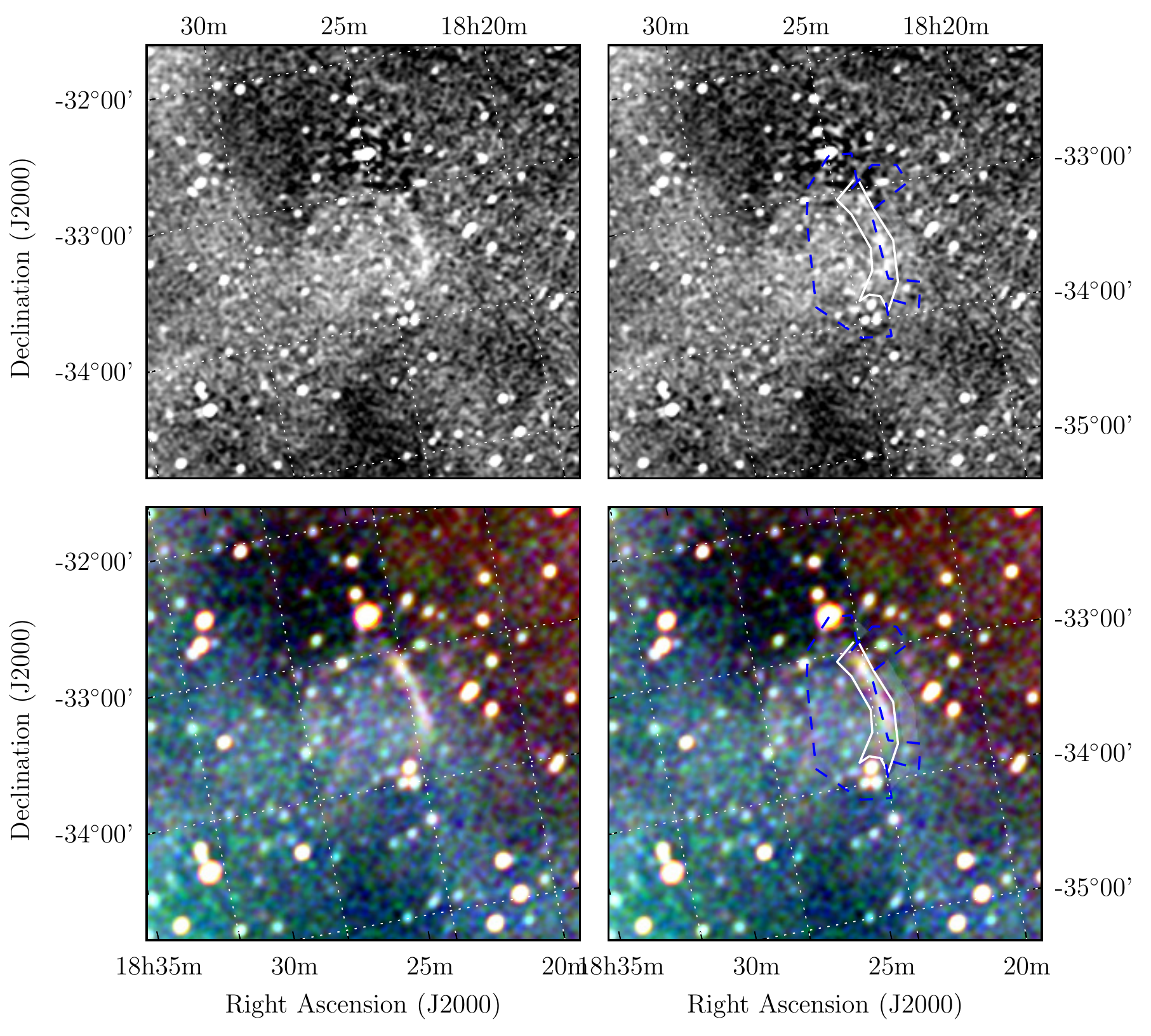}
    \caption{\polysummary G0.1-9.7. \polysuffix}
    \label{fig:SNR_G0.1-9.7_poly}
\end{figure}

\begin{figure}
    \centering
    \includegraphics[width=0.5\textwidth]{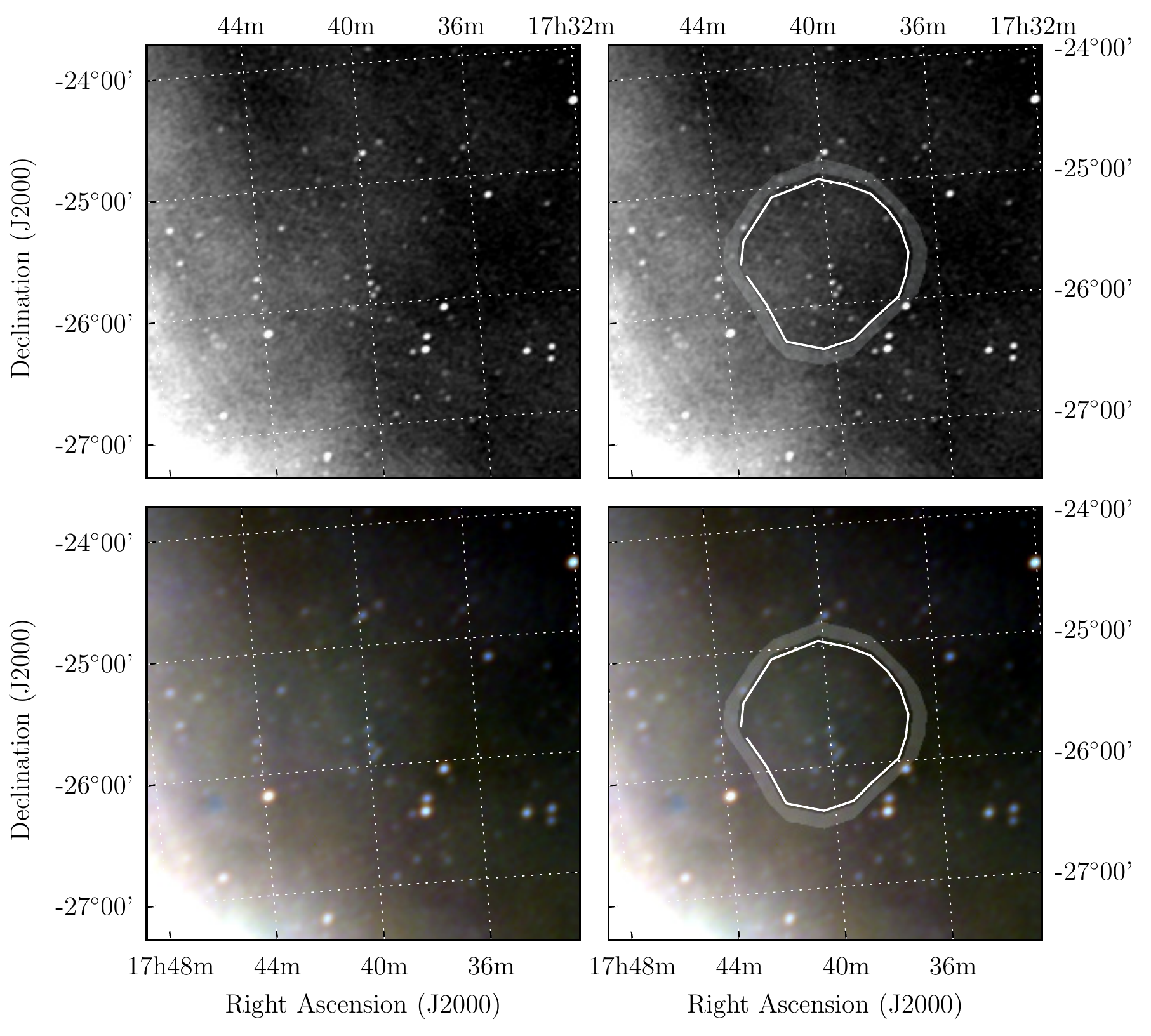}
    \caption{\polysummary G2.1+2.7. \polysuffix}
    \label{fig:SNR_G2.1+2.7_poly}
\end{figure}

\begin{figure}
    \centering
    \includegraphics[width=0.5\textwidth]{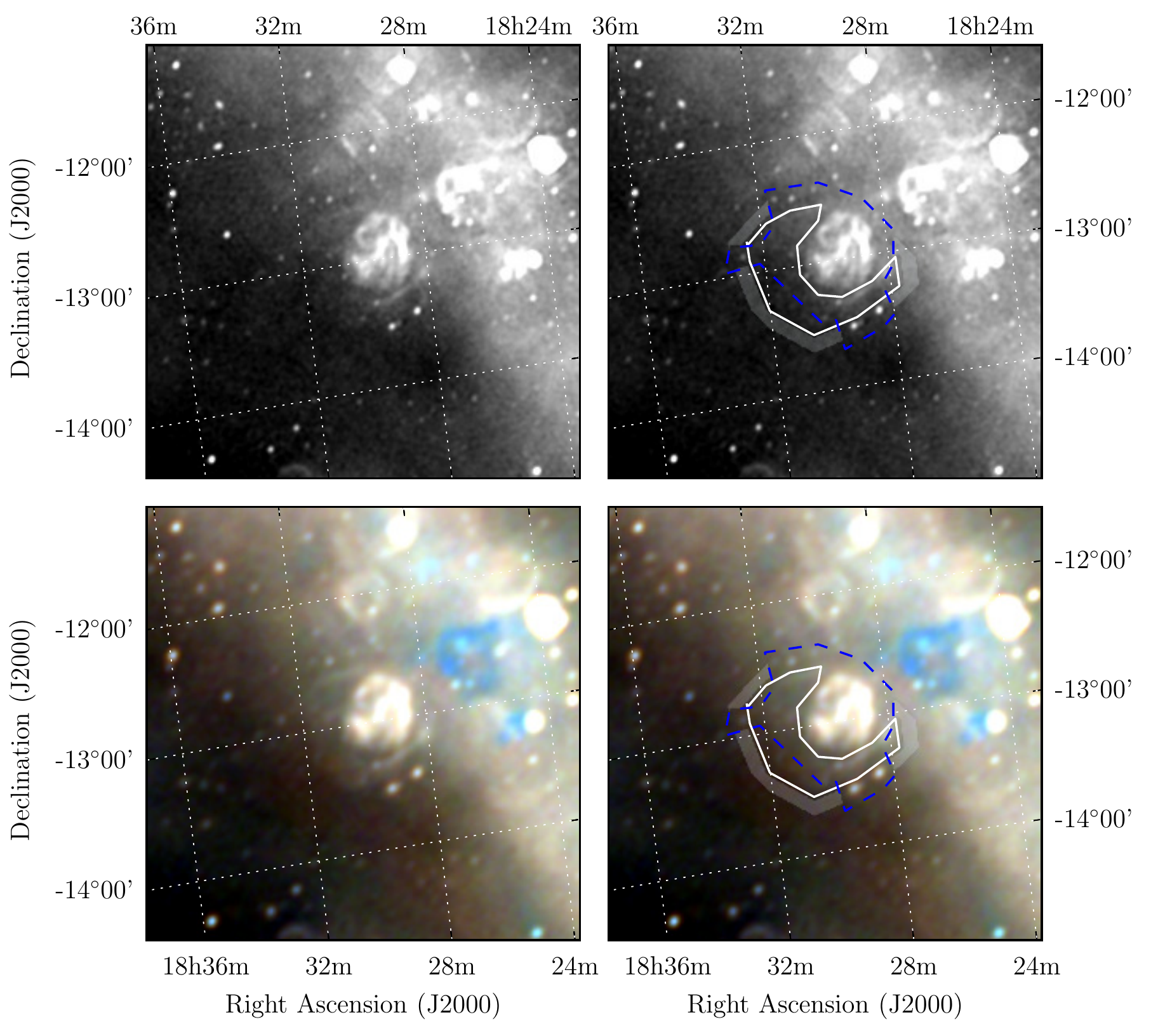}
    \caption{\polysummary G18.9-1.2. \polysuffix}
    \label{fig:SNR_G18.9-1.2_poly}
\end{figure}

\begin{figure}
    \centering
    \includegraphics[width=0.5\textwidth]{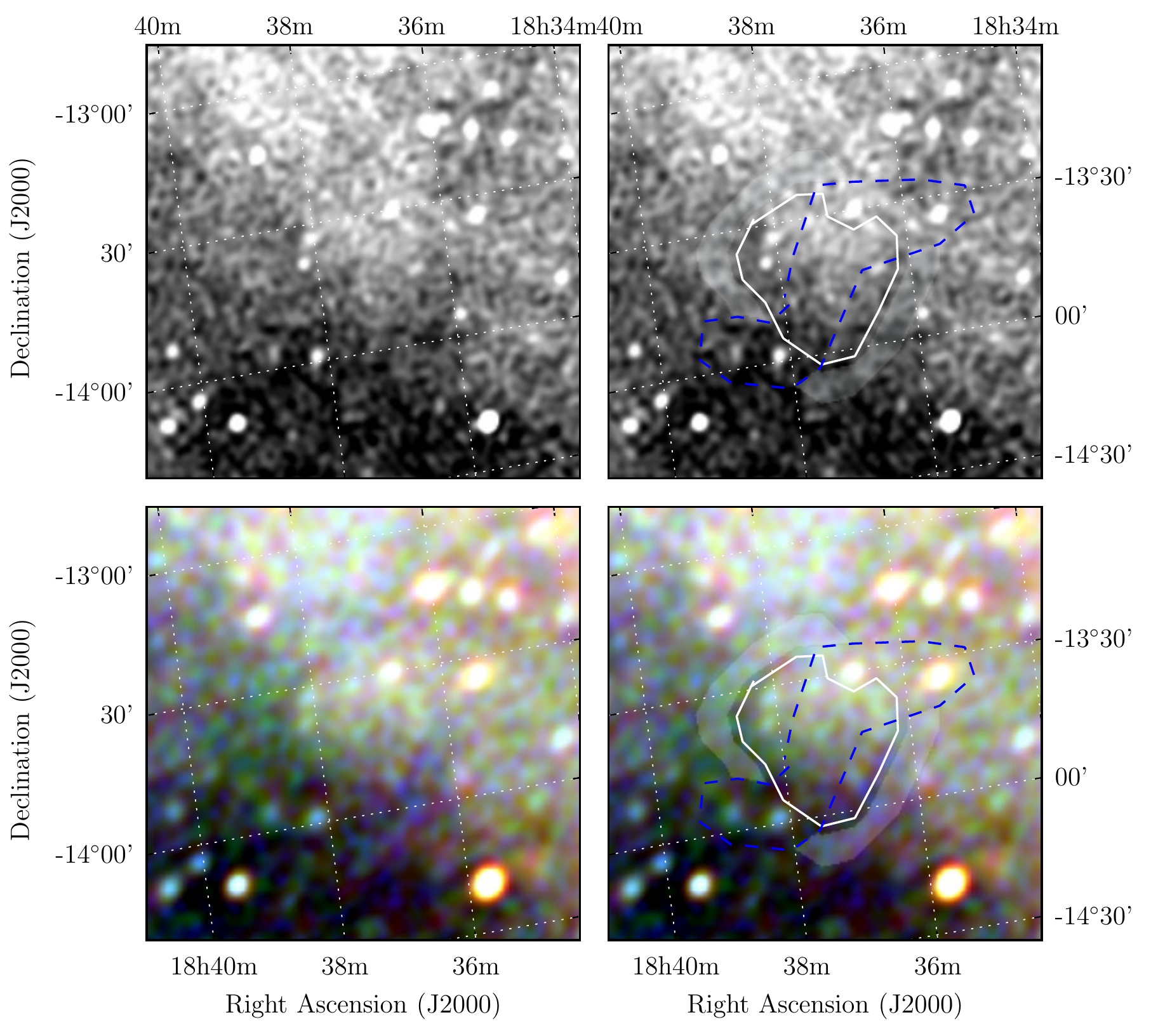}
    \caption{\polysummary G19.1-3.1. \polysuffix}
    \label{fig:SNR_G19.1-3.1_poly}
\end{figure}

\begin{figure}
    \centering
    \includegraphics[width=0.5\textwidth]{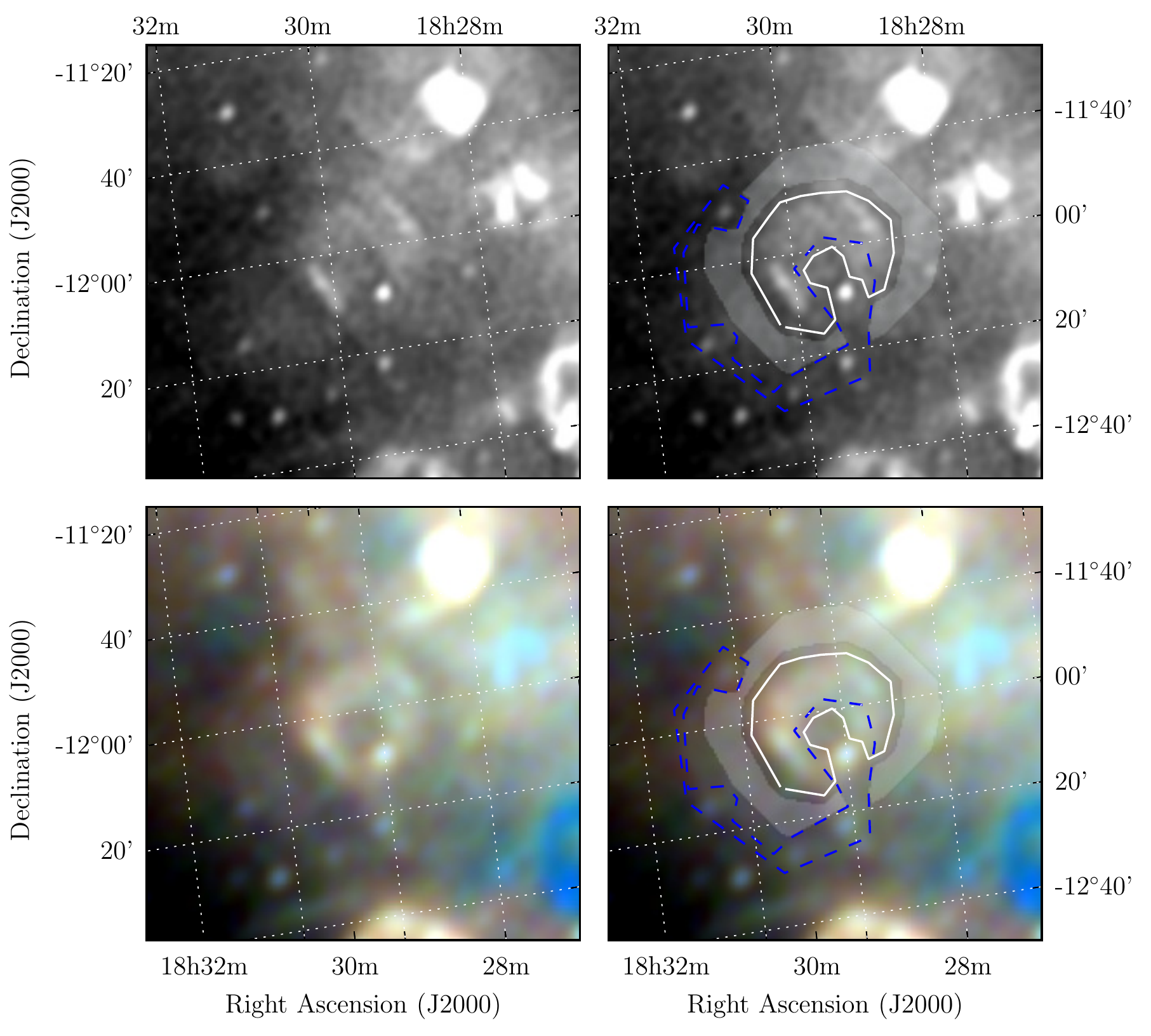}
    \caption{\polysummary G19.7-0.7. \polysuffix}
    \label{fig:SNR_G19.7-0.7_poly}
\end{figure}

\begin{figure}
    \centering
    \includegraphics[width=0.5\textwidth]{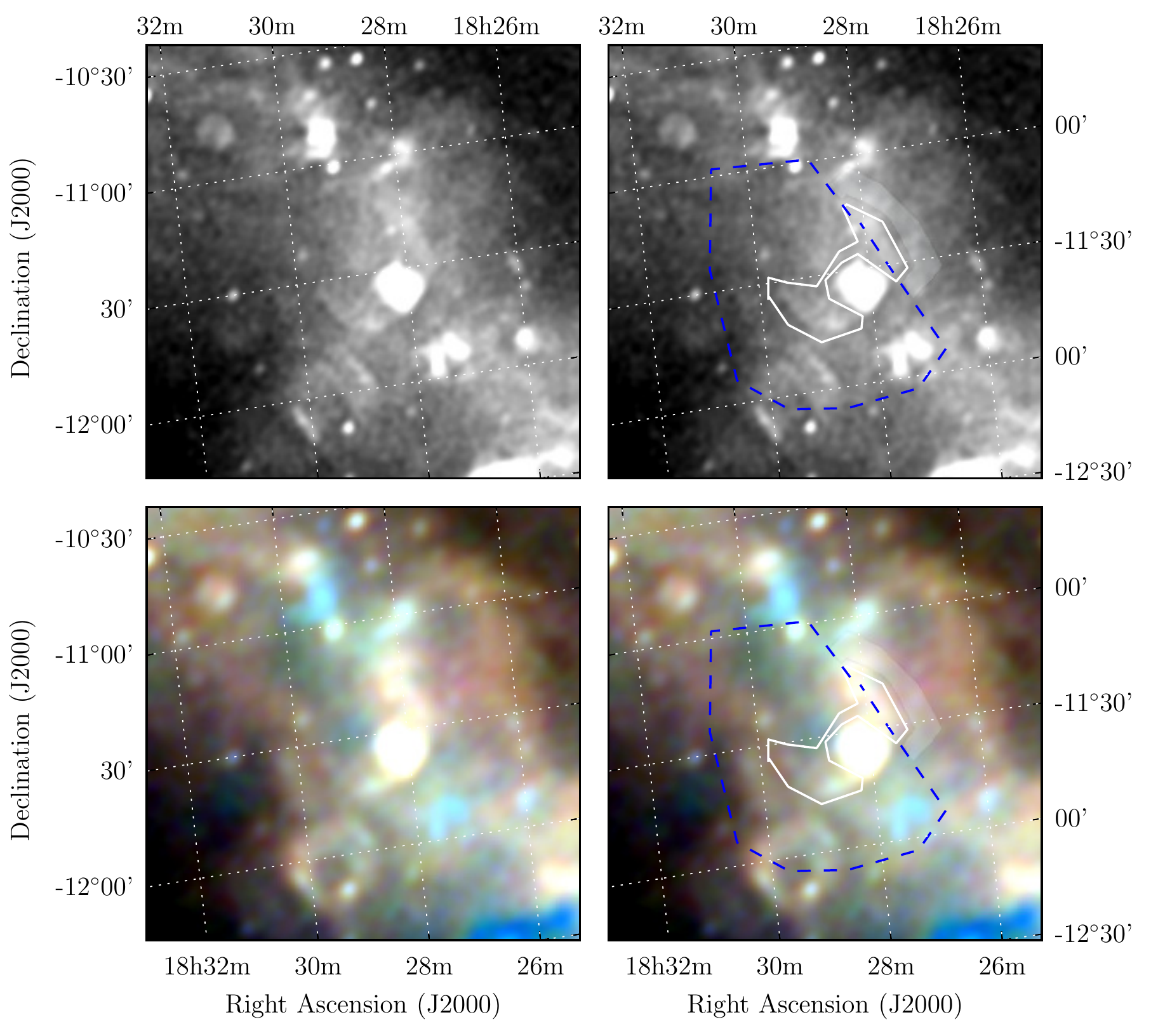}
    \caption{\polysummary G20.1-0.2. \polysuffix}
    \label{fig:SNR_G20.1-0.2_poly}
\end{figure}

\begin{figure}
    \centering
    \includegraphics[width=0.5\textwidth]{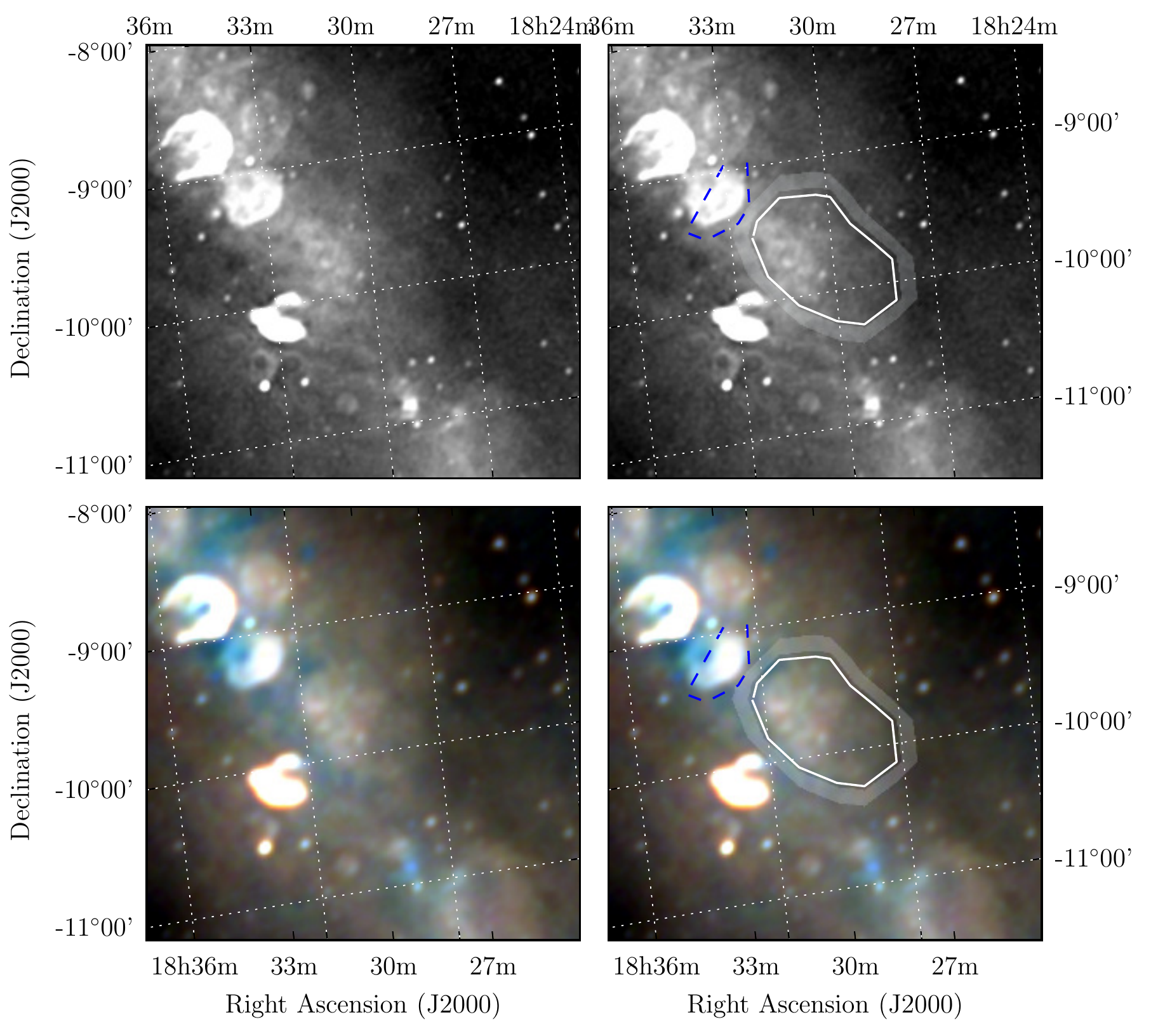}
    \caption{\polysummary G21.8+0.2. \polysuffix}
    \label{fig:SNR_G21.8+0.2_poly}
\end{figure}

\begin{figure}
    \centering
    \includegraphics[width=0.5\textwidth]{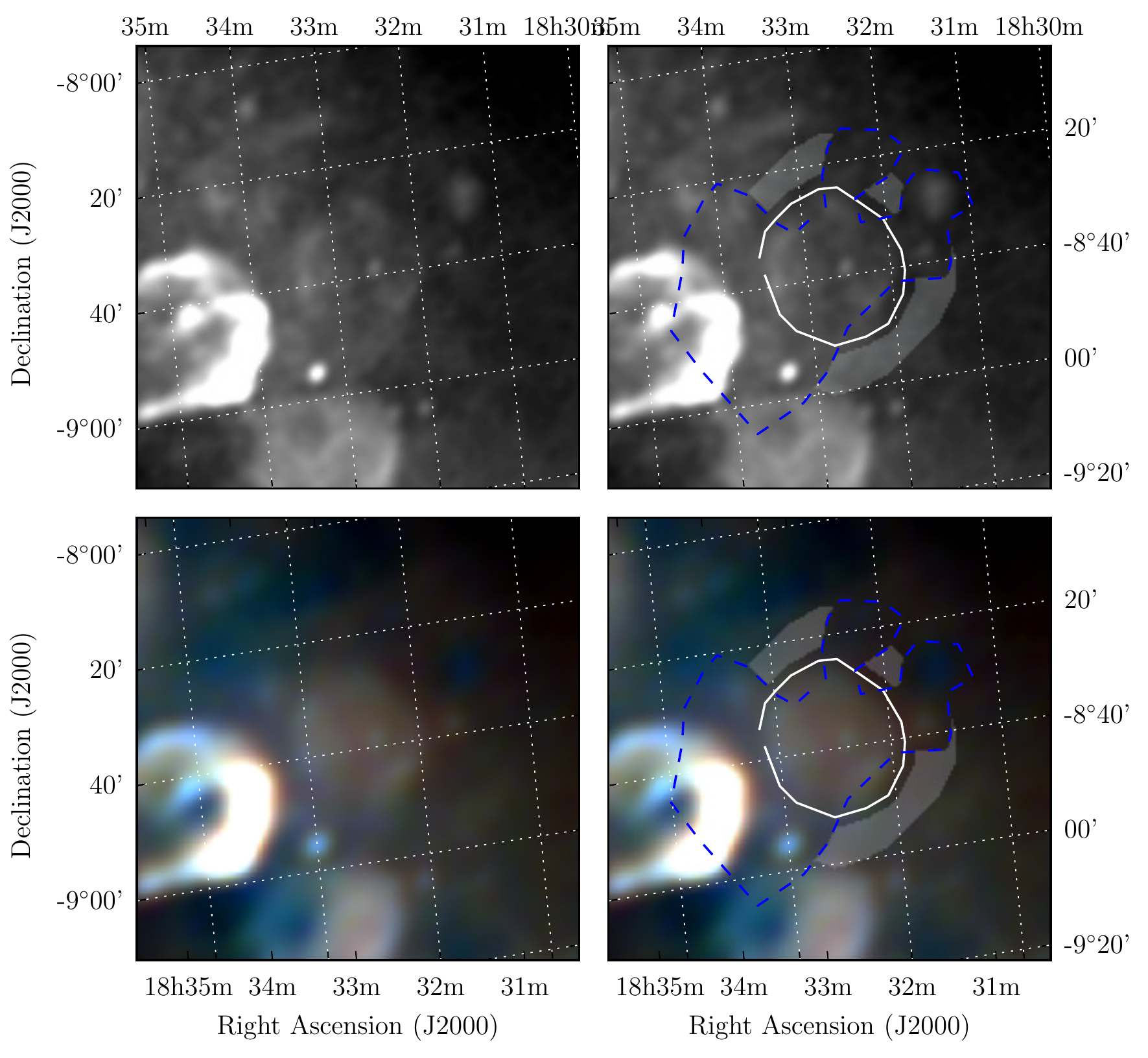}
    \caption{\polysummary G23.1+0.1. \polysuffix}
    \label{fig:SNR_G23.1+0.1_poly}
\end{figure}

\begin{figure}
    \centering
    \includegraphics[width=0.5\textwidth]{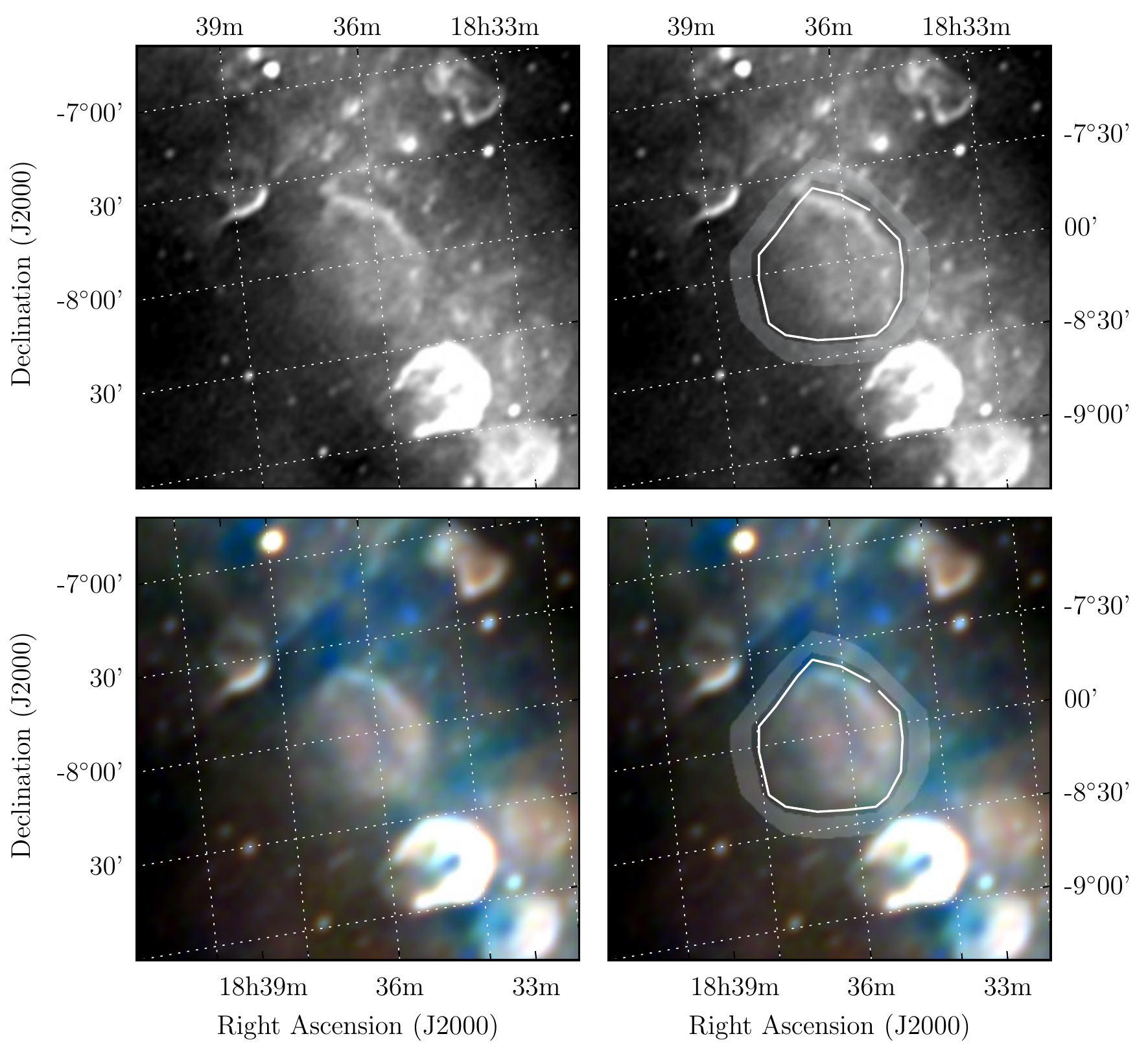}
    \caption{\polysummary G24.0-0.3. \polysuffix}
    \label{fig:SNR_G24.0-0.3_poly}
\end{figure}

\begin{figure}
    \centering
    \includegraphics[width=0.5\textwidth]{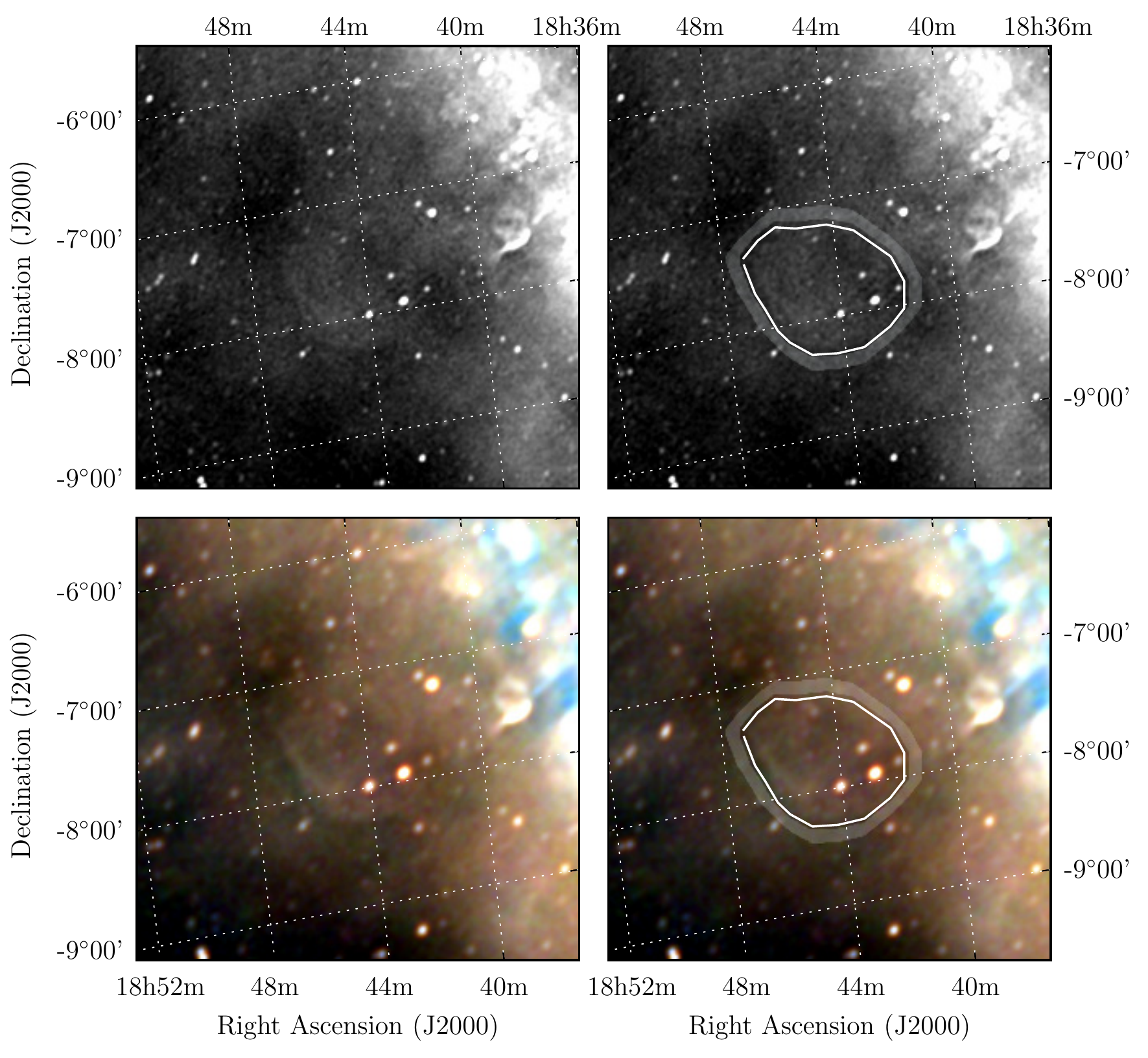}
    \caption{\polysummary G25.3-1.8. \polysuffix}
    \label{fig:SNR_G25.3-1.8_poly}
\end{figure}

\begin{figure}
    \centering
    \includegraphics[width=0.5\textwidth]{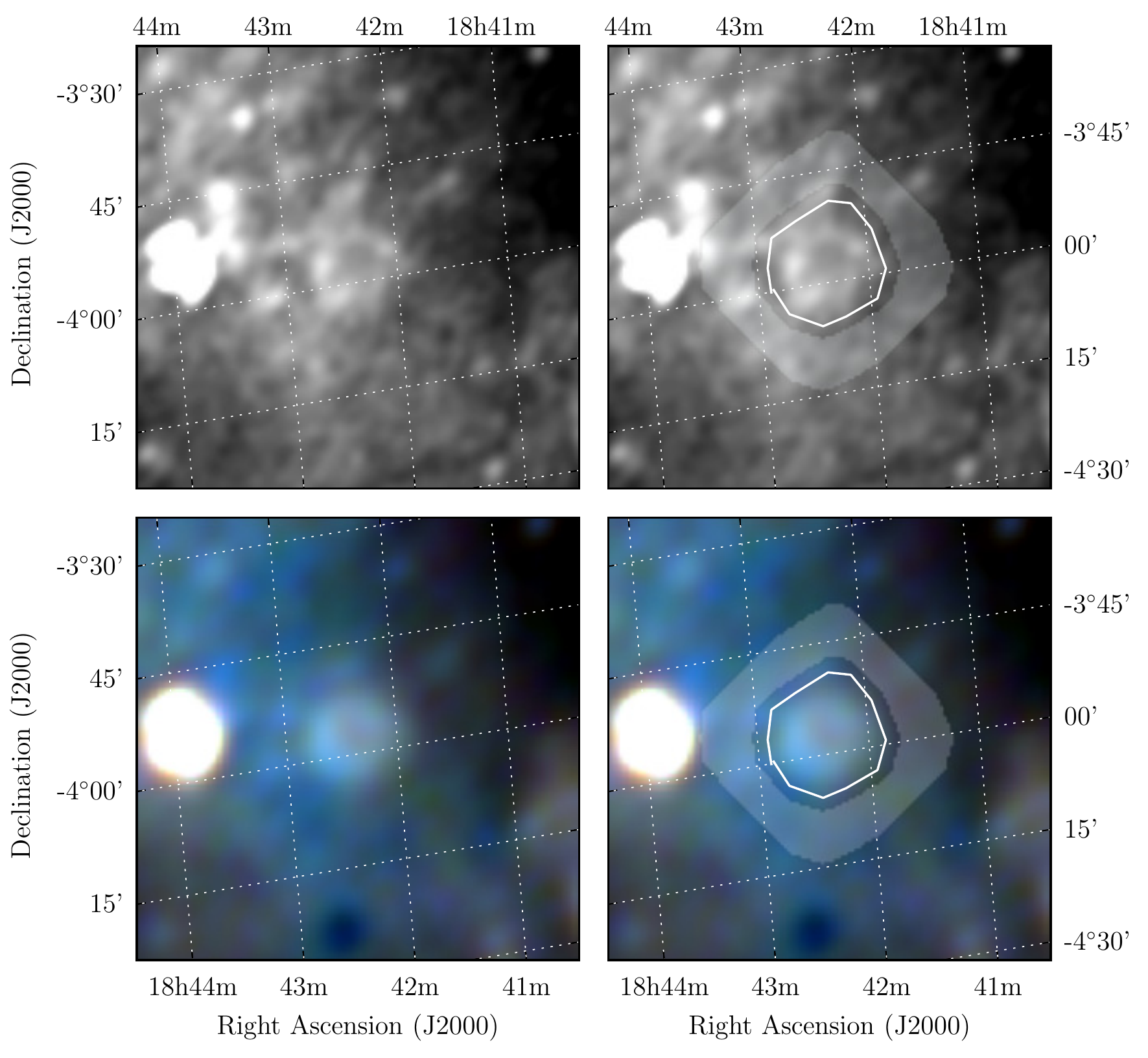}
    \caption{\polysummary G28.3+0.2. \polysuffix}
    \label{fig:SNR_G28.3+0.2_poly}
\end{figure}

\begin{figure}
    \centering
    \includegraphics[width=0.5\textwidth]{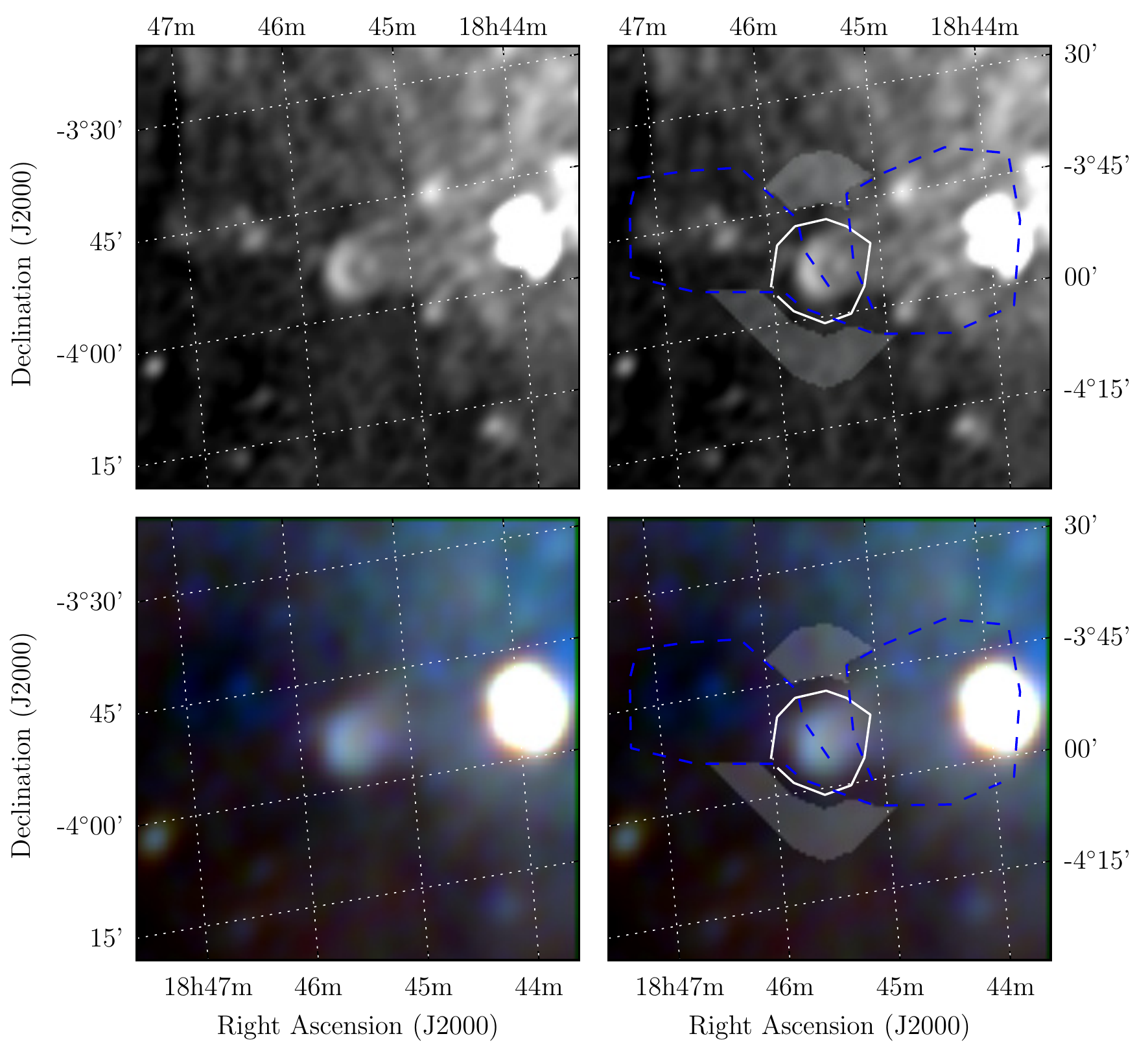}
    \caption{\polysummary G28.7-0.4. \polysuffix}
    \label{fig:SNR_G28.7-0.4_poly}
\end{figure}

\begin{figure}
    \centering
    \includegraphics[width=0.5\textwidth]{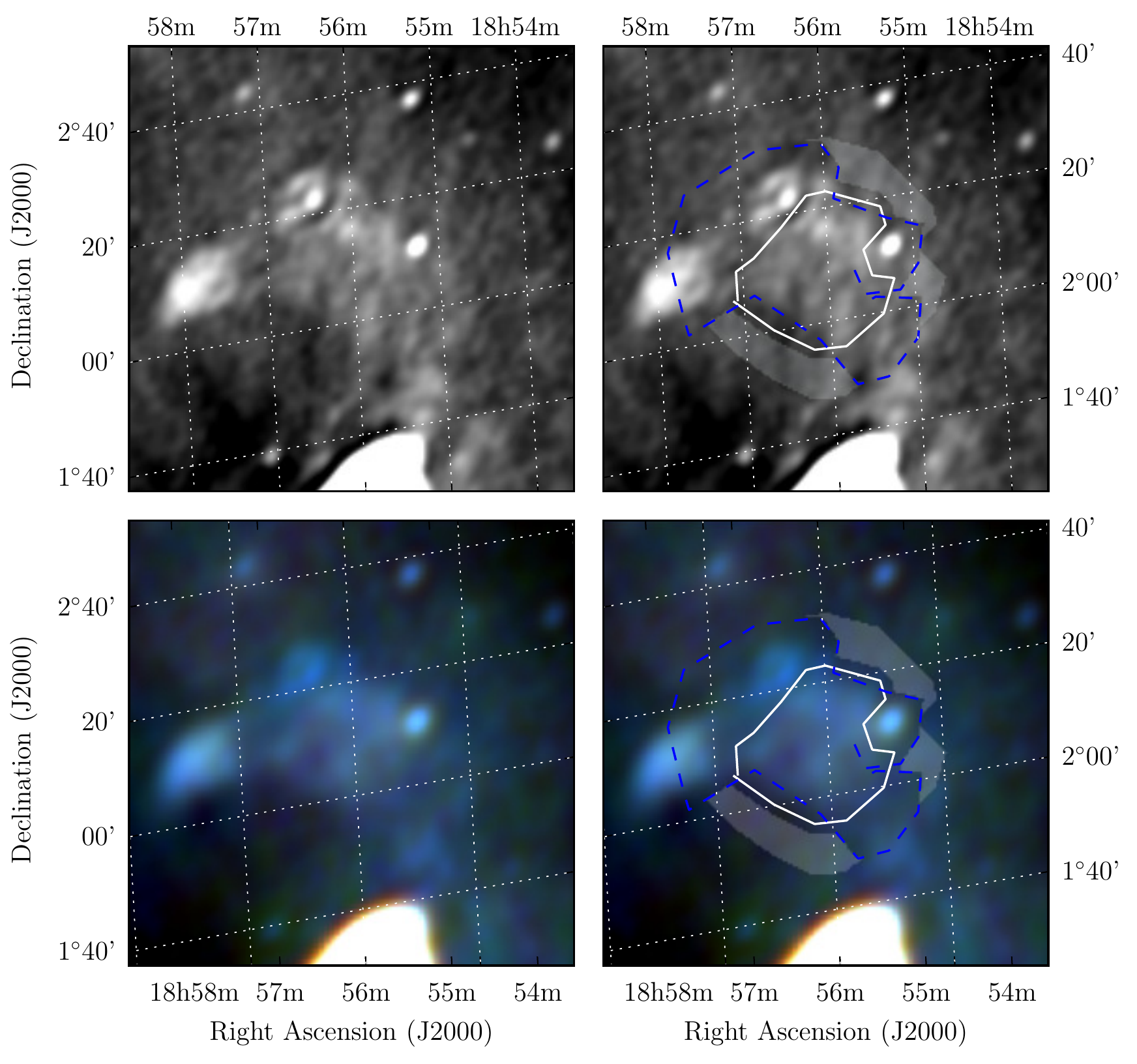}
    \caption{\polysummary G35.3-0.0. \polysuffix}
    \label{fig:SNR_G35.3-0.0_poly}
\end{figure}

\begin{figure}
    \centering
    \includegraphics[width=0.5\textwidth]{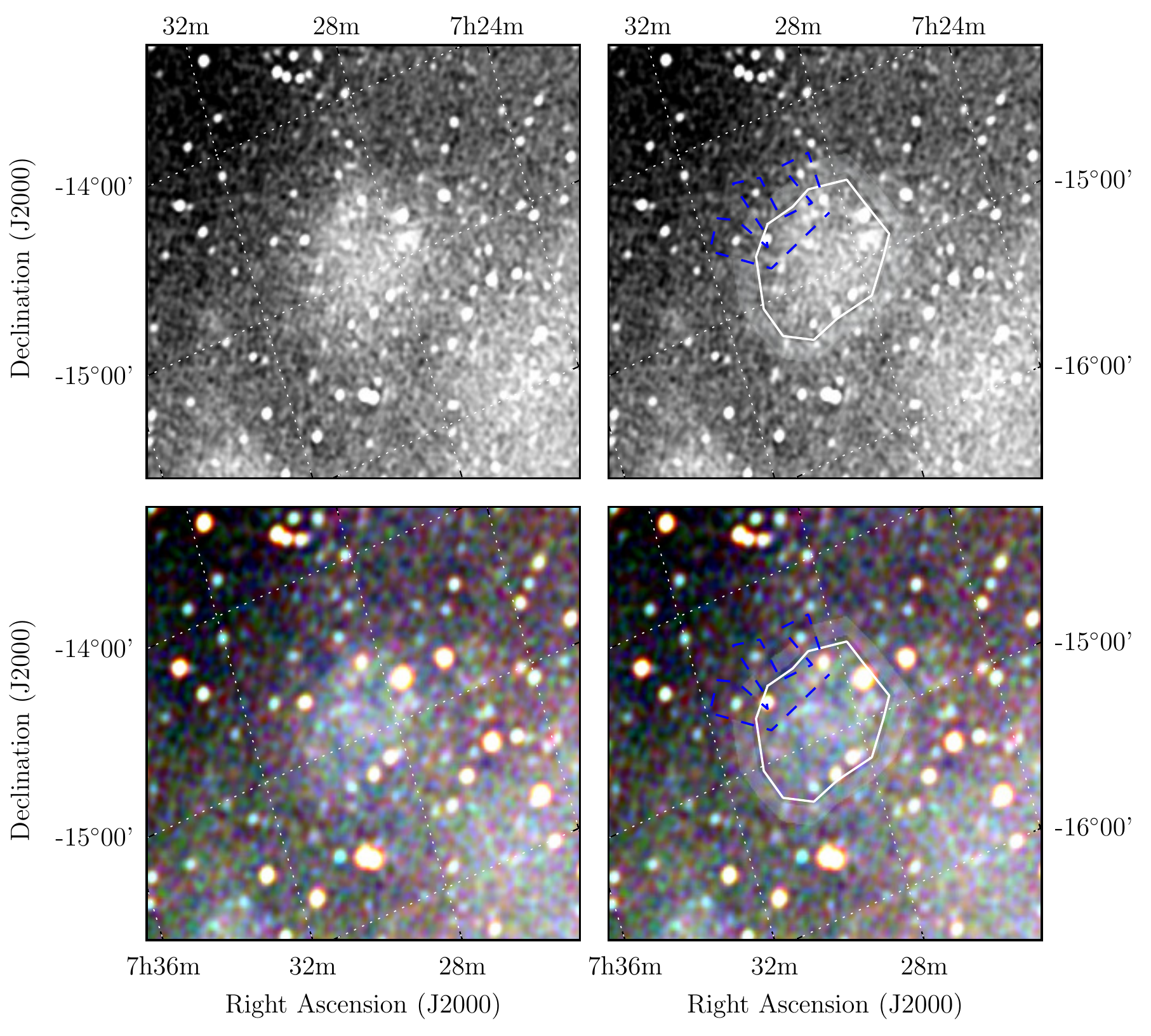}
    \caption{\polysummary G230.4+1.2. \polysuffix}
    \label{fig:SNR_G230.4+1.2_poly}
\end{figure}

\begin{figure}
    \centering
    \includegraphics[width=0.5\textwidth]{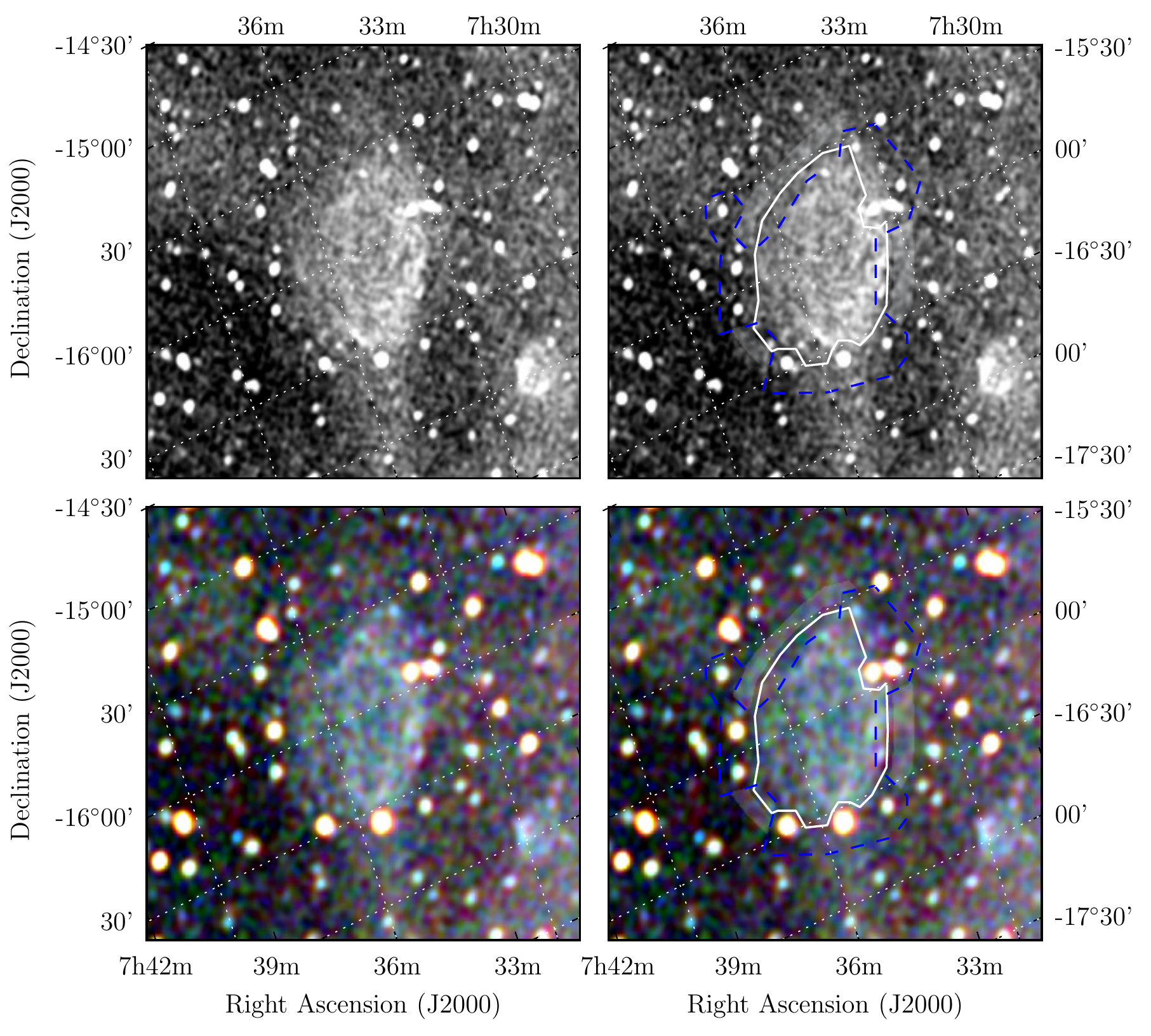}
    \caption{\polysummary G232.1+2.0. \polysuffix}
    \label{fig:SNR_G232.1+2.0_poly}
\end{figure}

\begin{figure}
    \centering
    \includegraphics[width=0.5\textwidth]{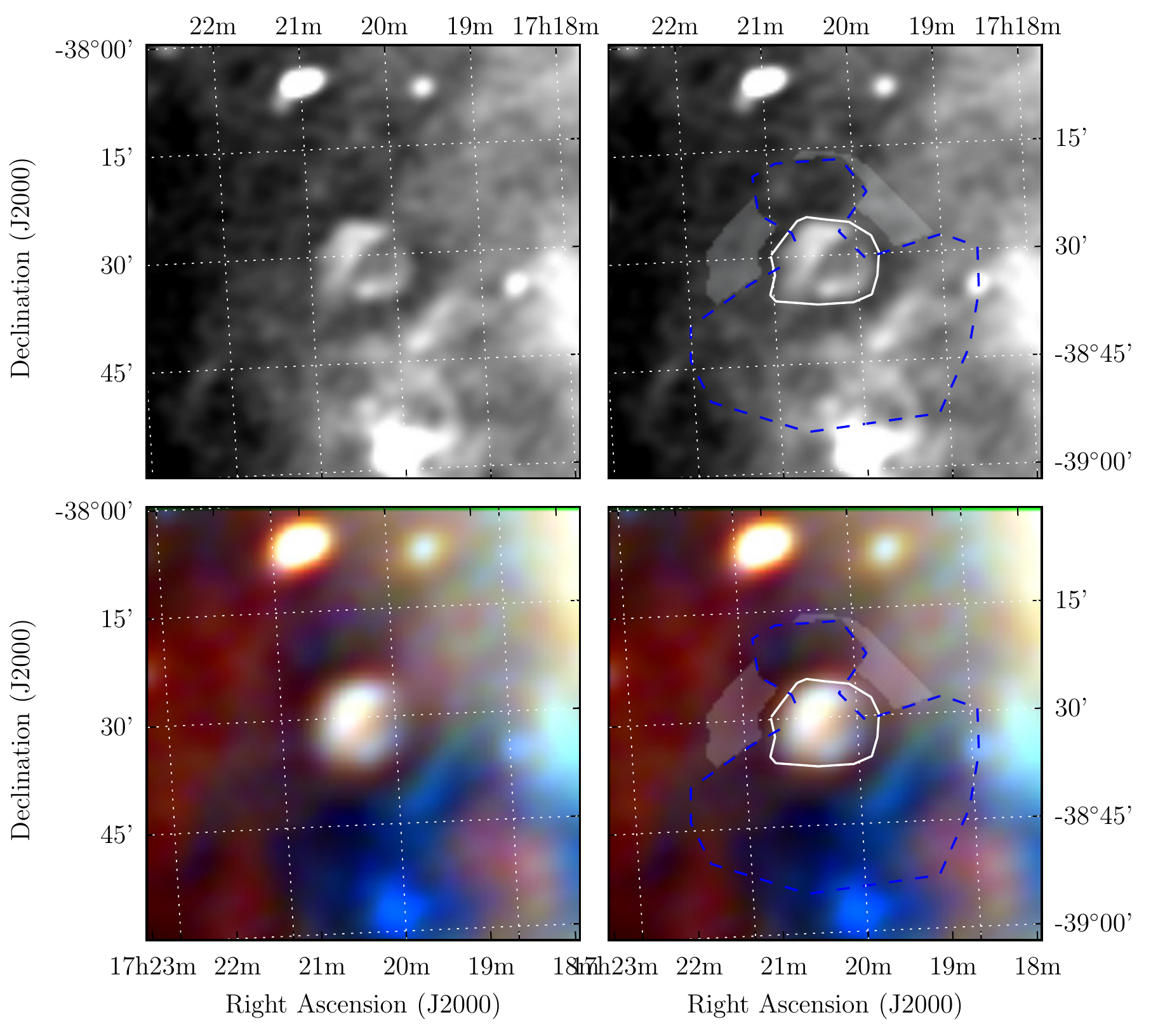}
    \caption{\polysummary G349.1-0.8. \polysuffix}
    \label{fig:SNR_G349.1-0.8_poly}
\end{figure}

\begin{figure}
    \centering
    \includegraphics[width=0.5\textwidth]{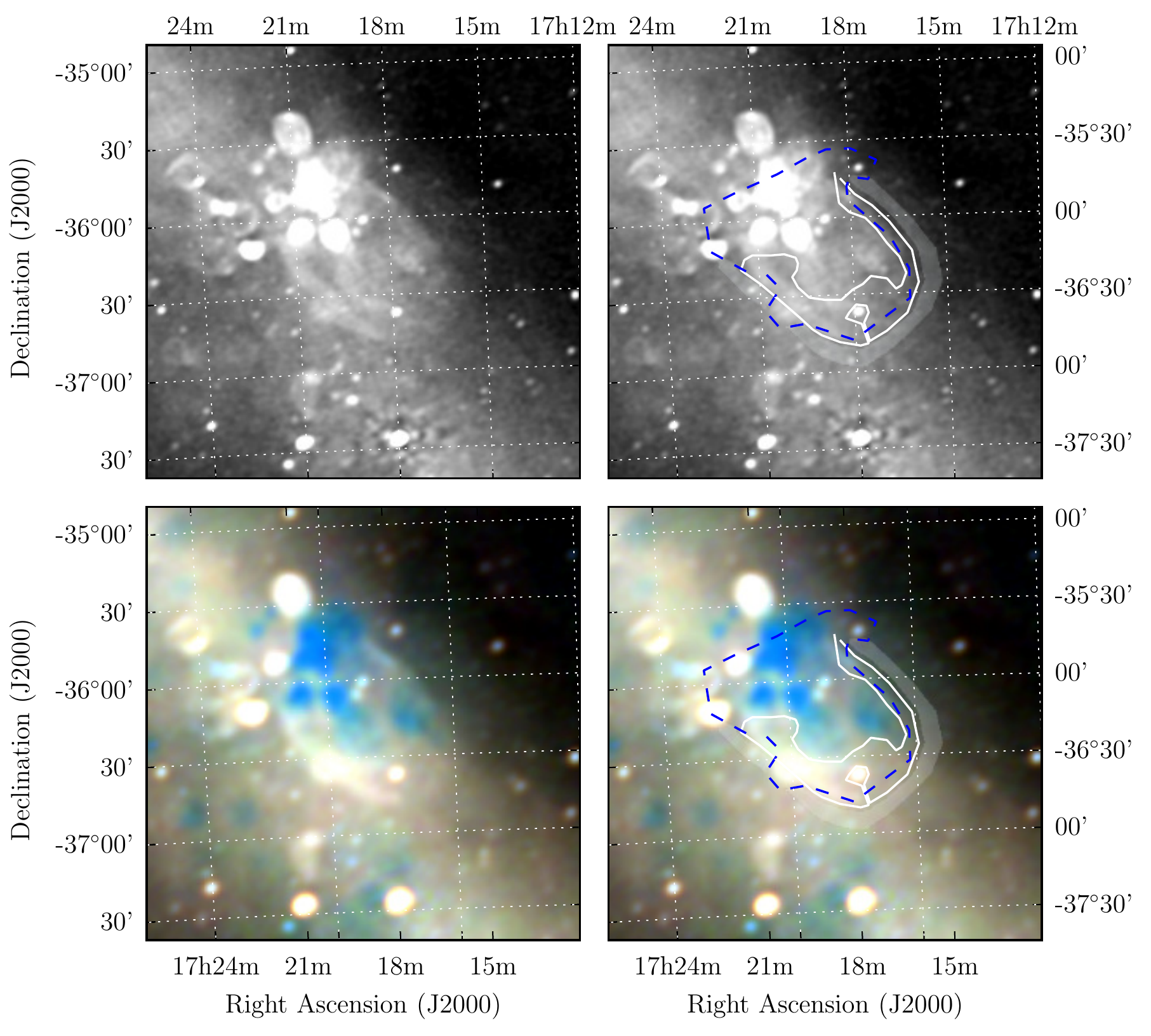}
    \caption{\polysummary G350.7+0.6. \polysuffix}
    \label{fig:SNR_G350.7+0.6_poly}
\end{figure}

\begin{figure}
    \centering
    \includegraphics[width=0.5\textwidth]{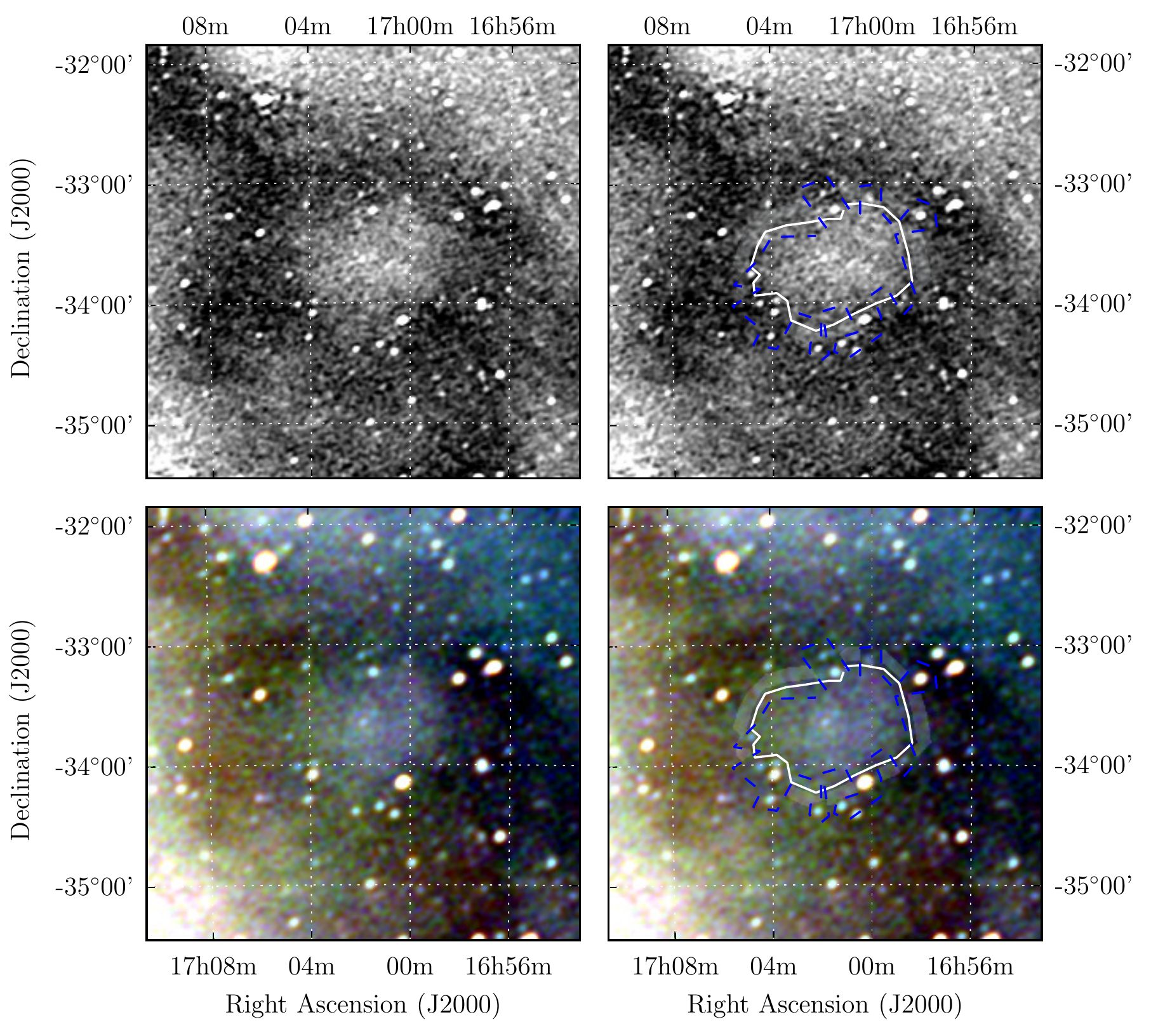}
    \caption{\polysummary G350.8+5.0. \polysuffix}
    \label{fig:SNR_G350.8+5.0_poly}
\end{figure}

\begin{figure}
    \centering
    \includegraphics[width=0.5\textwidth]{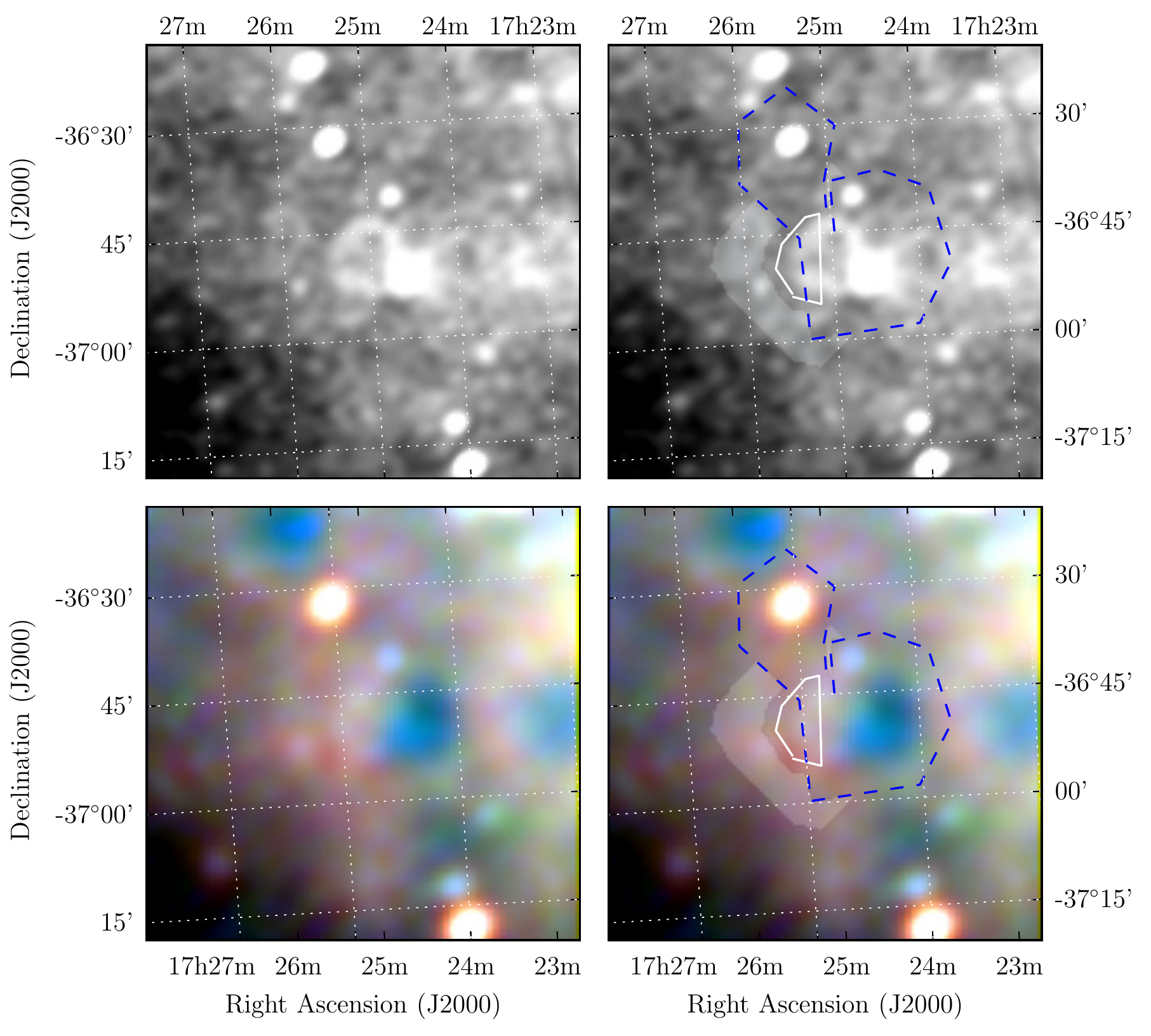}
    \caption{\polysummary G351.0-0.6. \polysuffix}
    \label{fig:SNR_G351.0-0.6_poly}
\end{figure}

\begin{figure}
    \centering
    \includegraphics[width=0.5\textwidth]{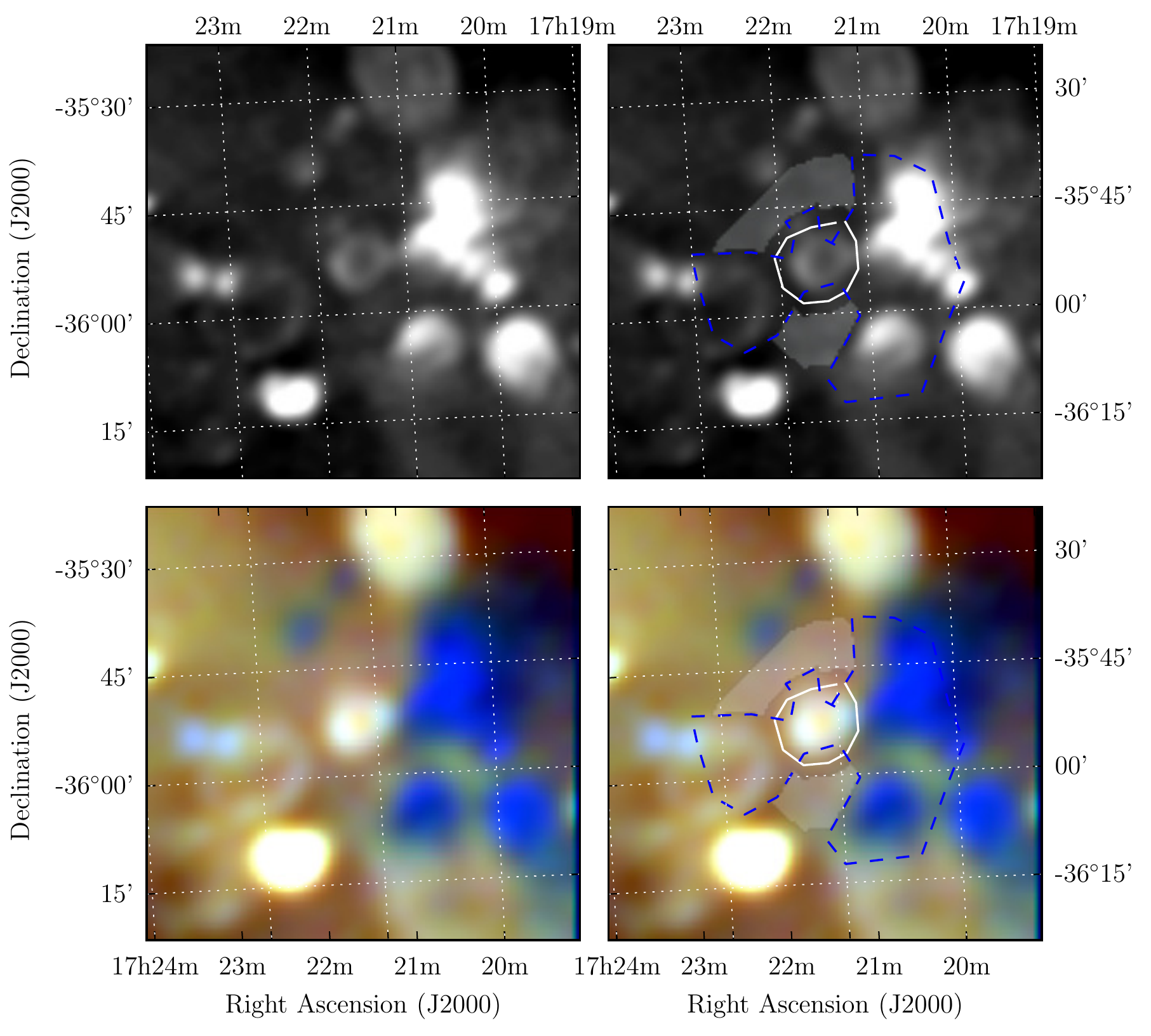}
    \caption{\polysummary G351.4+0.4. \polysuffix}
    \label{fig:SNR_G351.4+0.4_poly}
\end{figure}

\begin{figure}
    \centering
    \includegraphics[width=0.5\textwidth]{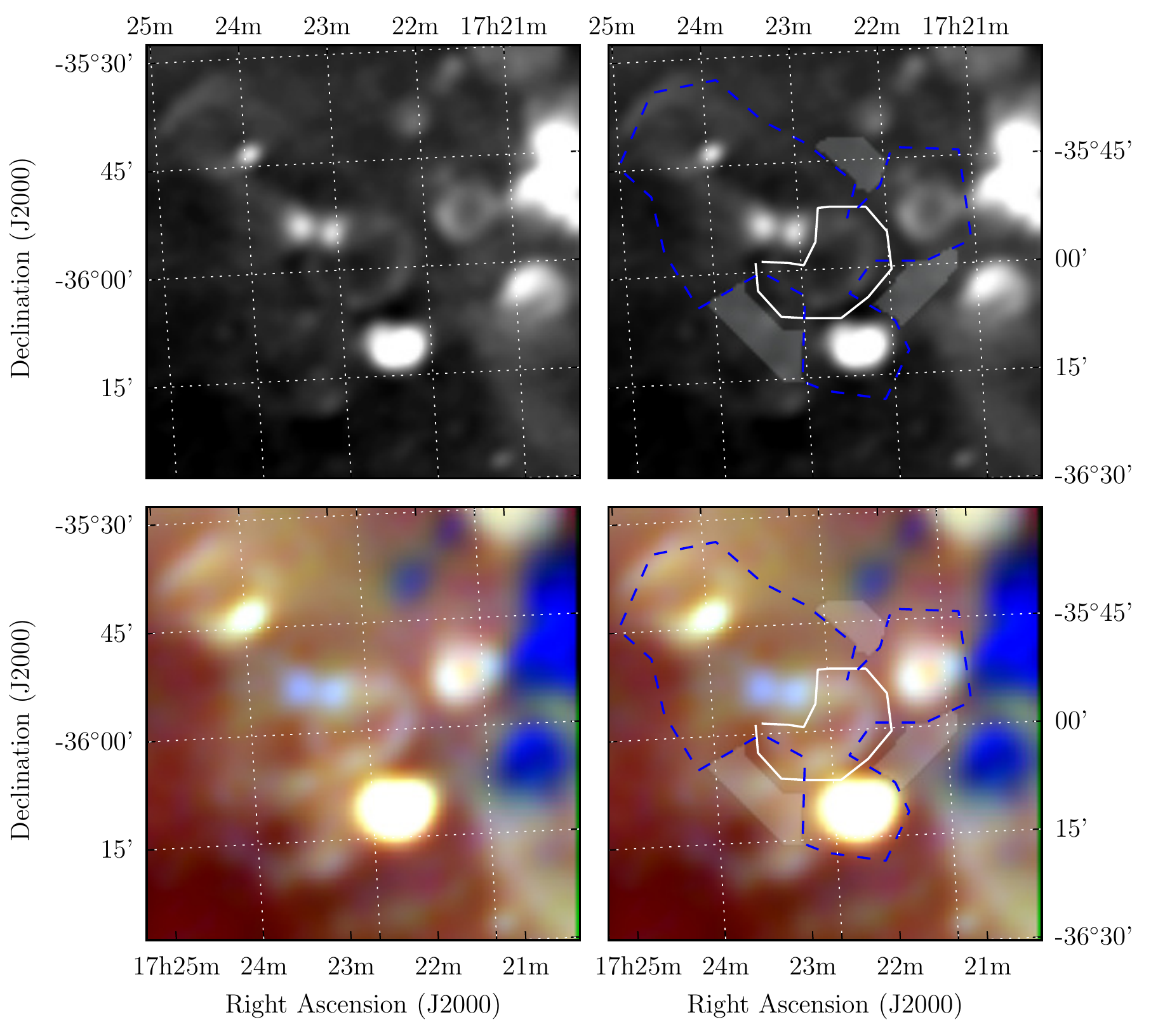}
    \caption{\polysummary G351.4+0.2. \polysuffix}
    \label{fig:SNR_G351.4+0.2_poly}
\end{figure}

\begin{figure}
    \centering
    \includegraphics[width=0.5\textwidth]{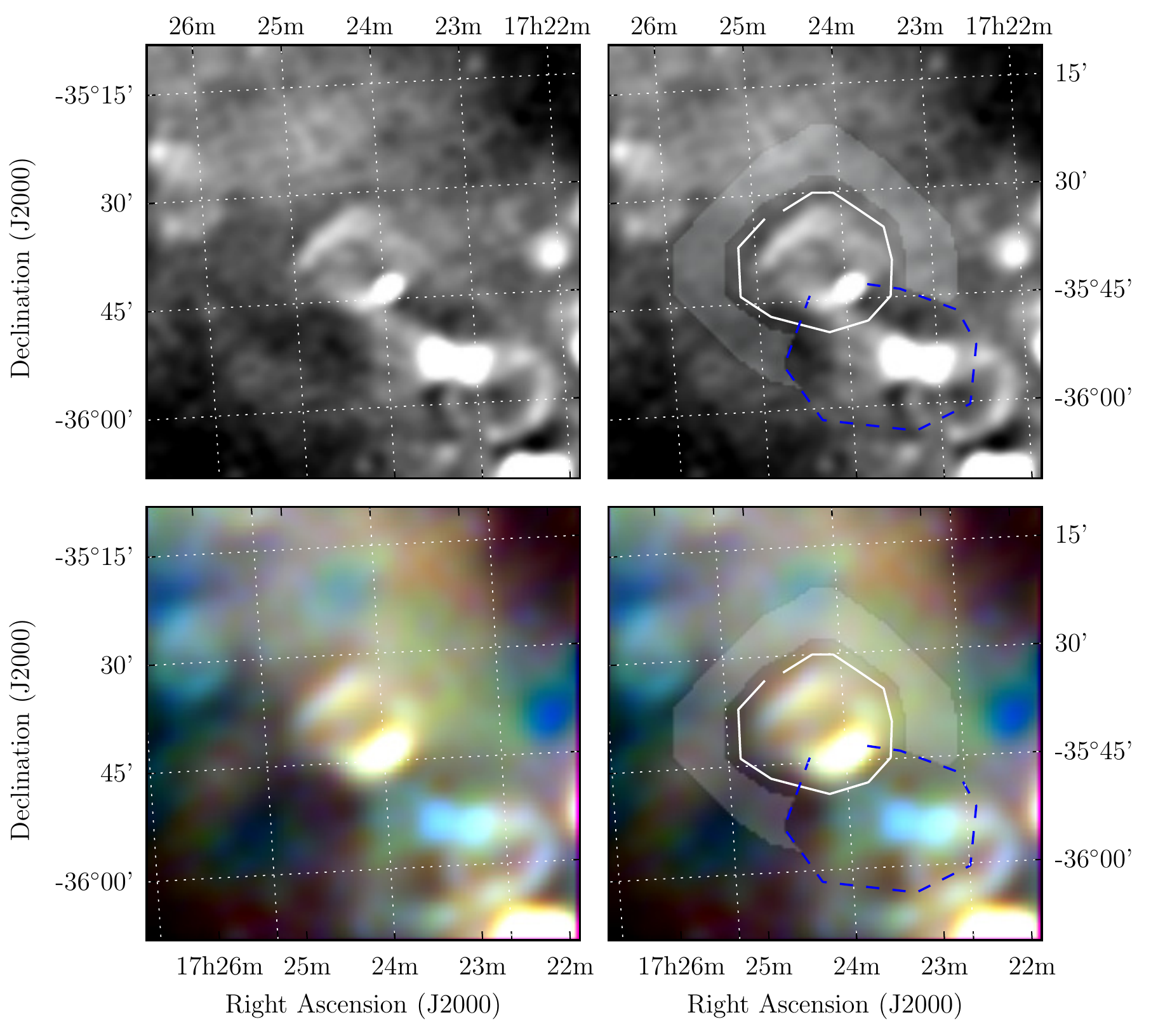}
    \caption{\polysummary G351.9+0.1. \polysuffix}
    \label{fig:SNR_G351.9+0.1_poly}
\end{figure}

\begin{figure}
    \centering
    \includegraphics[width=0.5\textwidth]{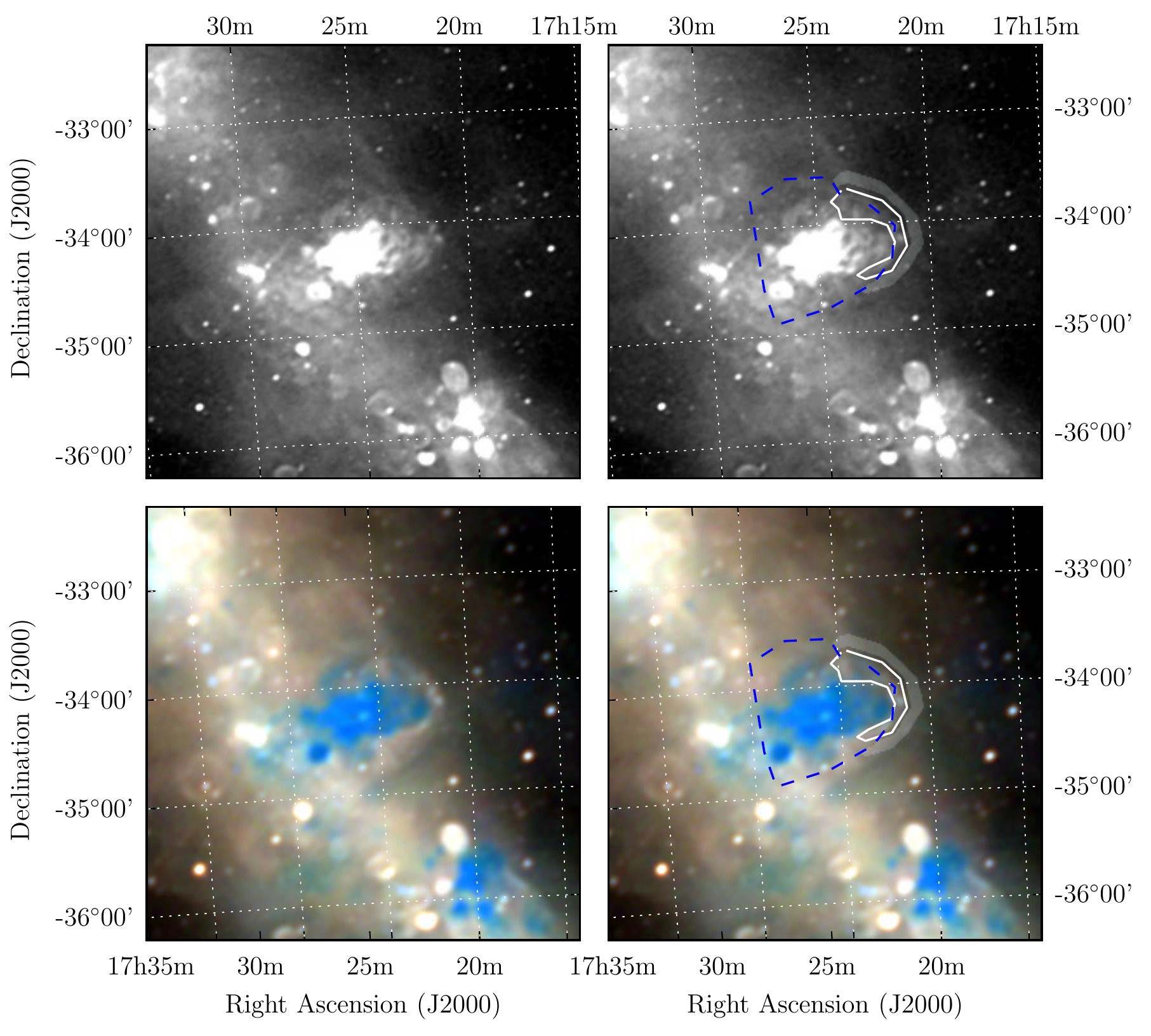}
    \caption{\polysummary G353.0+0.8. \polysuffix}
    \label{fig:SNR_G353.0+0.8_poly}
\end{figure}

\begin{figure}
    \centering
    \includegraphics[width=0.5\textwidth]{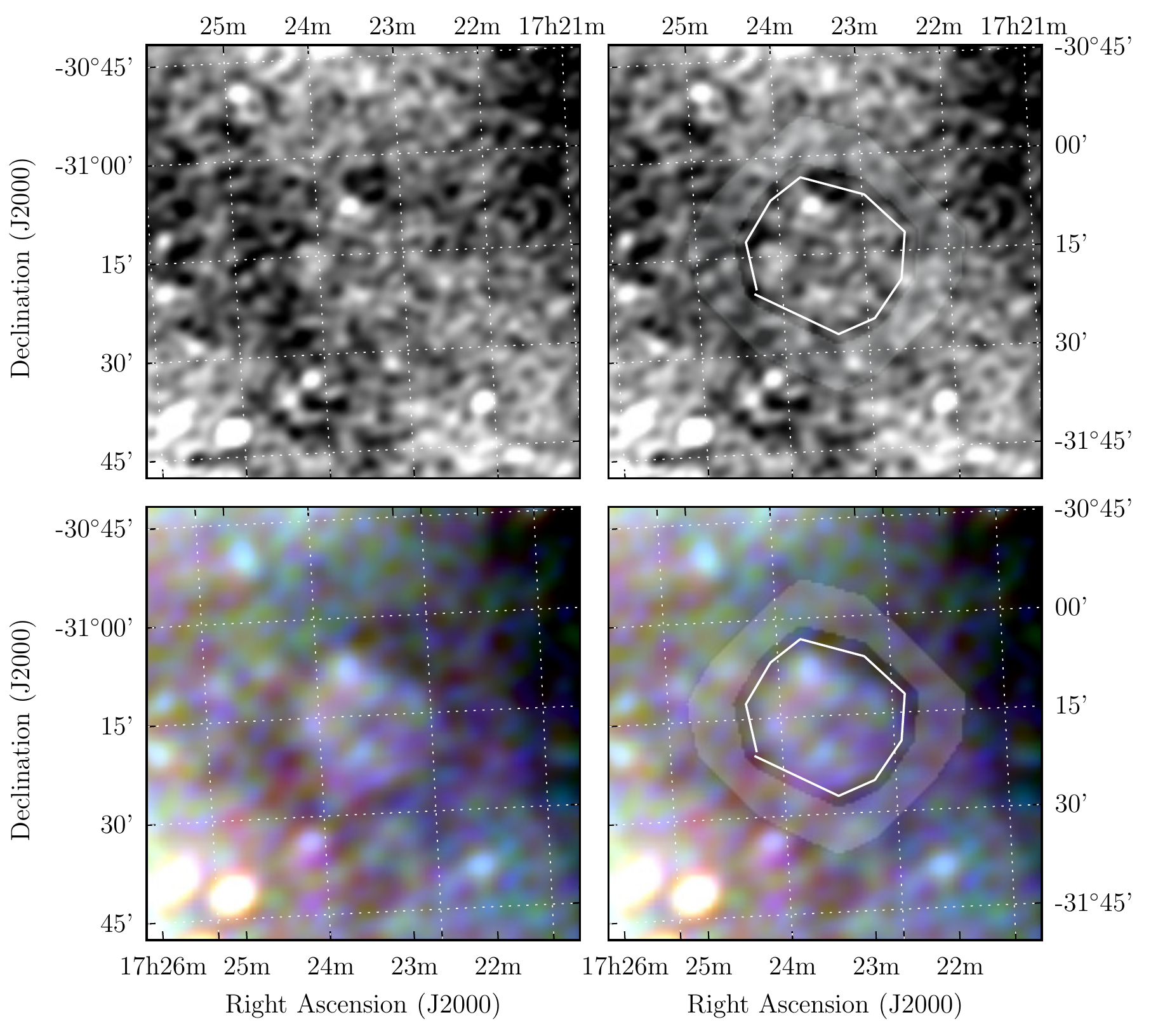}
    \caption{\polysummary G355.4+2.7. \polysuffix}
    \label{fig:SNR_G355.4+2.7_poly}
\end{figure}

\begin{figure}
    \centering
    \includegraphics[width=0.5\textwidth]{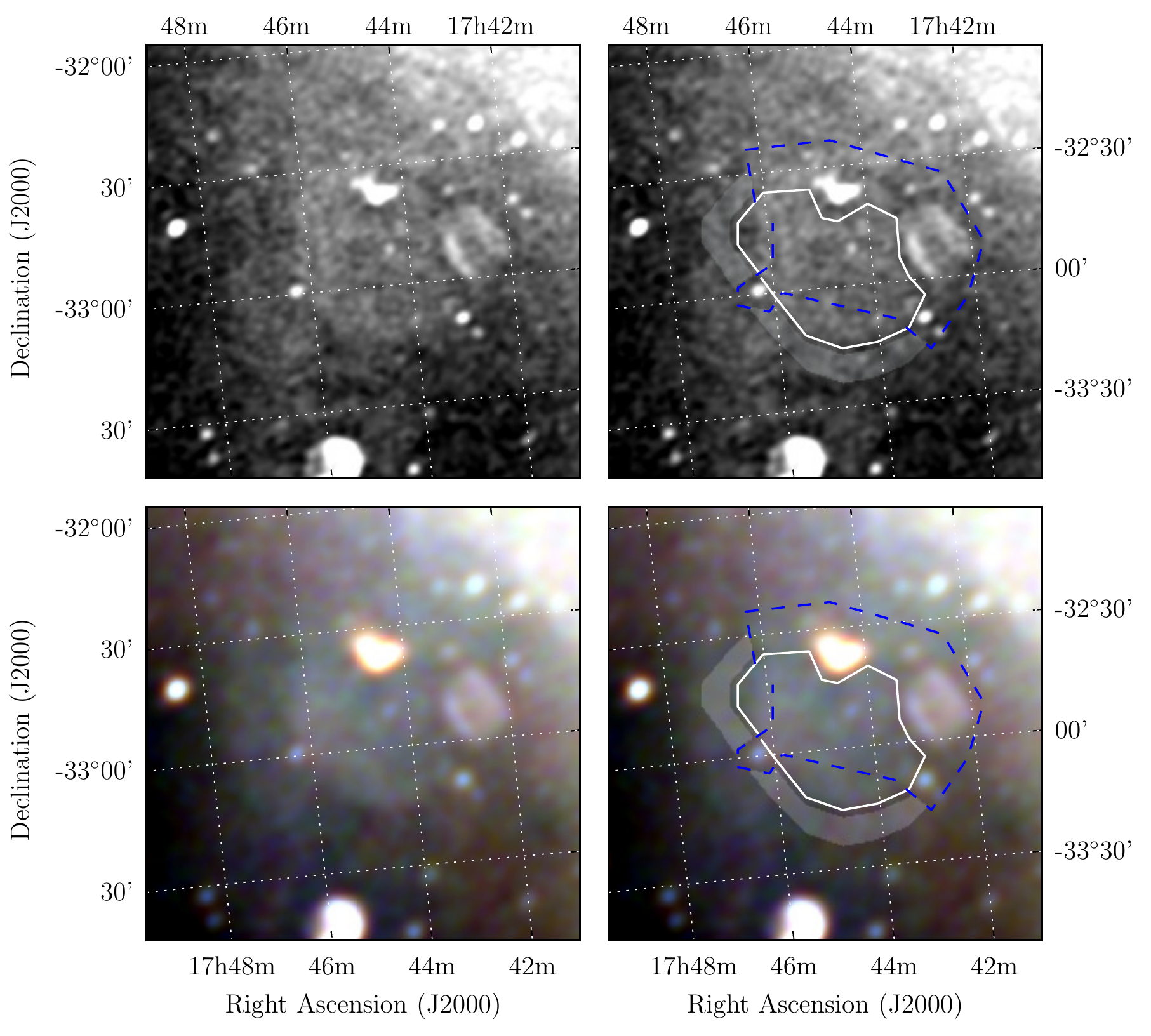}
    \caption{\polysummary G356.5-1.9. \polysuffix}
    \label{fig:SNR_G356.5-1.9_poly}
\end{figure}

\begin{figure}
    \centering
    \includegraphics[width=0.5\textwidth]{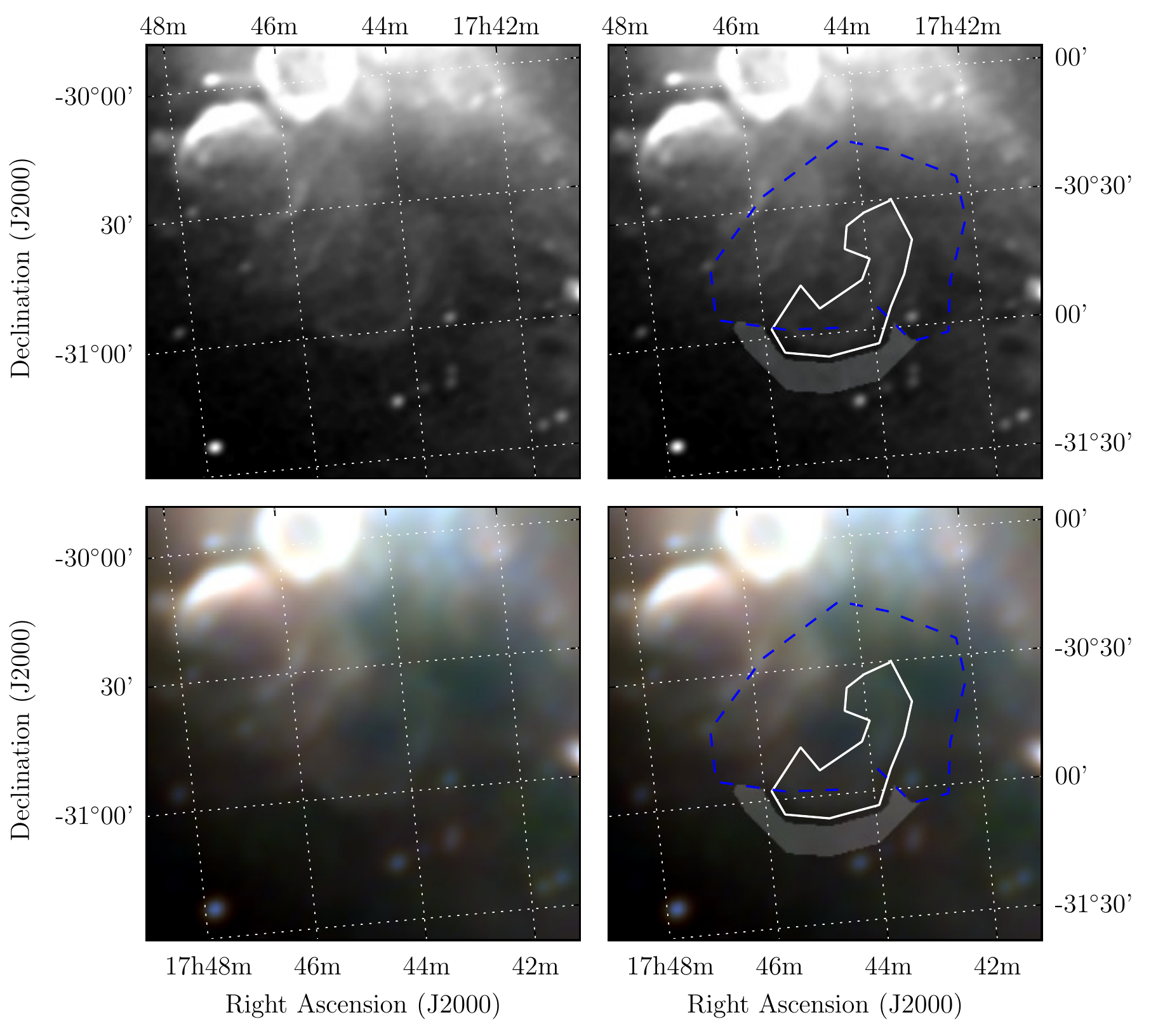}
    \caption{\polysummary G358.3-0.7. \polysuffix}
    \label{fig:SNR_G358.3-0.7_poly}
\end{figure}

\section{Spectra}

The spectra of the measured SNR using the backgrounding and flux summing technique described in \Sects~2.3 and 2.4 of Paper~\textsc{I}, for the 26 objects for which spectra could be derived. The left panels show flux density against frequency with linear axes while the right panels show the same data in log. (It is useful to include both when analysing the data as a log plot does not render negative data points, which occur for faint SNRs). The black points show the (background-subtracted) SNR flux density measurements, the red points show the measured background, and the blue curve shows a linear fit to the log-log data (i.e. $S_\nu \propto \nu^\alpha$). The fitted value of $\alpha$ is shown at the top right of each plot.

\begin{figure}
    \centering
    \includegraphics[width=0.5\textwidth]{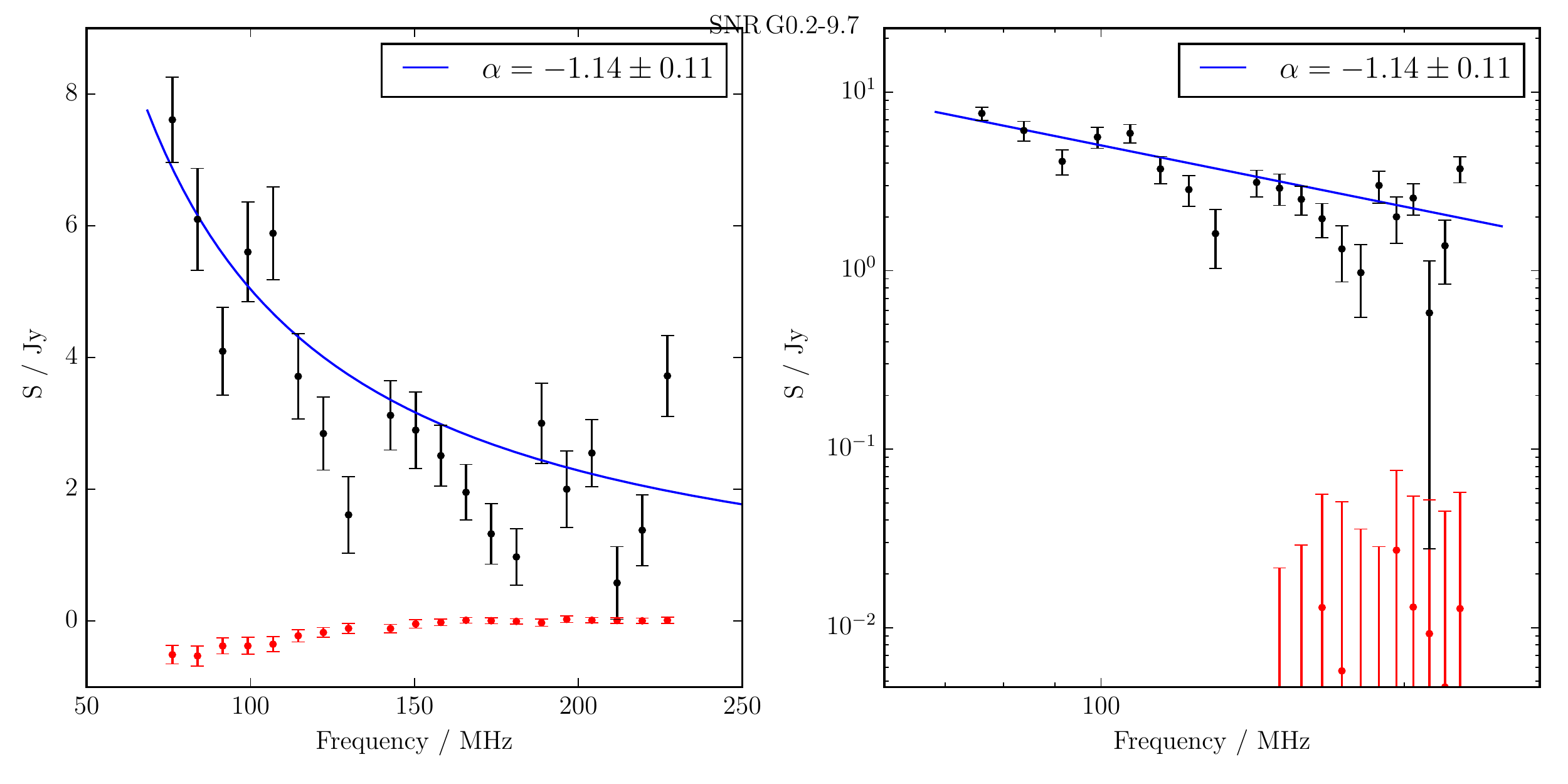}
    \caption{\spectrasummary\spectraN G\,$0.1-9.7$. \spectrasuffix}
    \label{fig:SNR_G0.1-9.7_spectrum}
\end{figure}

\begin{figure}
    \centering
    \includegraphics[width=0.5\textwidth]{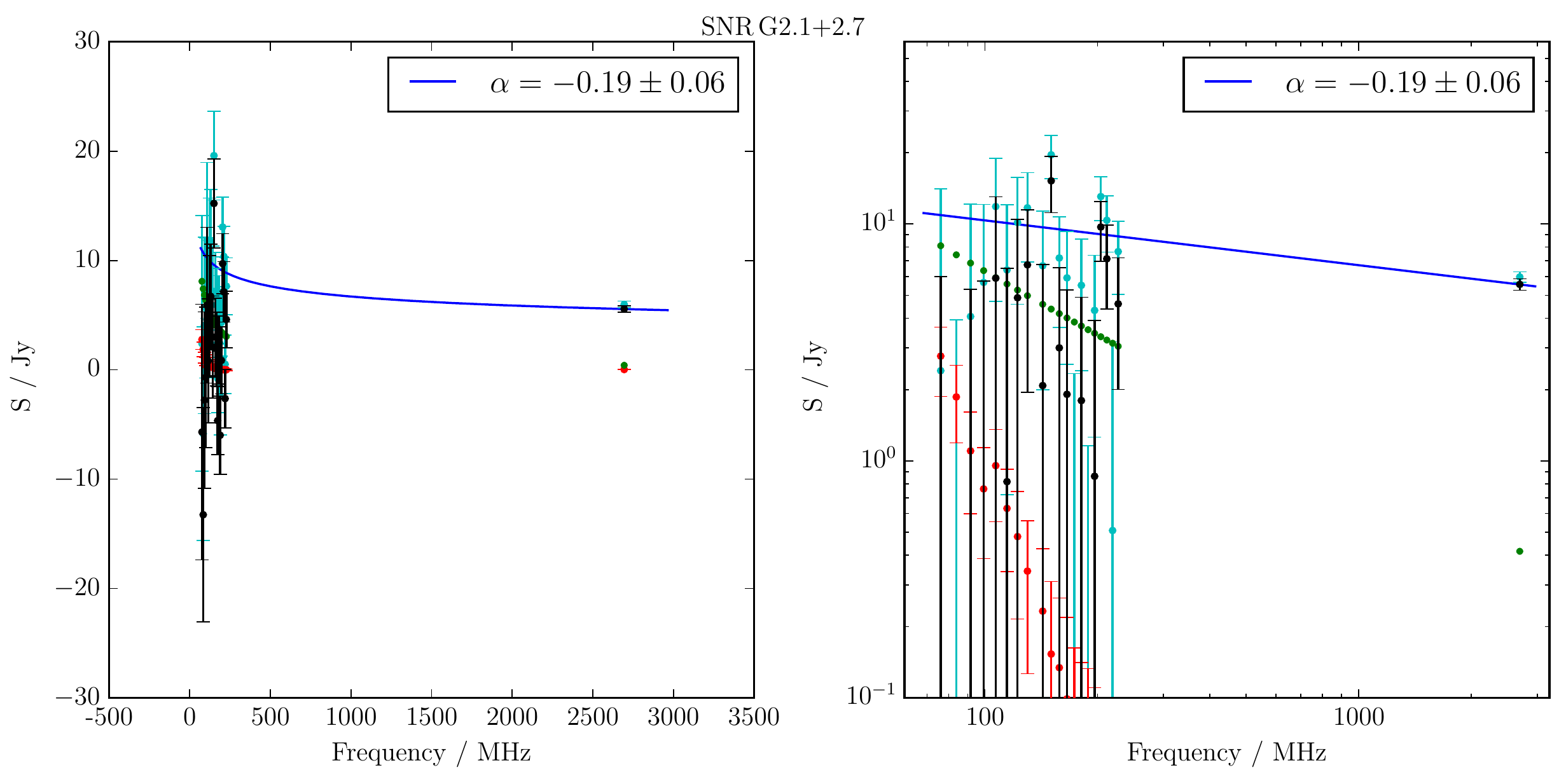}
    \caption{\spectrasummary\spectraE G\,$2.1+2.7$. Cyan points show the SNR flux densities before contaminating sources (green points) were subtracted. \spectrasuffix}
    \label{fig:SNR_G2.1+2.7_spectrum}
\end{figure}

\begin{figure}
    \centering
    \includegraphics[width=0.5\textwidth]{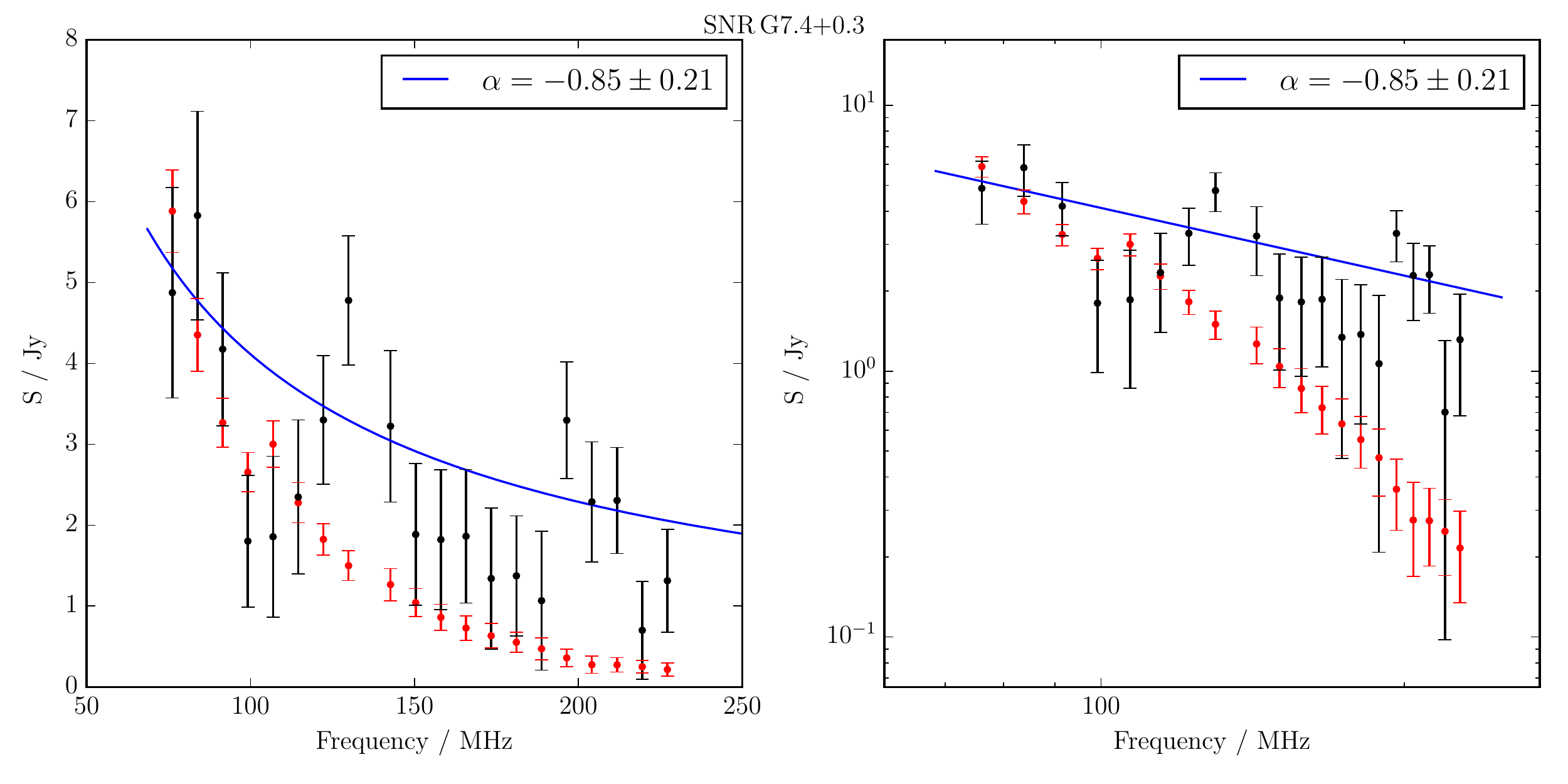}
    \caption{\spectrasummary\spectraN G\,$7.4+0.3$. \spectrasuffix}
    \label{fig:SNR_G7.4+0.3_spectrum}
\end{figure}

\begin{figure}
    \centering
    \includegraphics[width=0.5\textwidth]{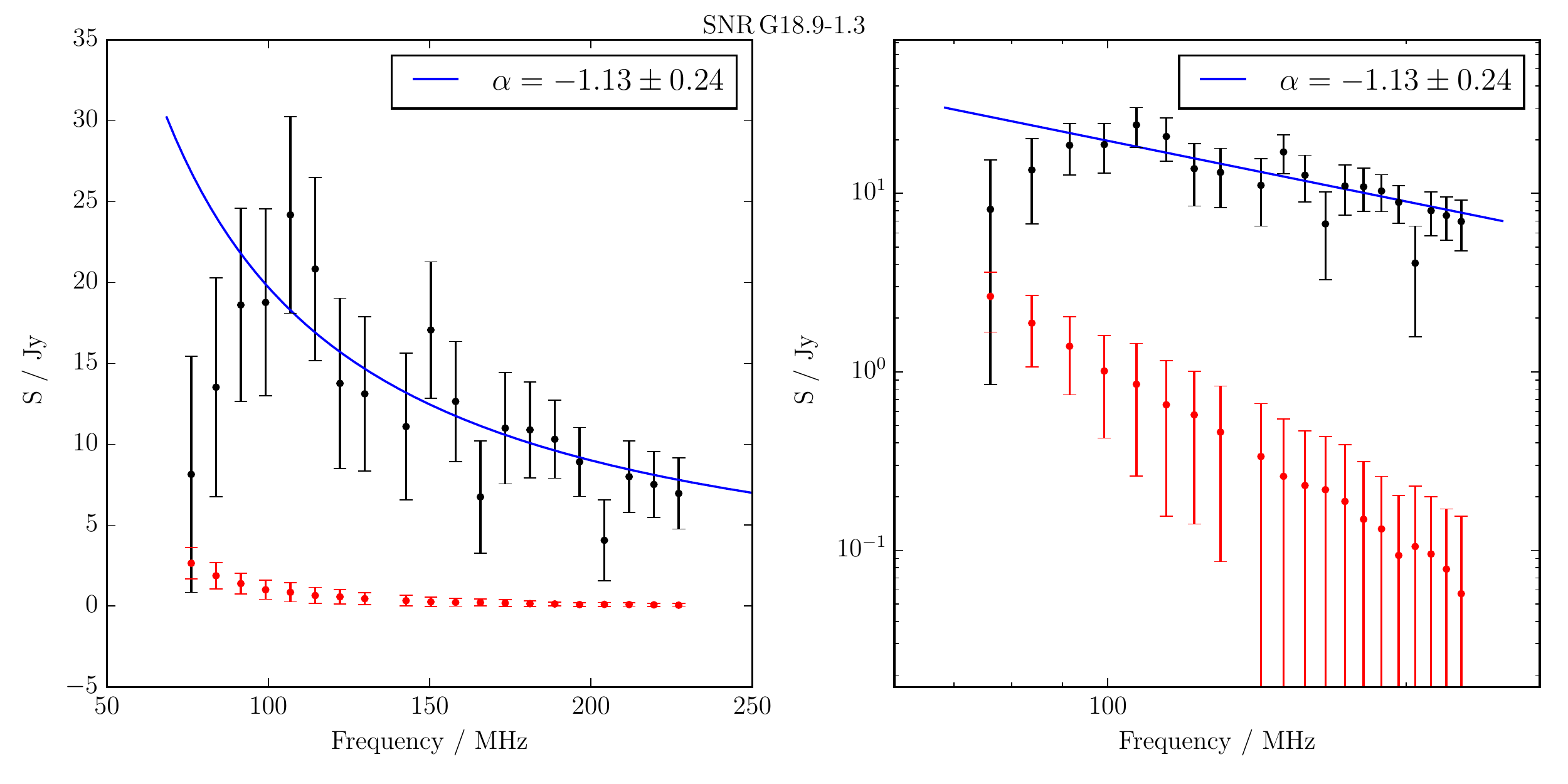}
    \caption{\spectrasummary\spectraN G\,$18.9-1.2$. \spectrasuffix}
    \label{fig:SNR_G18.9-1.2_spectrum}
\end{figure}

\begin{figure}
    \centering
    \includegraphics[width=0.5\textwidth]{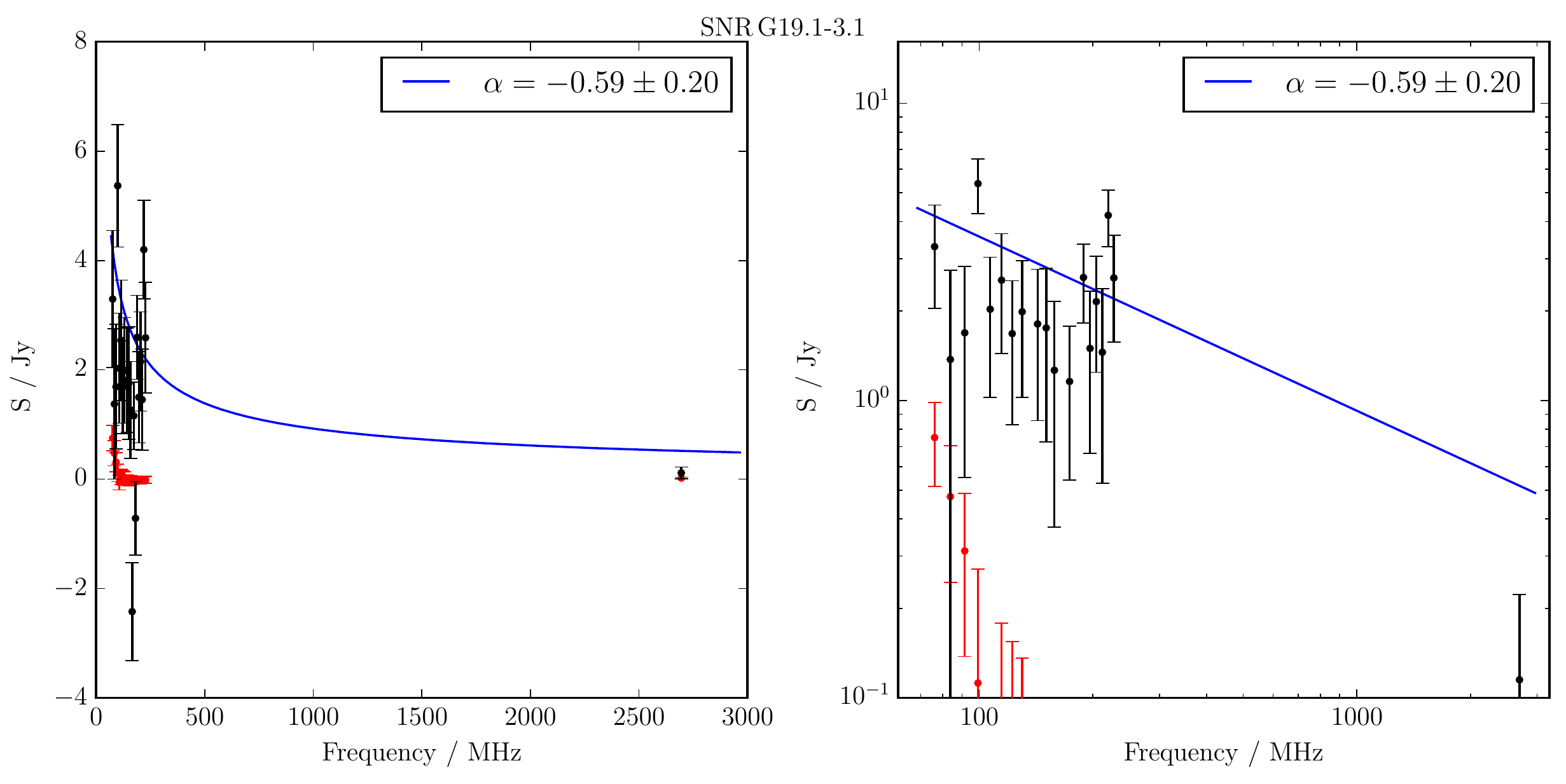}
    \caption{\spectrasummary\spectraE G\,$19.1-3.1$. \spectrasuffix}
    \label{fig:SNR_G19.1-3.1_spectrum}
\end{figure}

\begin{figure}
    \centering
    \includegraphics[width=0.5\textwidth]{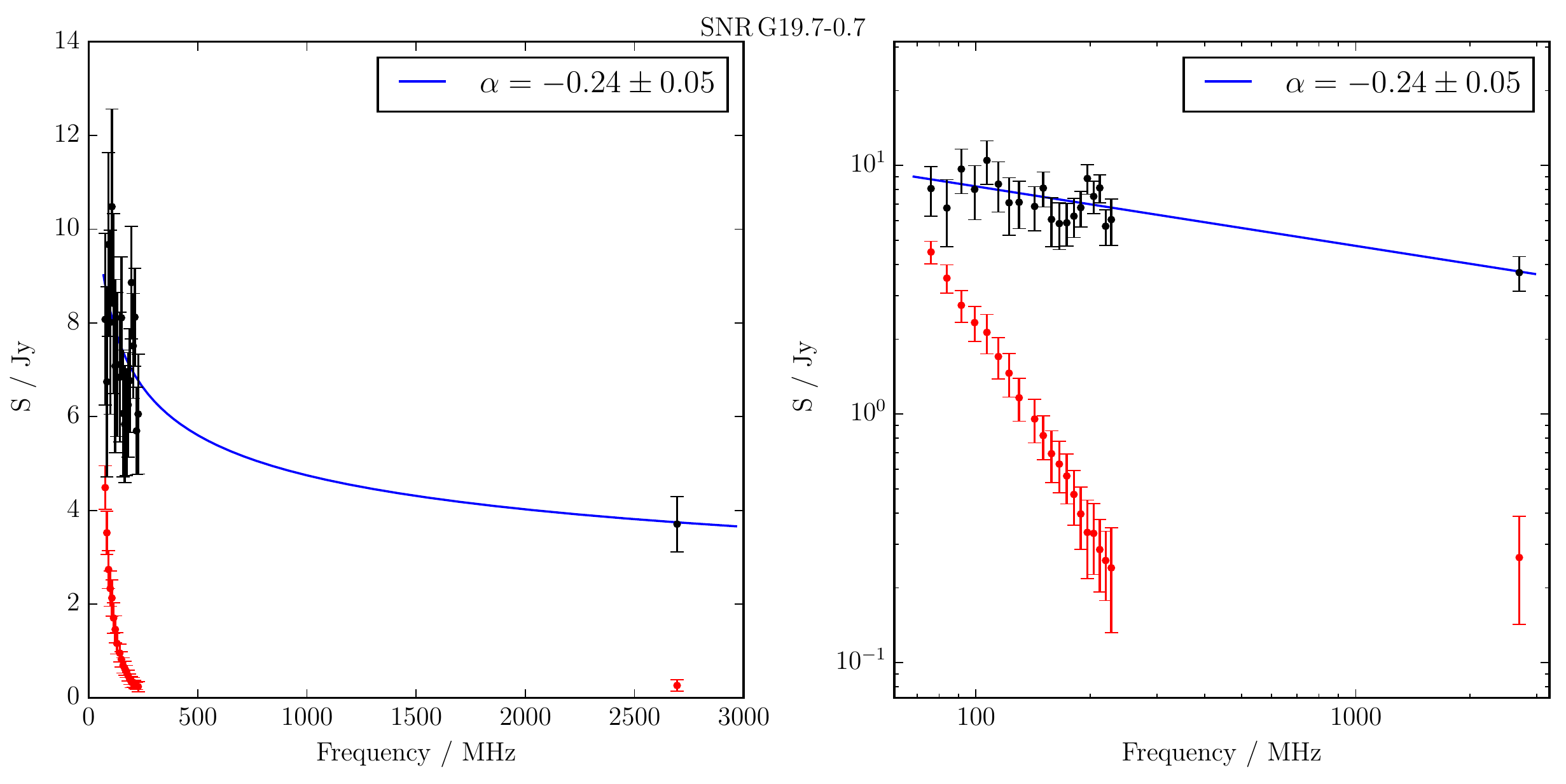}
    \caption{\spectrasummary\spectraE G\,$19.7-0.7$. \spectrasuffix}
    \label{fig:SNR_G19.7-0.7_spectrum}
\end{figure}

\begin{figure}
    \centering
    \includegraphics[width=0.5\textwidth]{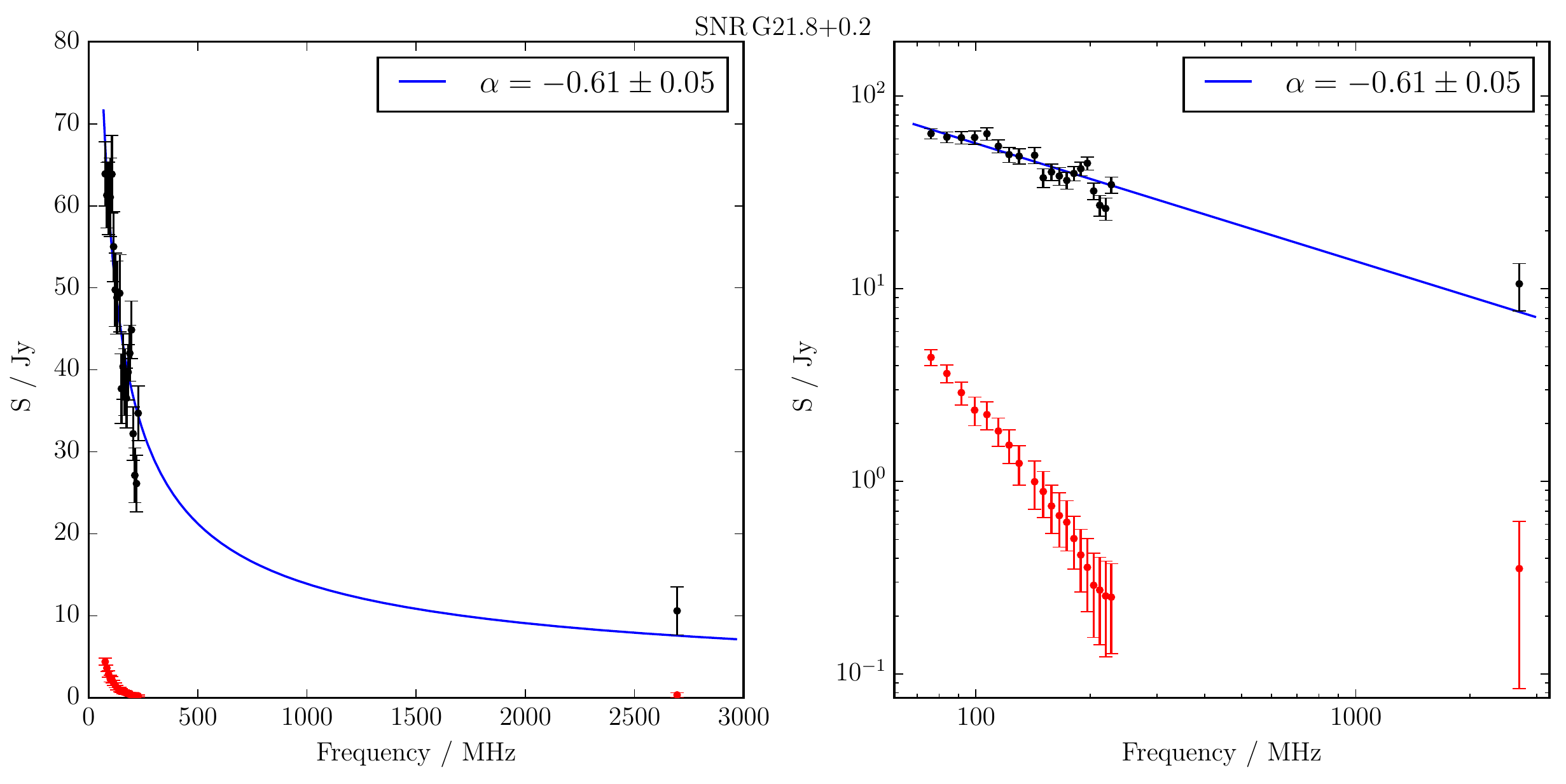}
    \caption{\spectrasummary\spectraE G\,$21.8+0.2$. \spectrasuffix}
    \label{fig:SNR_G21.8+0.2_spectrum}
\end{figure}

\begin{figure}
    \centering
    \includegraphics[width=0.5\textwidth]{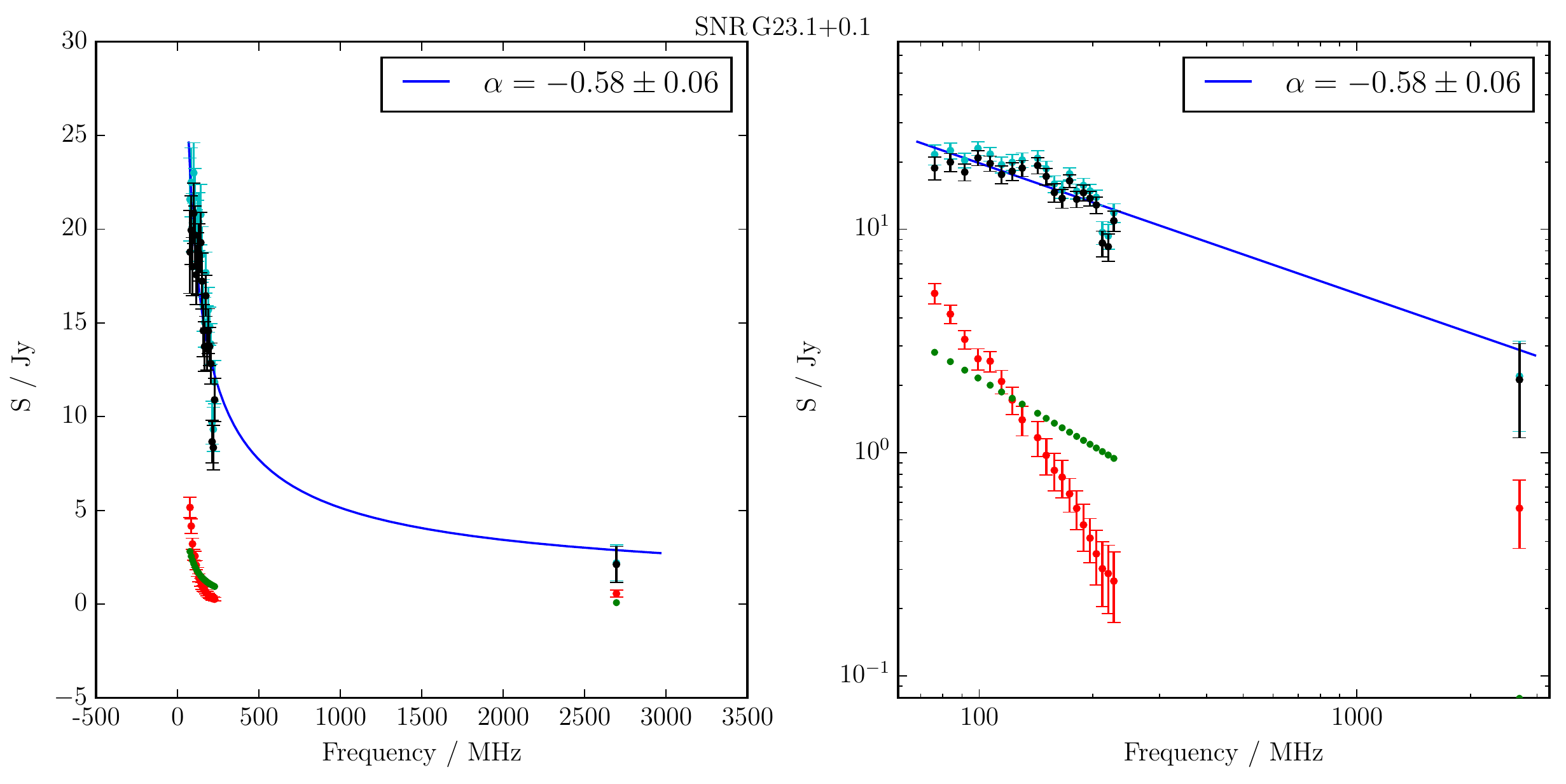}
    \caption{\spectrasummary\spectraE G\,$23.1+0.1$. Cyan points show the SNR flux densities before a contaminating source (green points) was subtracted. \spectrasuffix}
    \label{fig:SNR_G23.1+0.1_spectrum}
\end{figure}

\begin{figure}
    \centering
    \includegraphics[width=0.5\textwidth]{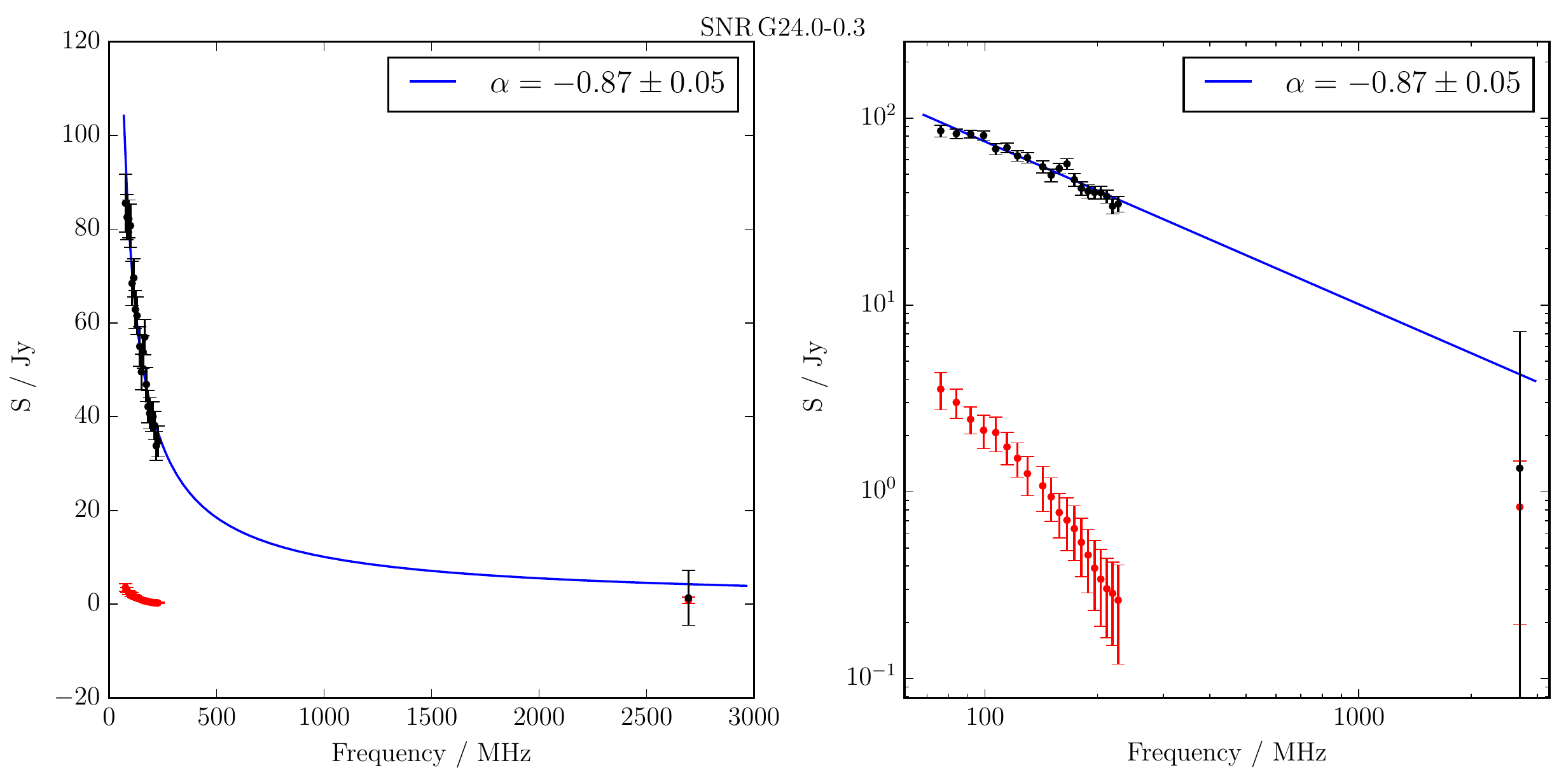}
    \caption{\spectrasummary\spectraE G\,$24.0-0.3$. \spectrasuffix}
    \label{fig:SNR_G24.0-0.3_spectrum}
\end{figure}

\begin{figure}
    \centering
    \includegraphics[width=0.5\textwidth]{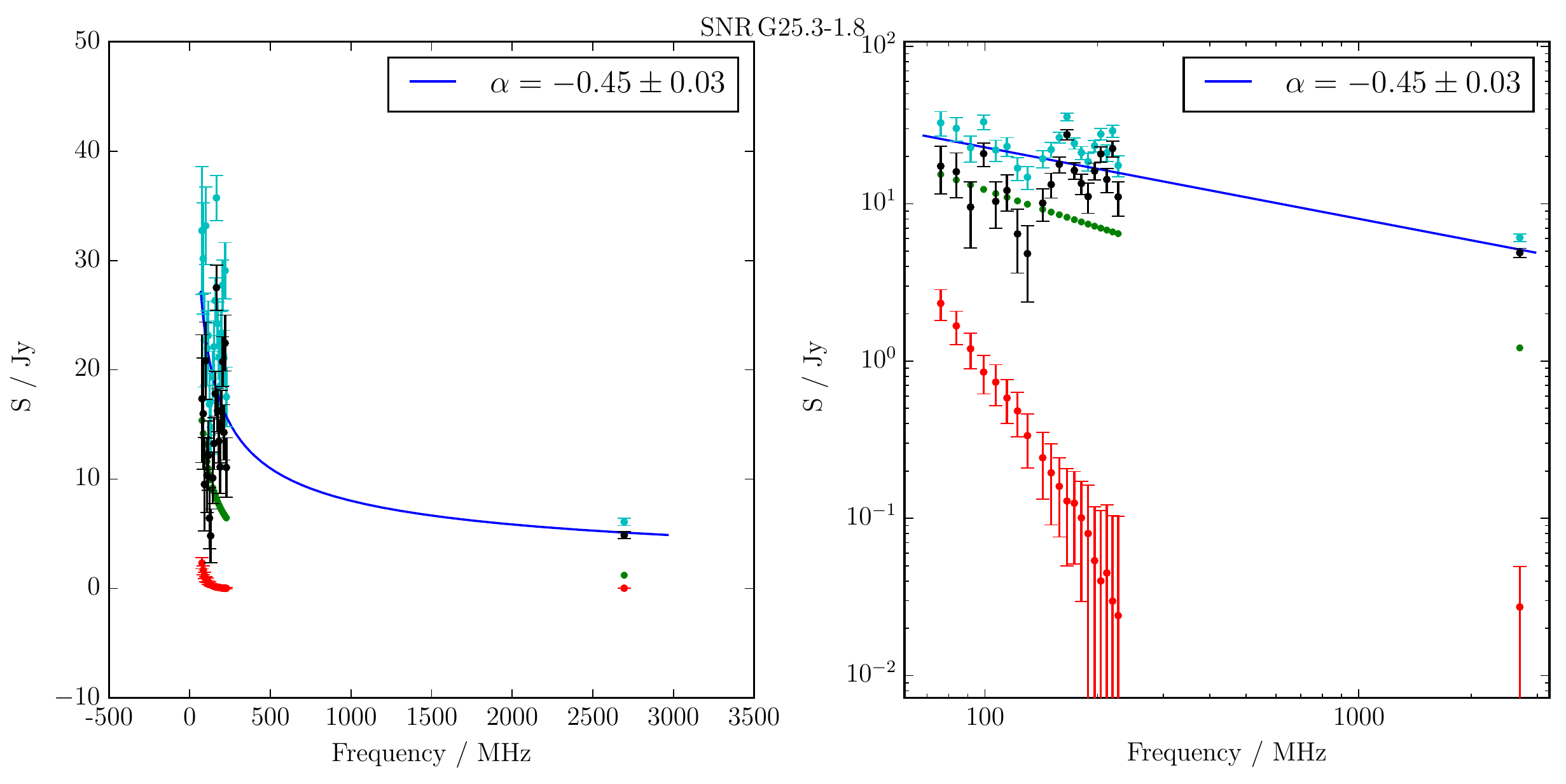}
    \caption{\spectrasummary\spectraE G\,$25.3-1.8$. Cyan points show the SNR flux densities before contaminating sources (green points) were subtracted. \spectrasuffix}
    \label{fig:SNR_G25.3-1.8_spectrum}
\end{figure}

\begin{figure}
    \centering
    \includegraphics[width=0.5\textwidth]{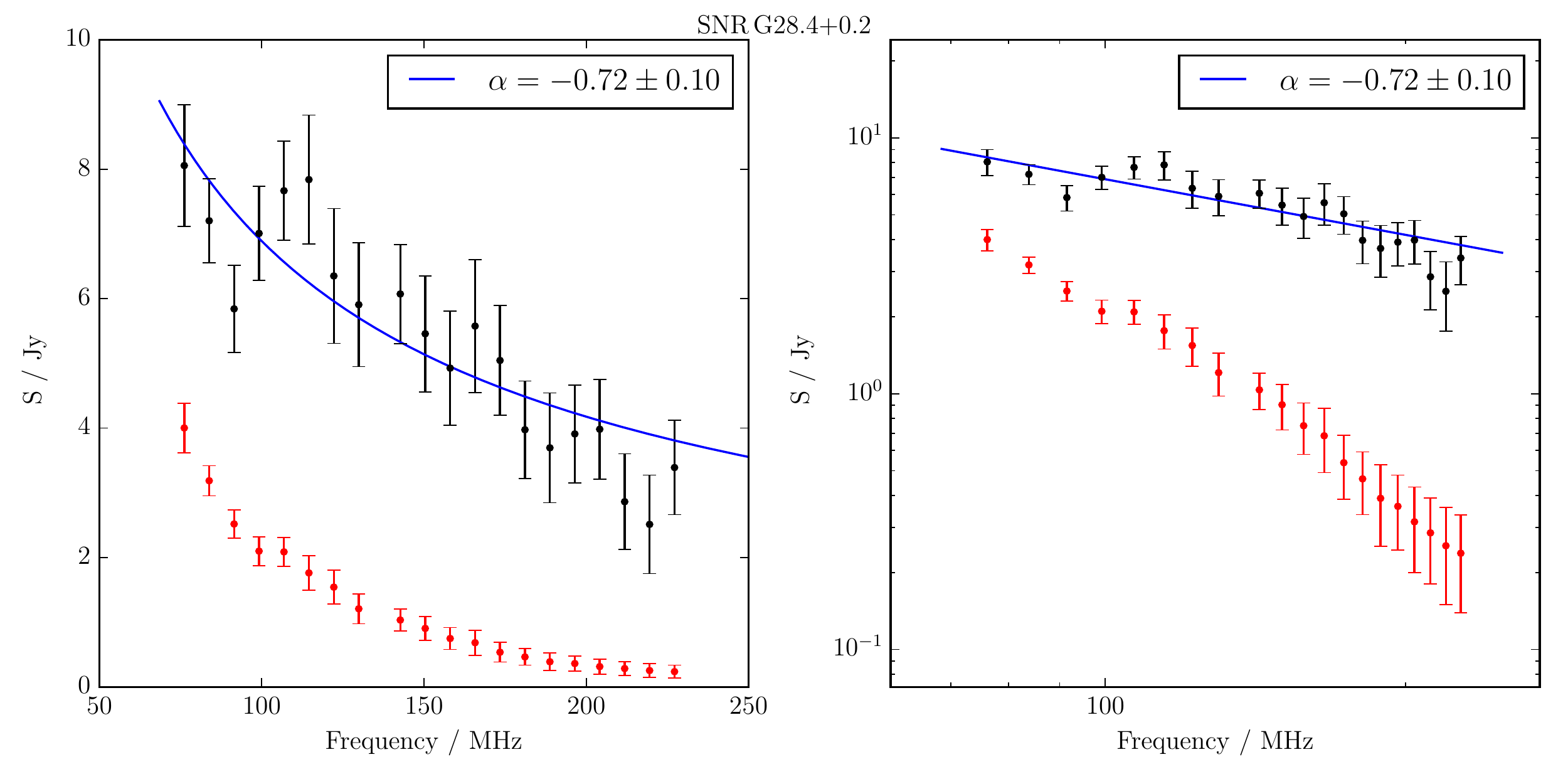}
    \caption{\spectrasummary\spectraN G\,$28.3+0.2$. \spectrasuffix}
    \label{fig:SNR_G28.3+0.2_spectrum}
\end{figure}

\begin{figure}
    \centering
    \includegraphics[width=0.5\textwidth]{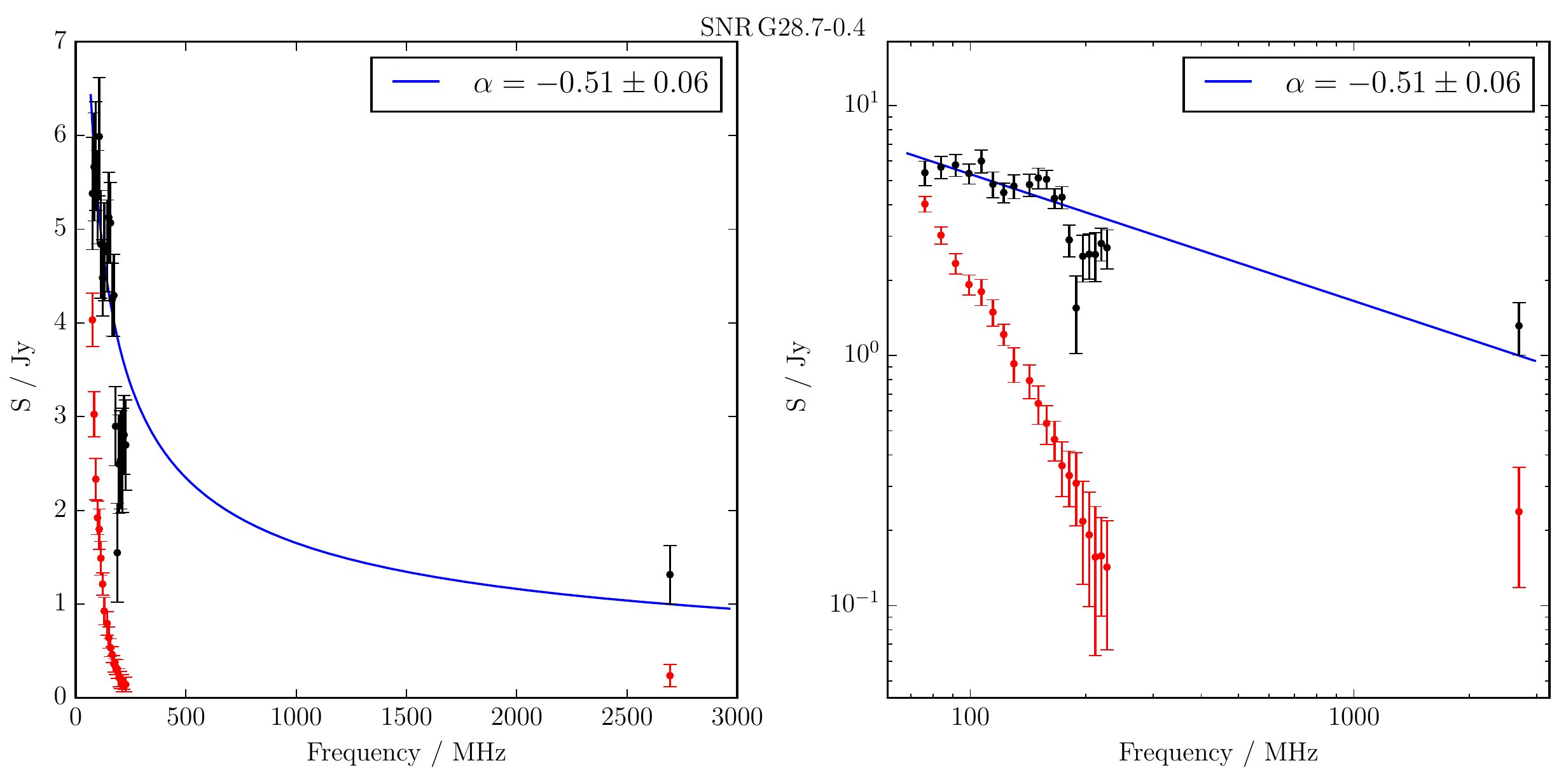}
    \caption{\spectrasummary\spectraE G\,$28.7-0.4$. \spectrasuffix}
    \label{fig:SNR_G28.7-0.4_spectrum}
\end{figure}

\begin{figure}
    \centering
    \includegraphics[width=0.5\textwidth]{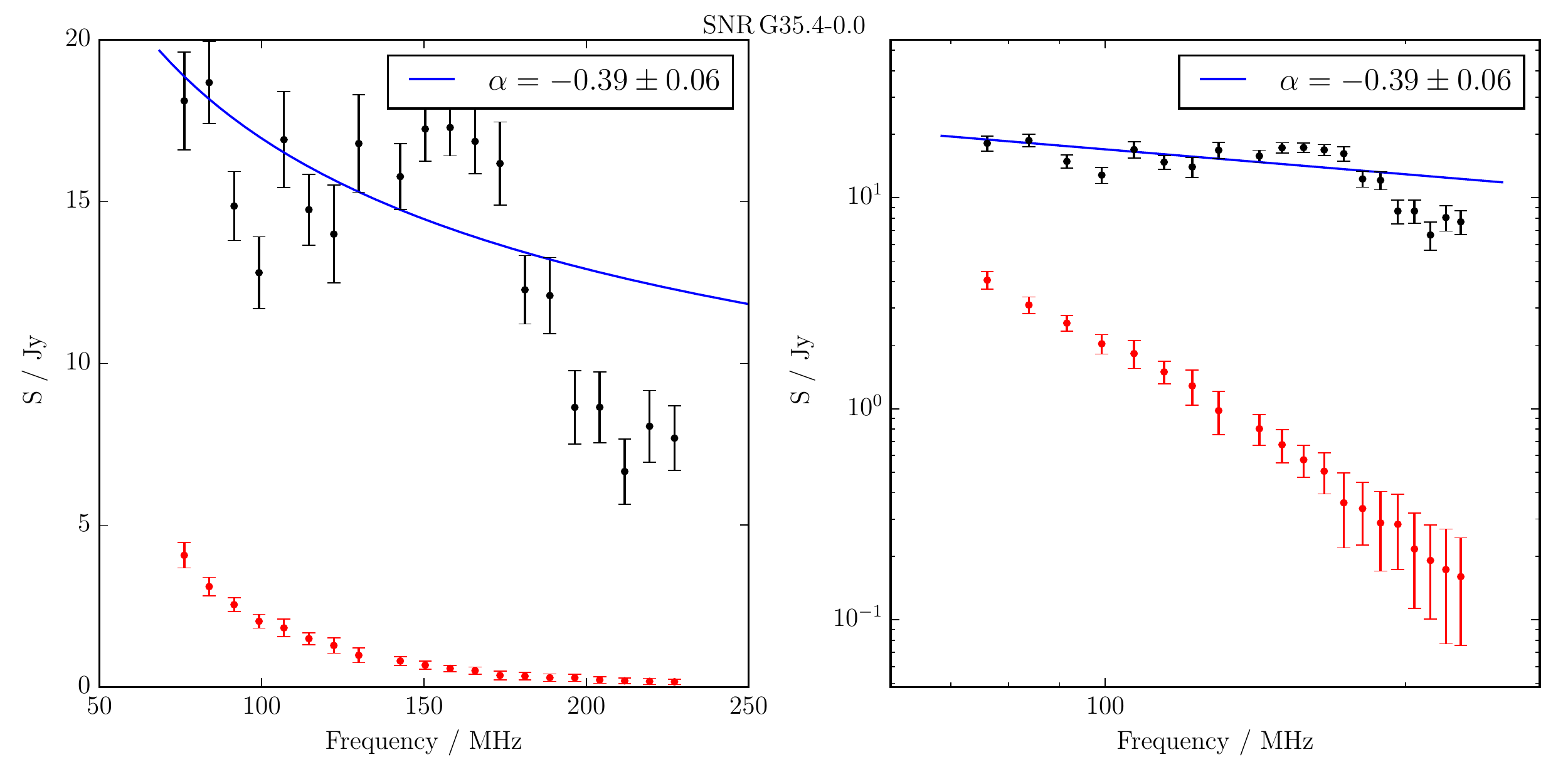}
    \caption{\spectrasummary\spectraN G\,$35.3-0.0$. \spectrasuffix}
    \label{fig:SNR_G35.3-0.0_spectrum}
\end{figure}

\begin{figure}
    \centering
    \includegraphics[width=0.5\textwidth]{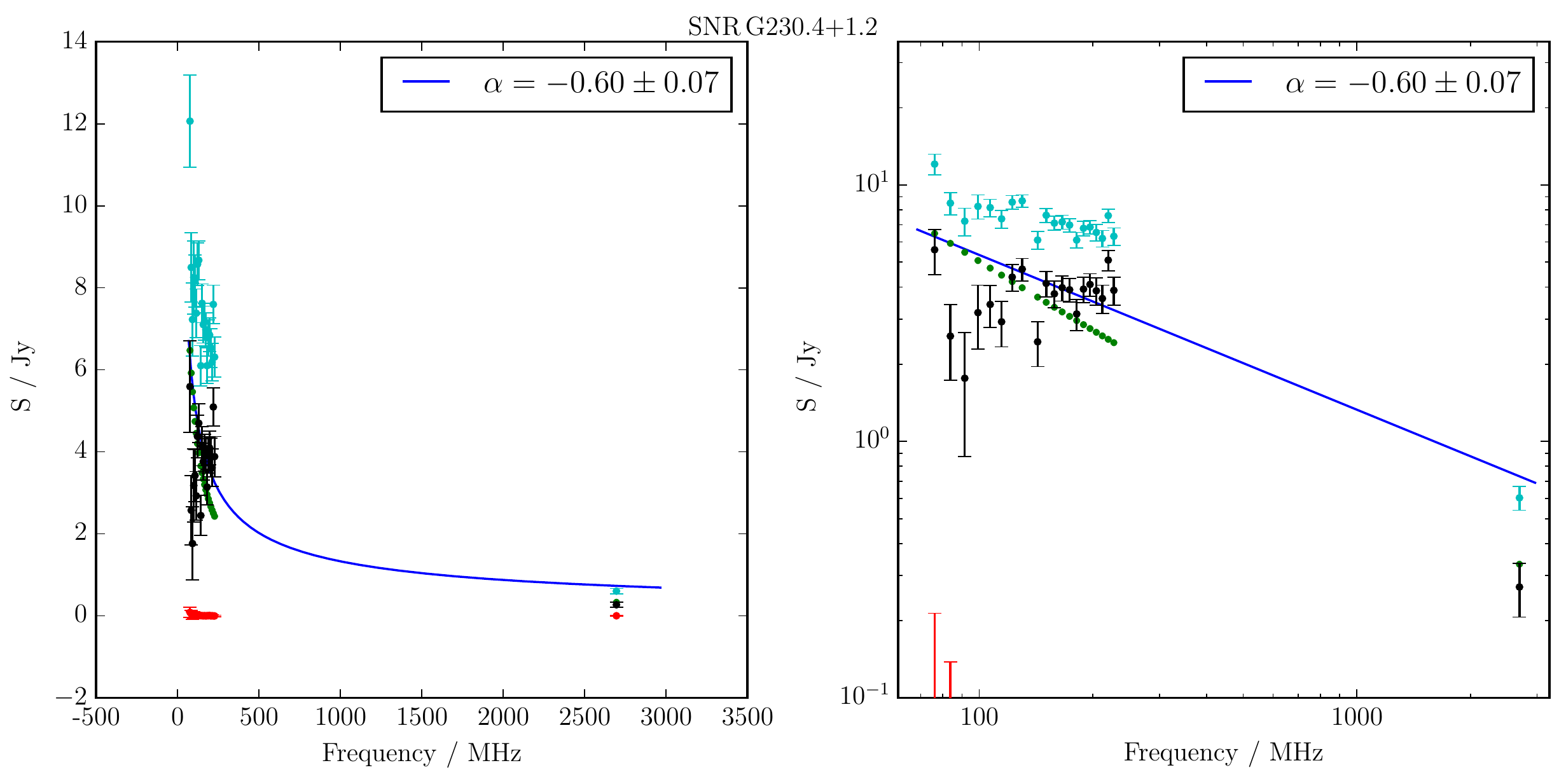}
    \caption{\spectrasummary\spectraE G\,$230.4+1.2$. Cyan points show the SNR flux densities before contaminating sources (green points) were subtracted. \spectrasuffix}
    \label{fig:SNR_G230.4+1.2_spectrum}
\end{figure}

\begin{figure}
    \centering
    \includegraphics[width=0.5\textwidth]{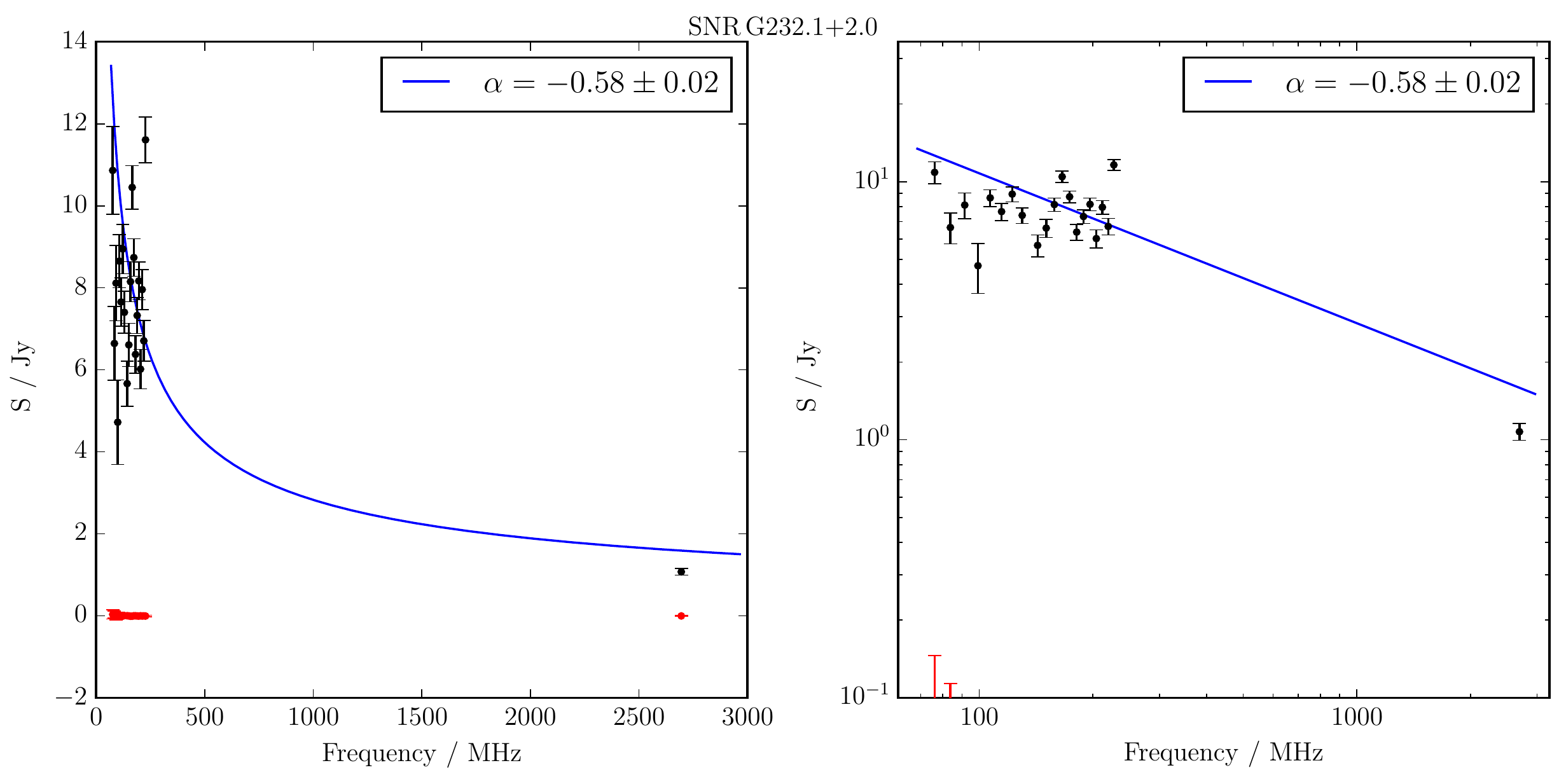}
    \caption{\spectrasummary\spectraE G\,$232.1+2.0$. \spectrasuffix}
    \label{fig:SNR_G232.1+2.0_spectrum}
\end{figure}

\begin{figure}
    \centering
    \includegraphics[width=0.5\textwidth]{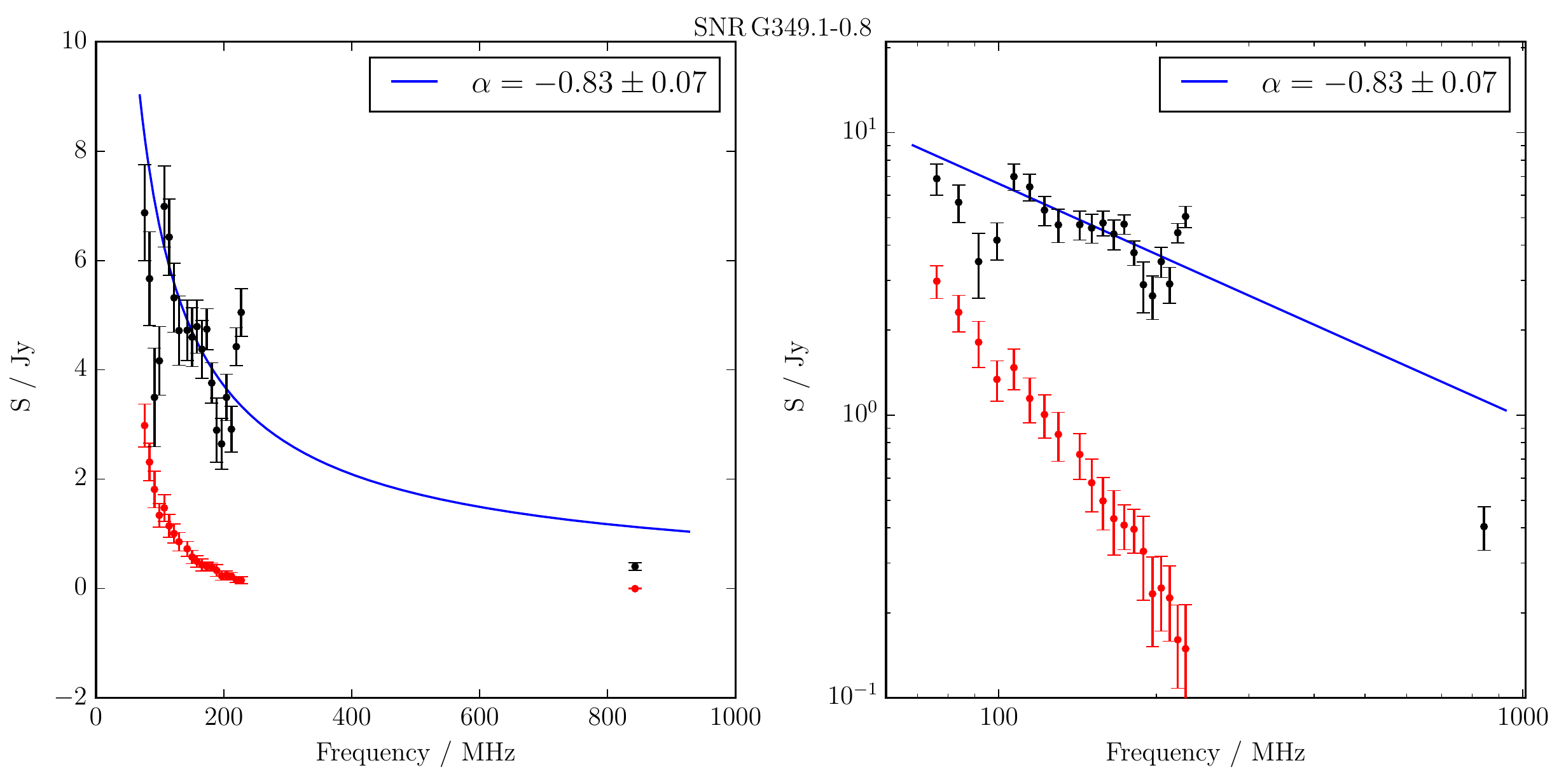}
    \caption{\spectrasummary\spectraM G\,$349.1-0.8$. \spectrasuffix}
    \label{fig:SNR_G349.1-0.8_spectrum}
\end{figure}

\begin{figure}
    \centering
    \includegraphics[width=0.5\textwidth]{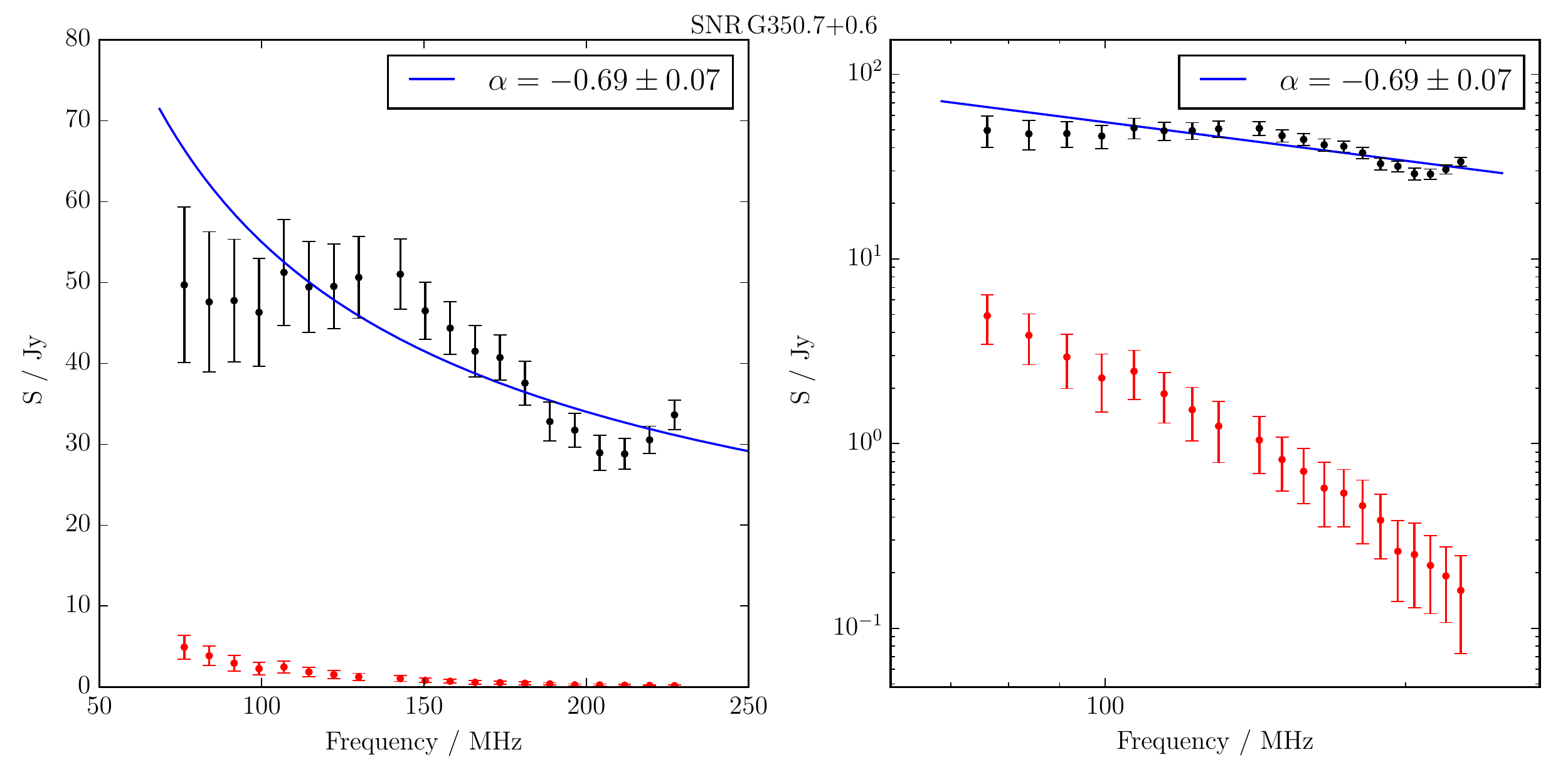}
    \caption{\spectrasummary\spectraA G\,$350.7+0.6$. \spectrasuffix}
    \label{fig:SNR_G350.7+0.6_spectrum}
\end{figure}

\begin{figure}
    \centering
    \includegraphics[width=0.5\textwidth]{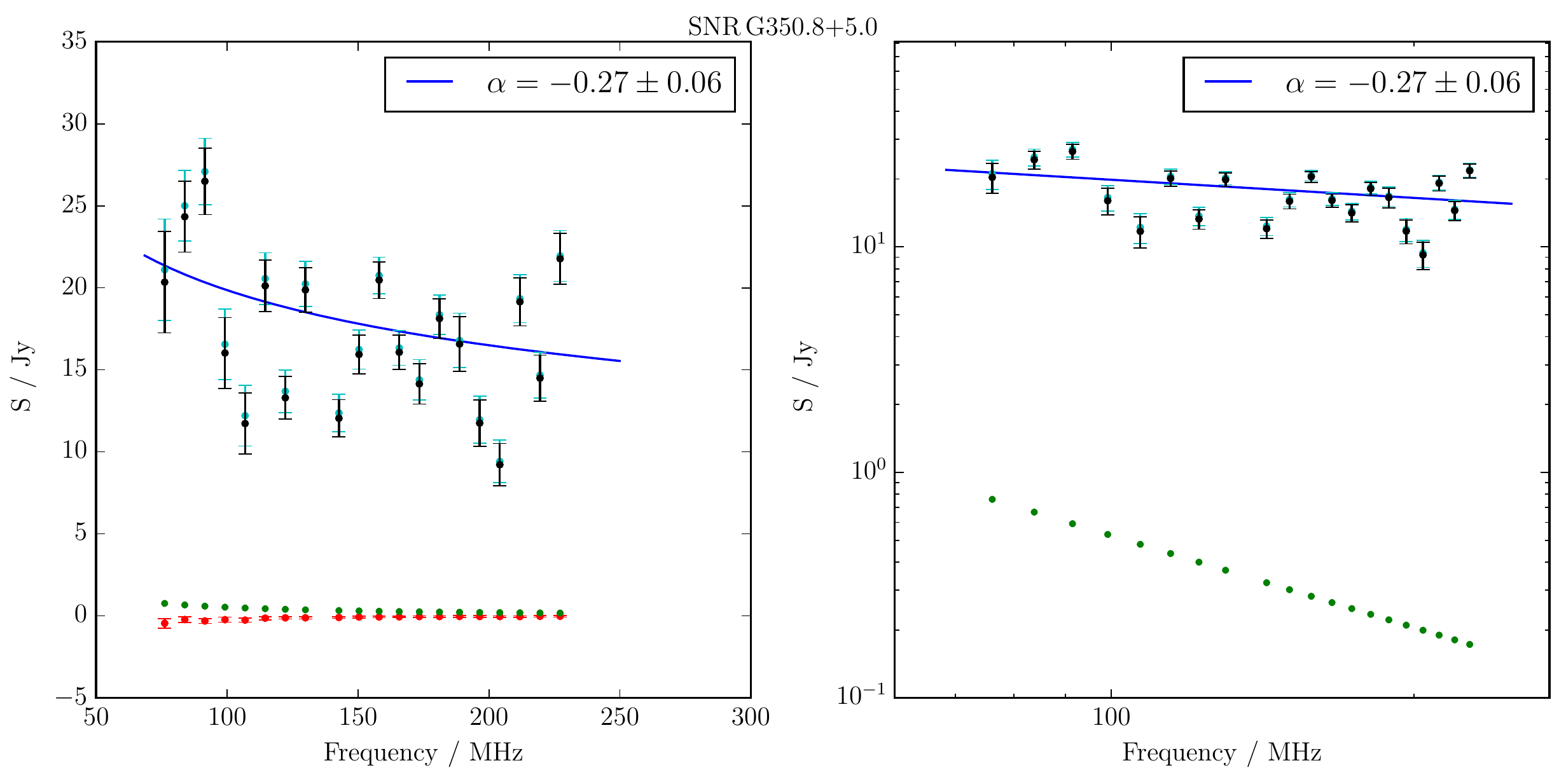}
    \caption{\spectrasummary\spectraN G\,$350.8+5.0$. Cyan points show the SNR flux densities before a contaminating source (green points) was subtracted. \spectrasuffix}
    \label{fig:SNR_G350.8+5.0_spectrum}
\end{figure}

\begin{figure}
    \centering
    \includegraphics[width=0.5\textwidth]{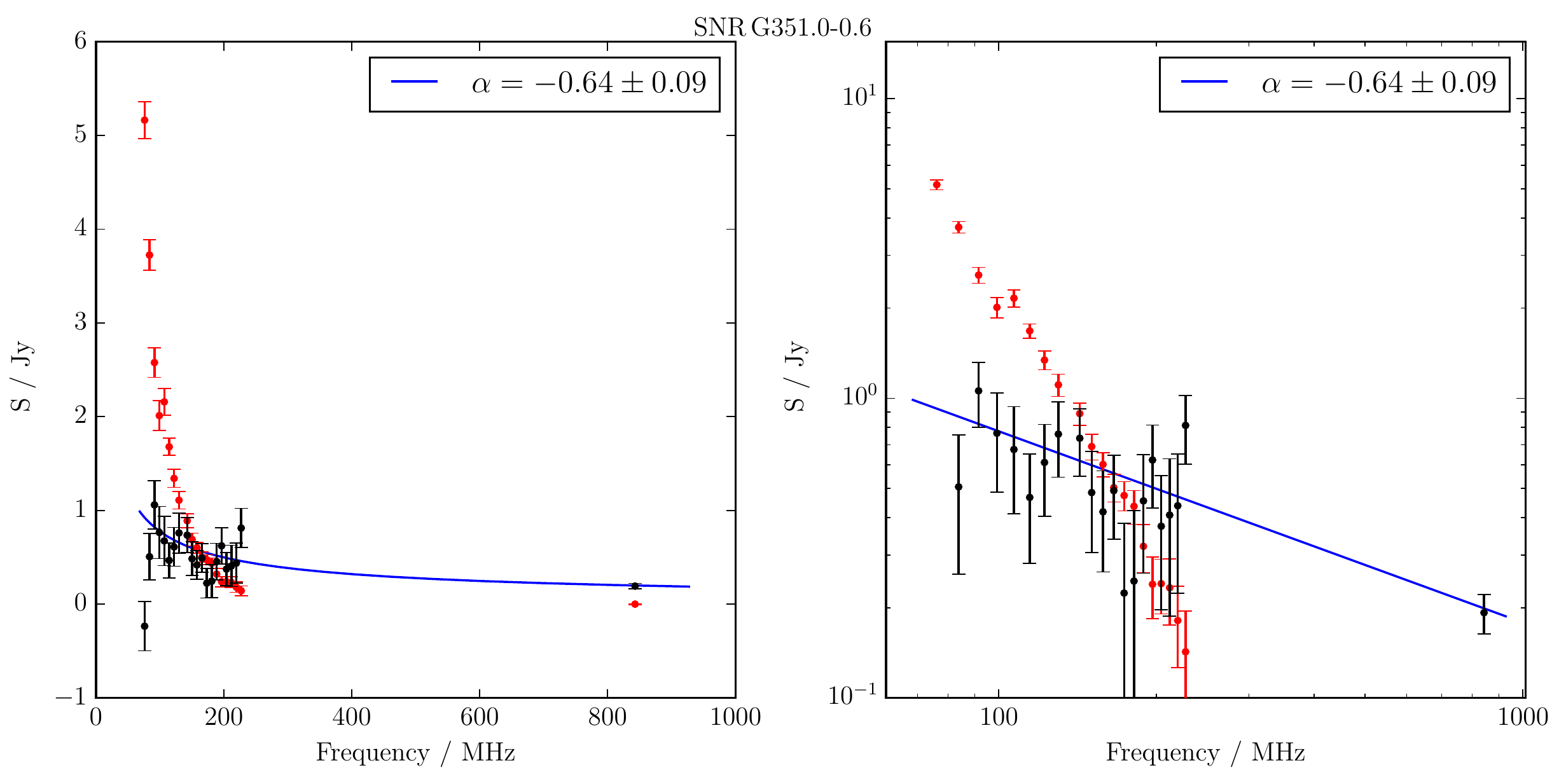}
    \caption{\spectrasummary\spectraM G\,$351.0-0.6$. \spectrasuffix}
    \label{fig:SNR_G351.0-0.6_spectrum}
\end{figure}

\begin{figure}
    \centering
    \includegraphics[width=0.5\textwidth]{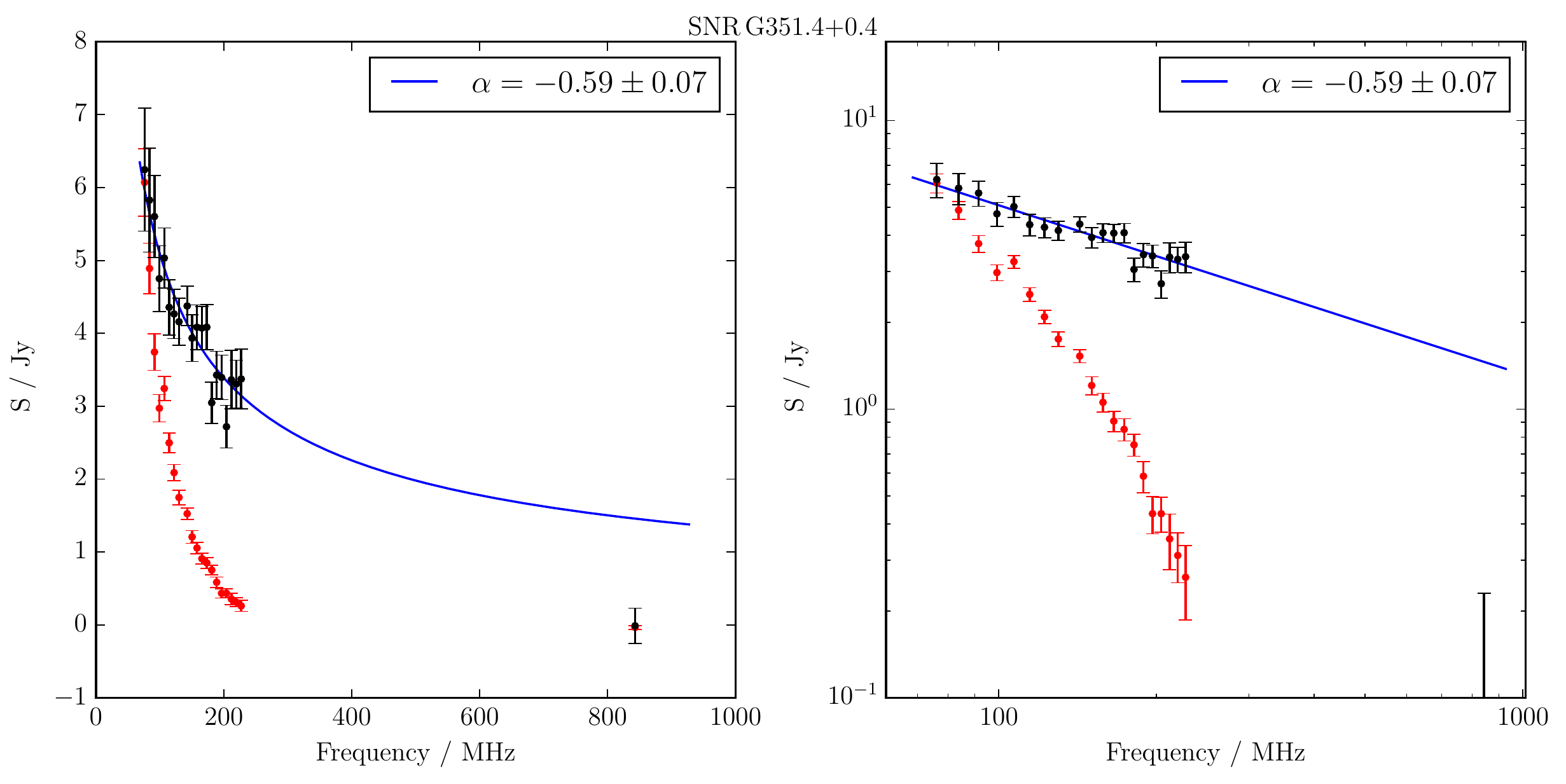}
    \caption{\spectrasummary\spectraM G\,$351.4+0.4$. \spectrasuffix}
    \label{fig:SNR_G351.4+0.4_spectrum}
\end{figure}

\begin{figure}
    \centering
    \includegraphics[width=0.5\textwidth]{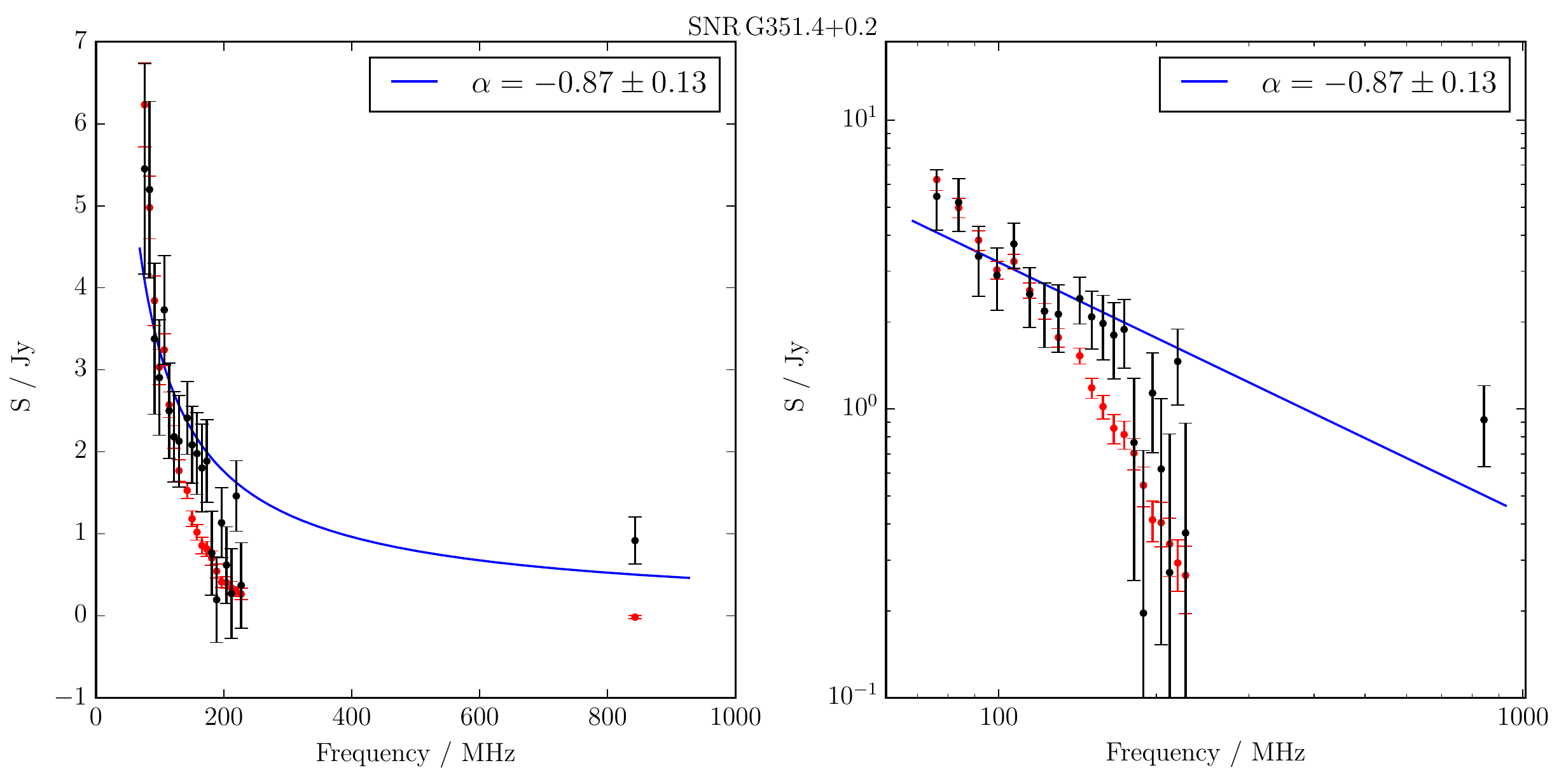}
    \caption{\spectrasummary\spectraM G\,$351.4+0.2$. \spectrasuffix}
    \label{fig:SNR_G351.4+0.2_spectrum}
\end{figure}

\begin{figure}
    \centering
    \includegraphics[width=0.5\textwidth]{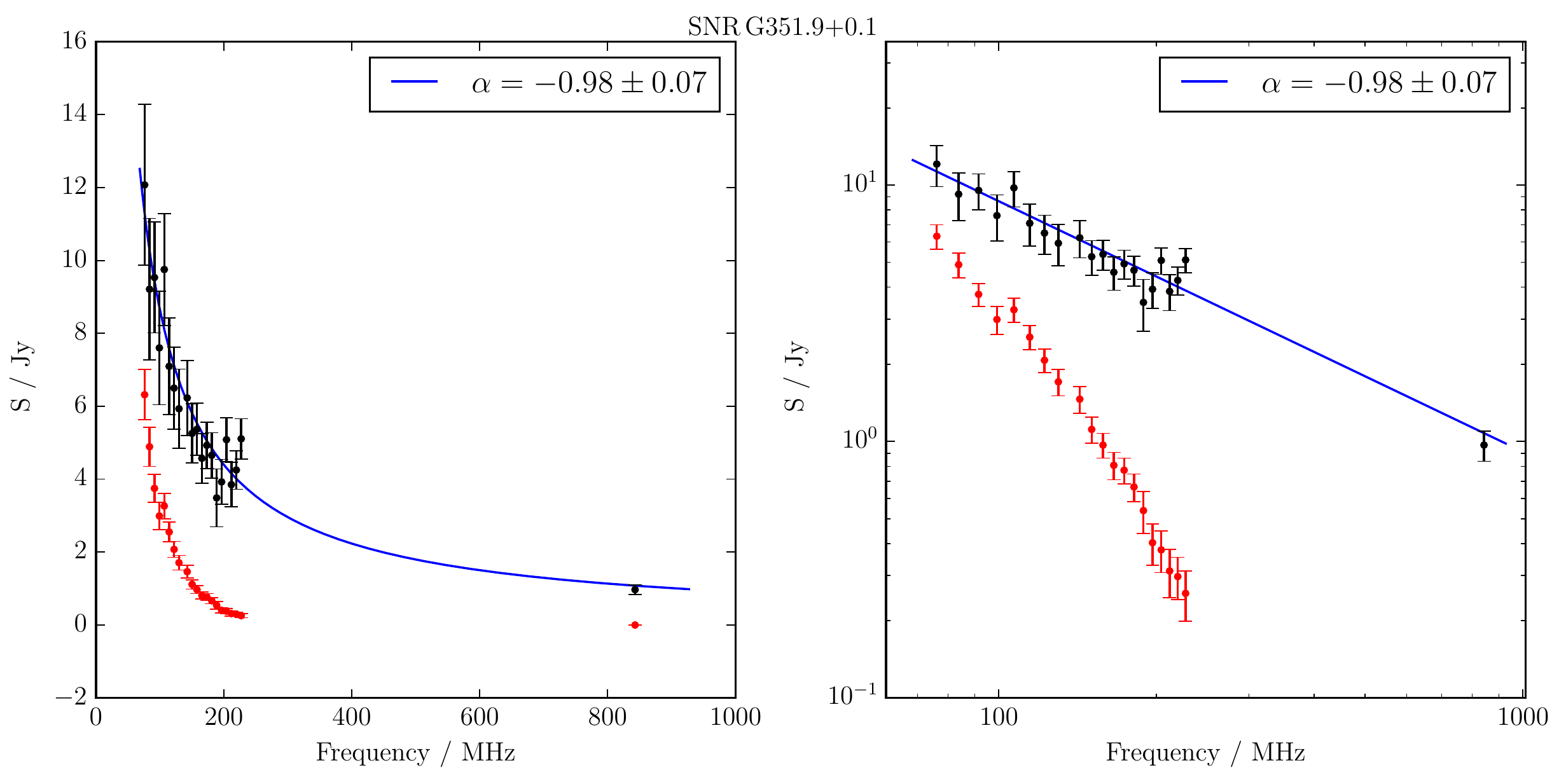}
    \caption{\spectrasummary\spectraM G\,$351.9+0.1$. \spectrasuffix}
    \label{fig:SNR_G351.9+0.1_spectrum}
\end{figure}

\begin{figure}
    \centering
    \includegraphics[width=0.5\textwidth]{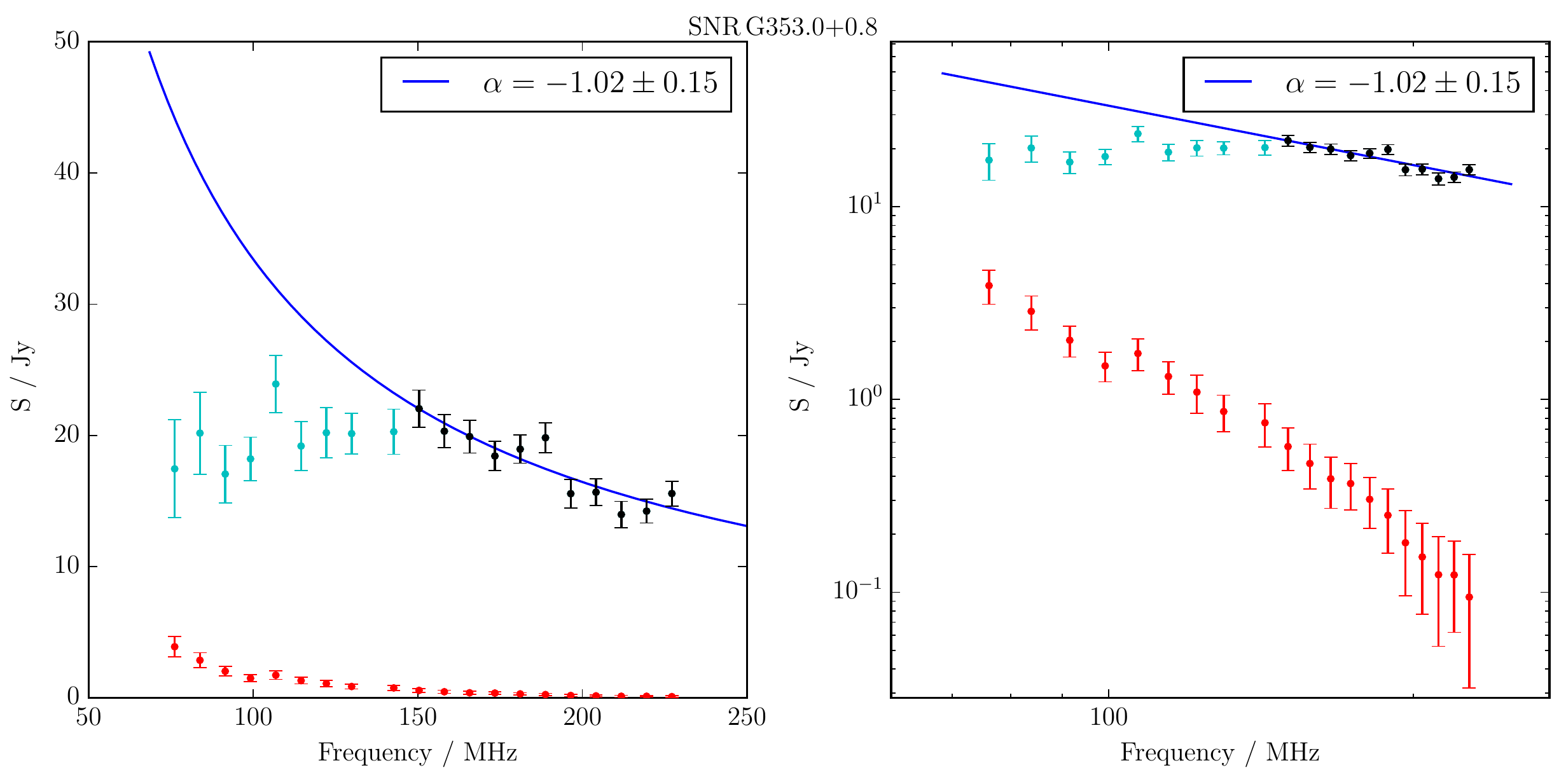}
    \caption{\spectrasummary\spectraA G\,$353.0+0.8$. \spectrasuffix}
    \label{fig:SNR_G353.0+0.8_spectrum}
\end{figure}

\begin{figure}
    \centering
    \includegraphics[width=0.5\textwidth]{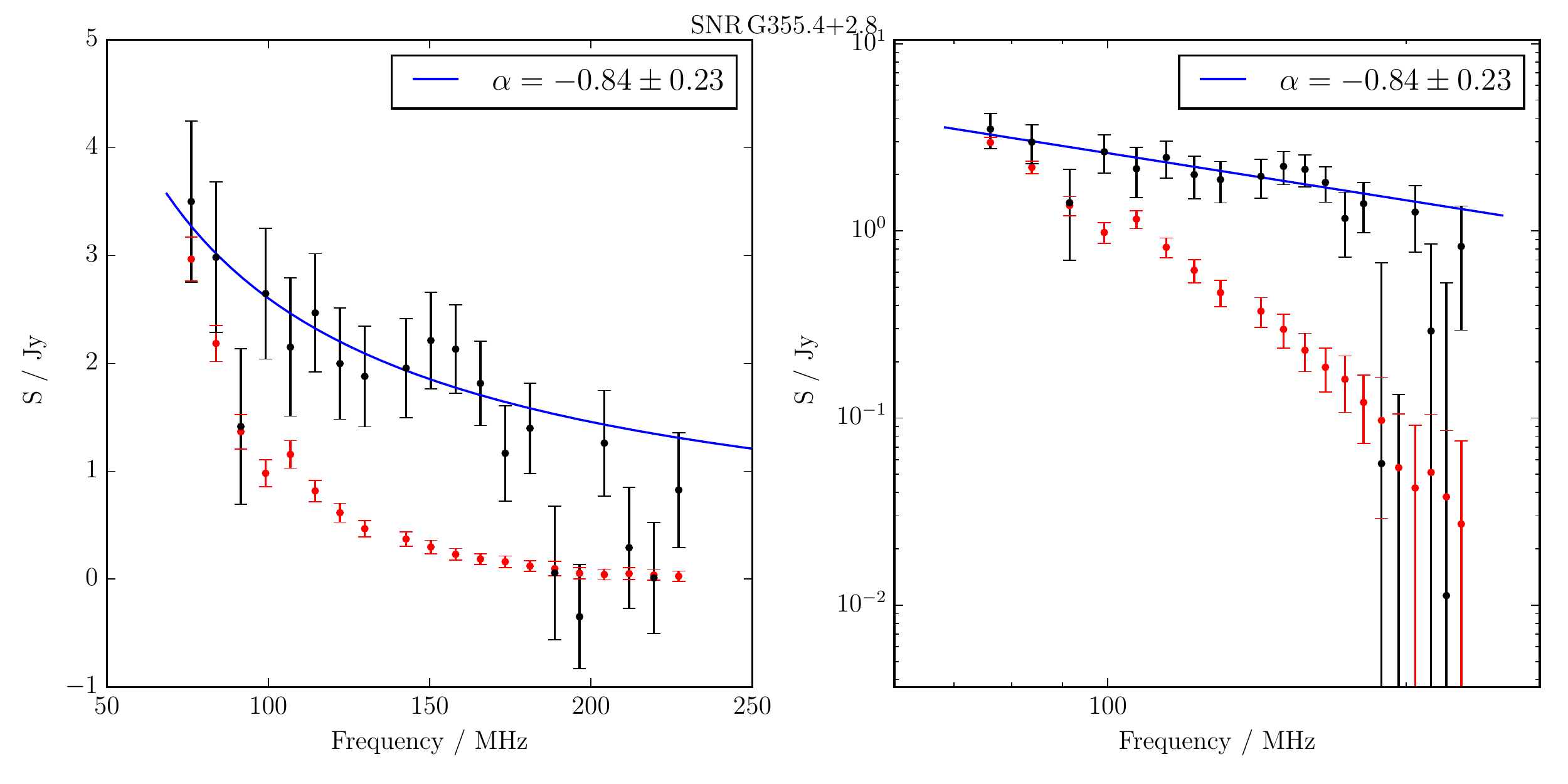}
    \caption{\spectrasummary\spectraN G\,$355.4+2.7$. \spectrasuffix}
    \label{fig:SNR_G355.4+2.7_spectrum}
\end{figure}

\begin{figure}
    \centering
    \includegraphics[width=0.5\textwidth]{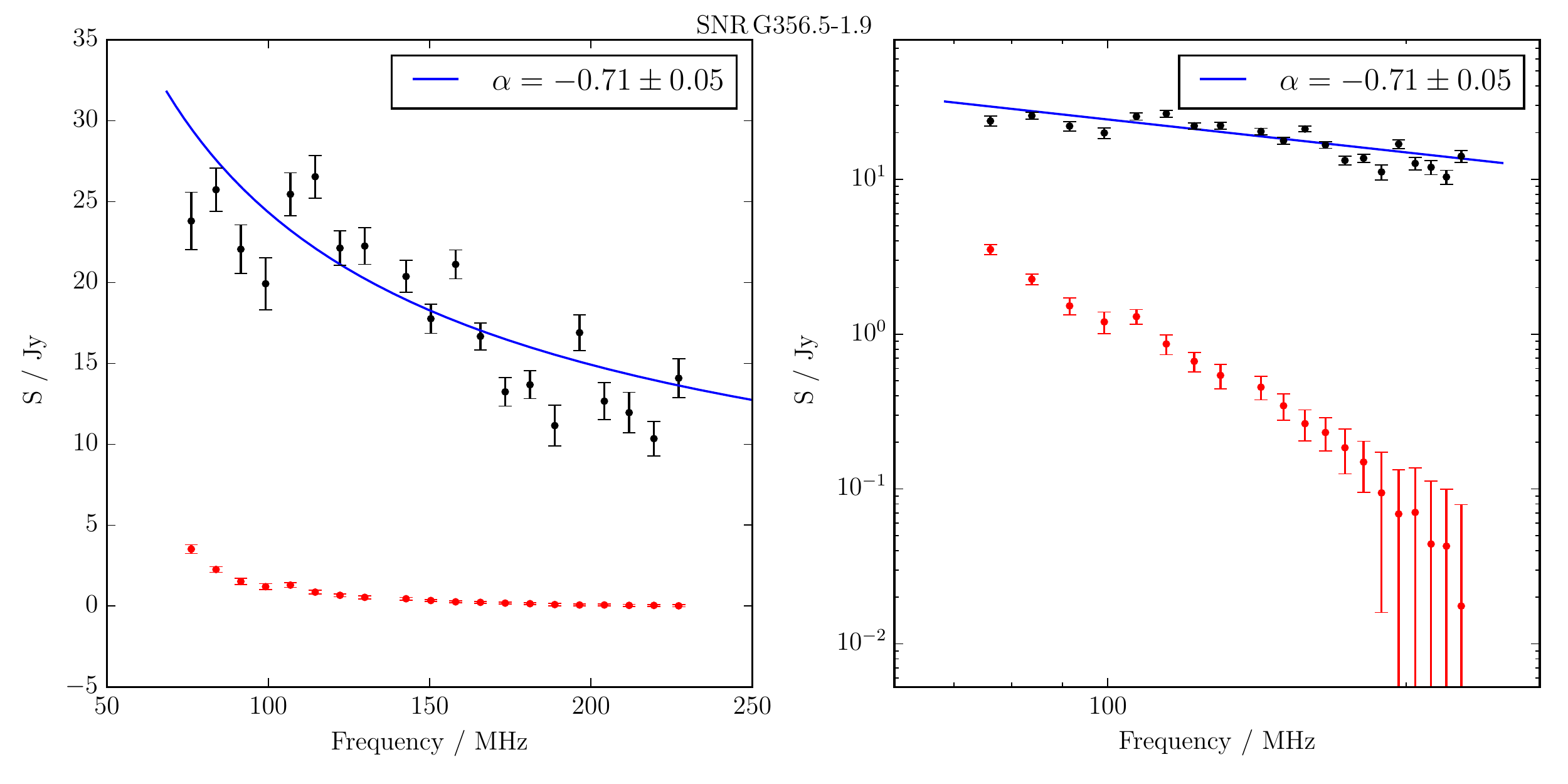}
    \caption{\spectrasummary\spectraN G\,$356.5-1.9$. \spectrasuffix}
    \label{fig:SNR_G356.5-1.9_spectrum}
\end{figure}

\begin{figure}
    \centering
    \includegraphics[width=0.5\textwidth]{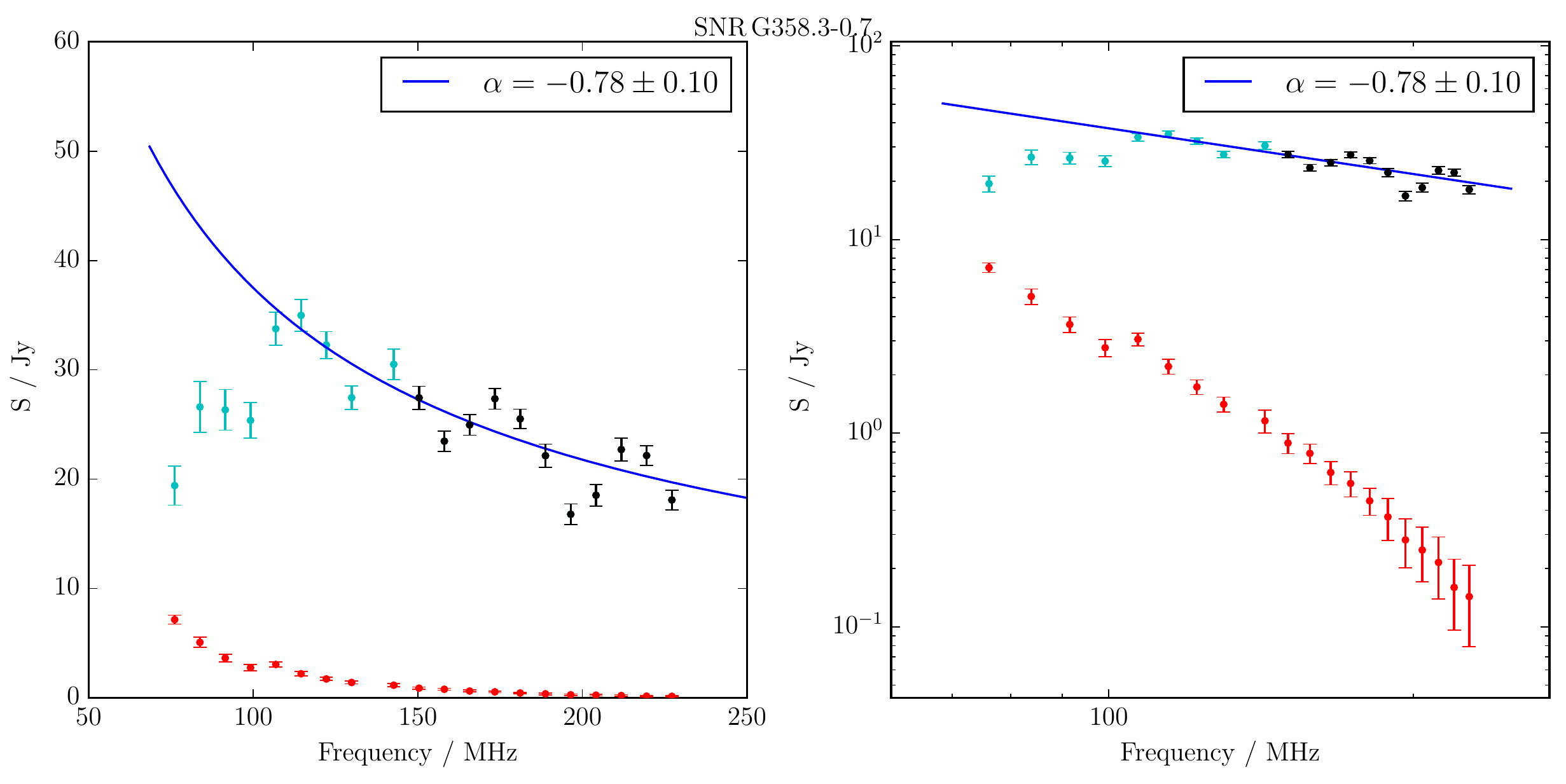}
    \caption{\spectrasummary\spectraA G\,$358.3-0.7$. \spectrasuffix}
    \label{fig:SNR_G358.3-0.7_spectrum}
\end{figure}

\end{appendix}

\bibliographystyle{pasa-mnras}
\bibliography{refs}

\end{document}